\documentclass[10pt,a4paper]{article}
\pdfoutput=1
\usepackage[english]{babel}
\usepackage{longtable}
\usepackage{amsmath,amsfonts,amssymb,bm}\interdisplaylinepenalty=2500
\usepackage{mathrsfs}	% give access to \mathscr{}

\newcommand{\ohm}{\ensuremath{\mathrm{\Omega}}}
\newcommand{\req}[1]{(\ref{#1})}
\newcommand{\sidenote}[1]{\marginpar{\sf\bfseries\small\flushleft#1}} \renewcommand{\sidenote}[1]{}
\usepackage{graphicx,color,rotating,epstopdf}
\graphicspath{{figs/}}					% user defined
\def\PrintGraphicFileName{1}			% 0/1, user defined
\newcommand{\namedgraphics}[3]{\parbox{#3}{\ifnum\PrintGraphicFileName>0\rotatebox{90}{\smash{\ttfamily\scriptsize\raisebox{0.8em}{#2}}}\fi\hspace*{\fill}\includegraphics[#1]{#2}\hspace*{\fill}}}
\newcommand{\ToDay}{April 30, 2010}
\newcommand{\TODAY}{April 30, 2010}
\title{A generalization of the Leeson effect}
\author{Enrico Rubiola and R\'{e}mi Brendel\\
\small web page \texttt{http://rubiola.org}
\\[4em]\includegraphics[width=0.35\textwidth]{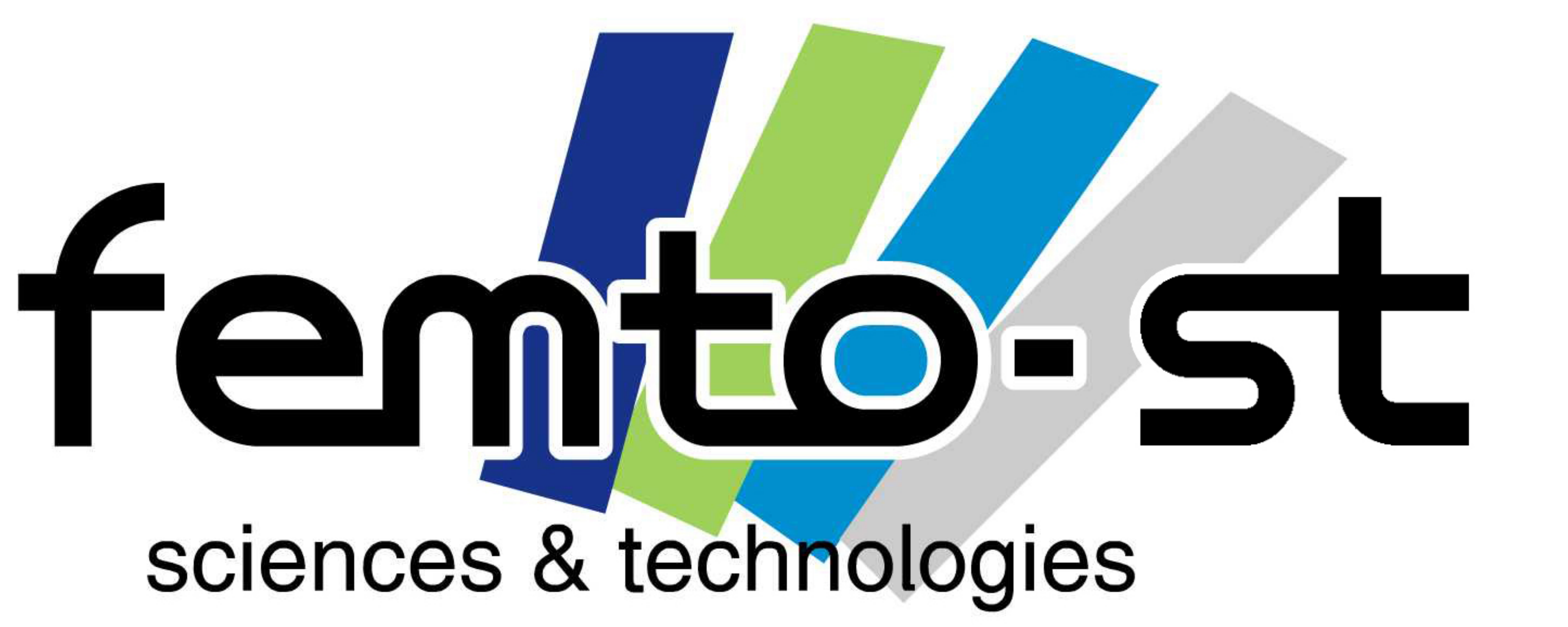}\\[0.5em]
\small FEMTO-ST Institute\\[-0.5ex]
\small CNRS UMR-6174, 
\small Besan\c{c}on, France\\[1.5em]}
\date{\small\TODAY}
\def\pageheads{E. Rubiola \& R. Brendel, Extended Leeson effect \hfill~\ToDay\hfill}
\def\pageheads{E. Rubiola \& R. Brendel, Generalized Leeson effect, \hfill~\ToDay\hfill}
\begin{document}
\maketitle

\begin{abstract}
The oscillator, inherently, turns the phase noise of its internal components into frequency noise, which results into a multiplication by $1/f^2$ in the phase-noise power spectral density.  This phenomenon is known as the \emph{Leeson effect}.  This report extends the Leeson effect to the analysis of amplitude noise.  This is done by analyzing the slow-varying complex envelope, after freezing the carrier.  In the case of amplitude noise, the classical analysis based on the frequency-domain transfer function is possible only after solving and linearizing the complete differential equation that describes the oscillator.  The theory predicts that AM noise gives an additional contribution to phase noise.   Beside the detailed description of the traditional oscillator, based on the resonator governed by a second-order differential equation (microwave cavity, quartz oscillators etc.), this report is a theoretical framework for the analysis of other oscillators, like for example the masers, lasers, and opto-electronic oscillators.

This manuscript is intended as a standalone report, and also as complement to the book E. Rubiola,
\emph{Phase Noise and Frequency Stability in Oscillators},
Cambridge University Press, 2008.
ISBN 978-0-521-88677-2 (hardback), 978-0-521-15328-7 (paperback).

\end{abstract}

\subsection*{Revision history}
\textbf{2010 April}. First submission on arXiv.
\clearpage
\tableofcontents

\clearpage
%---------------------------------------------------------------------
\section*{Notation}
\addcontentsline{toc}{section}{Notation}
\begin{center}
\begin{longtable}[t]{ll}
\bfseries Symbol	& \bfseries Meaning\\\hline
$A$	 & amplifier voltage gain
	(thus, the power gain is $A^2$)\\
$\mathcal{A}(s)$	& normalized-amplitude noise, see $\alpha(t)$\\
$b_i$ & coefficients of the power-law approximation of $S_\varphi(f)$,\\
$\mathrm{b}(t)\leftrightarrow \mathrm{B}(s)$	& resonator impulse response.  Also $\mathrm{b}_\varphi(t)\leftrightarrow \mathrm{B}_\varphi(s)$, etc.\\
$\mathcal{E}(s)$	& normalized-amplitude noise, see $\varepsilon(t)$\\
$f$		& Fourier frequency, Hz\\
$f_c$		& amplifier corner frequency, Hz\\
$f_L$		& Leeson frequency, Hz\\
$F$		& amplifier noise figure\\
$h(t)\leftrightarrow H(s)$& impulse response\\
$\mathrm{h}(t)\leftrightarrow \mathrm{H}(s)$ & phase or amplitude impulse response, 
	takes subscript $\varphi$ or $\alpha$\\
$h_i$	& coefficients of the power-law model of $S_\alpha(f)$ or $S_y(f)$\\
		& (in case of ambiguity use $h_i$ for $S_y(f)$ and $h'_i$ for $S_\alpha(f)$)\\
$j$		& imaginary unit, $j^2=-1$ \\
$k$		& Boltzmann constant, $k=1.381{\times}10^{-23}$ J/K\\
%$k_\text{(subscr)}$	& a constant, $k_d$, $k_o$, $k_L$, etc.\\
$\mathcal{L}\{\,\}$	& Laplace-transform operator\\ 
$\mathscr{L}(f)$	& single-sideband noise spectrum, dBc/Hz.\\
				& $\mathscr{L}(f)=\frac12S_\varphi(f)$, by definition \\ 
$\mathcal{N}(s)$	& gain fluctuation, see $\eta(t)$\\
%$\mathcal{N}$, $\mathcal{N}(s)$ &\\
$Q$		& resonator quality factor\\
$R$, $R_0$& resistance, load resistance (often, $R_0=50$ \ohm)\\
$s$		& Laplace complex variable, $s=\sigma+j\omega$\\
$s$		& derivative operator\\
$S_a(f)$	& one-sided power spectral density (PSD) of the quantity $a$\\ 
$t$		& time\\
$T$, $T_0$& absolute temperature, reference temperature $T_0=290$ K\\
$\mathfrak{u}(t)$	& Heaviside (step) function, $\mathfrak{u}(t)=\int\delta(t')\,dt'$\\
$u(t)\leftrightarrow U(s)$	& voltage (amplifier input)\\
$v(t)\leftrightarrow V(s)$	& voltage.  Also a dimensionless signal \\
$\tilde{v}(t)$	& complex-envelope associated to $v(t)$ \\
$V$, $V_0$ & dc or peak voltage\\
$\mathbf{V_o}$, $\mathbf{V_o}(t)$	& phasor associated with an ac signal $v(t)$\\
%$x(t)$	& phase time fluctuation.\\
$x(t)$	& generic function\\
$y(t)$	& generic function\\
$y(t)$	& fractional frequency fluctuation, $y(t)=[\nu(t)-\nu_0]/\nu_0$\\[1em]
$\alpha(t)\leftrightarrow\mathcal{A}(s)$	& normalized-amplitude noise, may take subscript $u$ or $v$\\
$\beta(s)$	& transfer function of the feedback path\\
$\gamma$& amplifier compression parameter ($0<\gamma<1$)\\
$\delta(t)$	& Dirac delta function\\
$\varepsilon(t)\leftrightarrow\mathcal{E}(s)$	& normalized-amplitude noise\\
$\eta(t)\leftrightarrow\mathcal{N}(s)$& amplifier gain fluctuation\\
$\kappa$	& small phase or amplitude step\\
$\nu$	& frequency (Hz), used for carriers\\
%$\sigma_y(\tau)$& Allan deviation, square root of the Allan variance.\\
%$\tau$	& measurement time, in $\sigma_y(\tau)$\\
$\sigma$	& real part of the Laplace variable $s=\sigma+j\omega$\\  
$\tau$	& resonator relaxation time\\
$\varphi(t)\leftrightarrow\Phi(s)$ & phase noise\\
$\chi$	& dissonance, $\chi=\omega/\omega_n-\omega_n/\omega$\\
$\psi(t)\leftrightarrow\Psi(s)$& phase noise (amplifier)\\
$\omega$	& imaginary part of the Laplace variable $s=\sigma+j\omega$\\  
$\omega$	& angular frequency, carrier or Fourier\\  
$\omega_0$& oscillator angular frequency\\  
$\omega_n$& resonator natural angular frequency\\  
$\omega_p$& resonator free-decay angular pseudo-frequency\\  
$\Omega$	& detuning angular frequency, $\Omega=\omega_0-\omega_n$	\\[1em]
\bfseries Subscript	& \bfseries Meaning\\\hline
$0$	& oscillator carrier, as in
	$\omega_0$, $P_0$, $V_0$, etc.\\
$i$	& input, as in $v_i(t)$, $\varphi_i(t)$, $\Phi_i(s)$\\
$i$	& electrical current, as in the shot noise $S_i(\omega)=2q\overline{i}$\\
$n$	& resonator natural frequency ($\omega_n$, $\nu_n$)\\
$o$	& output, as in $v_o(t)$, $\varphi_o(t)$, $\Phi_o(s)$\\
$p$	& resonator free-decay pseudo-frequency 
		($\omega_p$, $\nu_p$)\\[1em]
\bfseries Symbol	& \bfseries Meaning\\\hline
$<~>$	& mean.  Also $<~>_N$ mean of $N$ values\\
$\overline{~~}$	& time average, as in $\overline{x}$\\
$\leftrightarrow$& transform inverse-transform pair,
	as in $x(t)\leftrightarrow X(s)$\\
$\ast$	& convolution, as in $v(t)=h(t)\ast u\leftrightarrow V(s)=H(s)U(s)$\\
$\asymp$	& asymptotically equal\\
\boldmath{\footnotesize$\bigcirc$}\unboldmath
	& zero of a function (complex plane, Bode plot, or spectrum)\\
\boldmath$\times$\unboldmath 
	& pole of a function (complex plane, Bode plot, or spectrum)\\[1em]
\bfseries Acronym	& \bfseries Meaning\\\hline
AM	& Amplitude Modulation (often `AM noise')\\
CAD	& Computer-Aid Design (software)\\
FM	& Frequency Modulation (often `FM noise')\\
PM	& Phase Modulation (often `PM noise')\\
PSD	& (single-side) Power Spectral Density\\
RF	& Radio Frequency 
	
\end{longtable}
\end{center}

\clearpage
%---------------------------------------------------------------------
\section{Introduction}\label{sec:ele-introduction}
%==================================================
The oscillator noise, which in the absence of environmental or aging effect is cyclostationary, is best described as a baseband process, after freezing the periodic oscillation.  
The polar-coordinate representation of the limit cycle splits the model of the oscillator into two subsystems, in which all signals are the amplitude and the phase of the main system, respectively.  Putting things simply, these two subsystems are (almost) decoupled and all the nonlinearity goes to amplitude.  This occurs because amplitude nonlinearity is necessary for the oscillation to be stationary.  Conversely the phase, which ultimately is time, cannot be stretched.

The baseband equivalent of a resonator, either for phase or amplitude is a lowpass filter whose time constant is equal to the resonator's relaxation time.  Hence, the phase model of the oscillator consists of an amplifier of gain exactly equal to one and the lowpass filter in the feedback path, as extensively discussed in \cite{Rubiola2008cambridge-leeson-effect}.  The amplitude model is a nonlinear amplifier, whose gain is equal to one at the stationary amplitude and decreases with power, and the lowpass filter in the feedback path.  In the baseband representation both AM and PM perturbations map into additive noise, even in the case of flicker and other parametric noises.  
This model gives a new perspective on the classical van der Pol oscillator.

The elementary theory of nonlinear differential equations tells us that nonlinearity stretches the feedback time constant.  Asymptotically, the time constant is split into two constants, one at the oscillator startup and one in stationary conditions.  If the gain varies linearly with amplitude, which is always true for small perturbations, the oscillator can be solved in closed form. 

After the pioneering work of D. B. Leeson \cite{Leeson1966pieee}, a number of different analyses has been published.  Sauvage derived a formula that has the same behavior of the Leeson formula using the autocorrelation functions \cite{Sauvage1977im-Leeson}.   Hajimiri and Lee \cite{Hajimiri-Lee1998jssc-oscillator-noise,Hajimiri-Lee1998jssc-errata,Lee00jssc-oscillator-noise-tutorial} proposed a model based on the ``impulse-sensitivity function'' (ISF), which emphasizes that the impulse has the largest effect on phase noise if it occurs at the zero-crossing of the carrier.  This model, mainly oriented to the description of phase noise in CMOS circuits, is extended in \cite{Hegazi-Rael-Abidi:high-purity-oscillators}.
Demir \& al.\ proposed a theory based on the stochastic calculus \cite{Demir2000tcs1}, in which they introduce a decomposition of phase and amplitude noise through a projection onto the periodic time-varying eigenvectors (the Floquet eigenvectors), by which they analyze the oscillator phase noise as a stochastic-diffusion problem.  This theory was extended to the case of $1/f$ noise \cite{Demir2002tcs1}.
Other articles are mainly oriented to the microwave oscillators \cite{Nallatamby2003mtt-Leeson,Nallatamby2003mtt-Leeson-extension,Nallatamby2005mtt-nonlinear-oscillator}.
Demir inspired a work on the phase noise in opto-electronic oscillators \cite{Chembo2009jqe-oeo}.  Some of the articles cited make use of  sophisticated mathematics, as compared to our simple methods.  All give little or no  attention to amplitude noise.  

It is well known that the instability of the resonator natural frequency contributes to the oscillator fluctuations, which in some cases turns out to be the most important source of frequency fluctuations.  Almost nothing is known about the amplitude fluctuation of the resonator.   That said, the resonator instability is not considered here.  This article stands upon our earlier works \cite{Rubiola2008cambridge-leeson-effect} and \cite{brendel1999uffc}.  The latter is mainly oriented to the ultra-stable quartz oscillator.  Here, we present an unified approach to AM and PM noise in oscillators by analyzing the mechanism with which the noise of the oscillator internal components is transferred to the output.
This theory turns out to be particularly suitable to high stability oscillators, based on high quality-factor quartz resonators, microwave whispering gallery resonators, etc.  This work is easily extended to the delay-line oscillator, including the opto-electronic oscillator \cite{Yao1996josab-oeo,volyanskiy2008josab-optical-fiber}.

\section{Basics}\label{sec:ele-basics}
%==================================================
\subsection{Phase noise}
%---------------------------------------------------------------------------------------
Phase noise is a well established subject, clearly explained in classical references, among which we prefer \cite{rutman78pieee,kimball:precise-frequency,ccir90rep580-3,ieee99std1139} and \cite[vol.\,1, chap.\,2]{vanier:frequency-standards}.  The reader may also find useful \cite[chap.\,1]{Rubiola2008cambridge-leeson-effect}.

The quasi-perfect sinusoidal signal of angular frequency\footnote{We use interchangeably 
$\omega$ as a shorthand for $2\pi\nu$ for the carrier frequency, and as a shorthand for  $2\pi f$ for the offset (Fourier) frequency, making the meaning clear with appropriate subscripts when needed but omitting the word `angular.'}  $\omega_0$, of random amplitude fluctuation $\alpha(t)$, and of random phase fluctuation $\varphi(t)$ is
\begin{align}
v(t)=\left[1+\alpha(t)\right]\cos\left[\omega_0t+\varphi(t)\right]
\qquad\text{(clock signal)}~.
\label{eqn:ele-clock-signal}
\end{align}
We may need that $|\alpha(t)|\ll1$ and $|\varphi(t)|\ll1$ or $|\dot{\varphi}(t)|\ll1$ during the measurement.
The phase noise is generally measured as the average PSD (power spectral density)
\begin{align}
S_{\varphi}(f) = \left<|\Phi(jf)|^2\right>_m
&&\text{(avg, $m$ spectra)},
\end{align}
The uppercase denotes the Fourier transform, so $\varphi(t)\leftrightarrow\Phi(jf)$ form a transform inverse-transform pair.
In experimental science, the single-sided PSD is preferred to the two-sided PSD because the negative frequencies are redundant.
It has been found that the power-law model describes accurately the oscillator phase noise
\begin{align}
&S_{\varphi}(f) = \sum_{n=-4}^{0}b_nf^n \qquad\qquad\text{(power law)}
\label{eqn:ddl-power-law-Sphi}\\
&\text{\begin{tabular}{cl}
coefficient & noise type\\\hline
$b_{-4}$ & frequency random walk\\
$b_{-3}$ & flicker of frequency\\
$b_{-2}$ & white frequency noise, or phase random walk\\
$b_{-1}$ & flicker of phase\\
$b_{0}$ & white phase noise
\end{tabular}}\nonumber
\end{align}
The power law relies on the fact that white ($f^0$) and flicker ($1/f$) noises exist per-se, and that phase integration is present in oscillators, which multiplies the spectrum $\times1/f^2$.  This will be discussed extensively in Section~\ref{sec:ele-leeson}.  Additional terms with $n<-4$ show up at very low frequency.

The two-port components are also described by the power-law \req{eqn:ddl-power-law-Sphi}.  Yet at very low frequency the dominant terms cannot be steeper than $f^{-1}$, otherwise the input-output delay would diverge.  Nonetheless, these steeper terms may show up in some regions of the spectrum, due to a variety of phenomena.

\subsection{Amplitude noise}
%---------------------------------------------------------------------------------------

Amplitude noise, far less studied than phase noise, can be described in very similar way.  The reader may find useful the report \cite{rubiola2005arxiv-am-noise}, which describes in depth the experimental aspects.

The amplitude noise is generally measured as the average PSD
\begin{align}
S_{\alpha}(f) = \left<|\mathcal{A}(jf)|^2\right>_m
&&\text{(avg, $m$ spectra)}~,
\end{align}
where $\alpha(t)\leftrightarrow\mathcal{A}(j\omega)$ form a transform-inverse-transform pair, and described in terms of power law
\begin{align}
&S_{\alpha}(f) = \sum_{n=-4}^{0}h_nf^n \qquad\qquad\text{(power law)}
\label{eqn:ddl-power-law-Salpha}\\
&\text{\begin{tabular}{cl}
coefficient & noise type\\\hline
$h_{-3}$ & (integrated flicker)\\
$h_{-2}$ & amplitude random walk\\
$h_{-1}$ & flicker of amplitude\\
$h_{0}$ & white amplitude noise
\end{tabular}}\nonumber
\end{align}
In the current literature the coefficients $h_i$ are used to model $S_y(f)$, i.e., the PSD of the fractional-frequency fluctuation $y(t)$.  We use the same coefficients for simplicity, because the relationships between spectrum and two-sample (Allan) variance are the same.

At very low frequency, the amplitude noise cannot be steeper than $f^{-1}$ otherwise the amplitude would diverge.  This applies to both two-port components and oscillators.
Yet, steeper terms can show up in some regions of the spectrum.

\subsection{Oscillator fundamentals}
\begin{figure*}[t]
\centering\namedgraphics{scale=0.55}{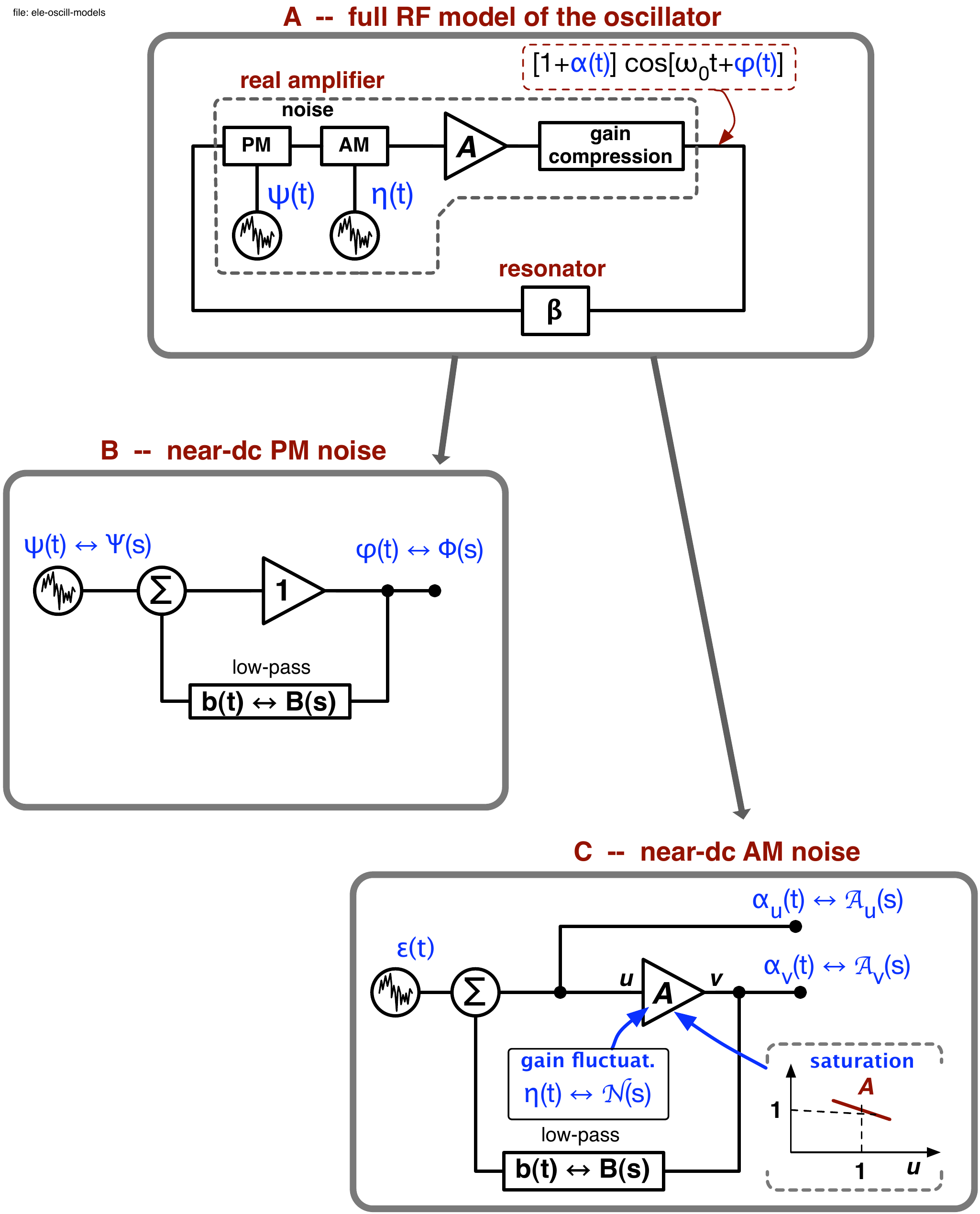}{\textwidth}
\caption{Feedback oscillator and its decomposition in PM and AM models.}
\label{fig:ele-oscill-models}
\end{figure*}

%---------------------------------------------------------------------------------------
The simplest form of oscillator is a resonator with an amplifier of gain $A$ in closed loop that compensates for the resonator loss\footnote{Since the quantity $\beta$ is the resonator gain, $1/\beta$ is  the loss.} $1/\beta$.  Stationary oscillation takes place at the frequency $\omega_0$ that verify $A\beta=1$.  This is known as the Barkhausen condition.   The actual oscillator can be represented with the scheme of Fig.~\ref{fig:ele-oscill-models}, which includes a gain compression mechanism and noise.  For our purposes the noise is represented in polar coordinates as amplitude noise and phase noise.  The gain compression is necessary for the amplitude not to decay or diverge.   We assume that $A$ is independent of frequency, at least in a range sufficiently larger than the resonator bandwidth.
For the sake of simplicity we normalize the loop elements so that $A=1$ and $\beta=1$ at the oscillation frequency $\omega=\omega_0$ and at the nominal output amplitude $v=1$.

The resonator has narrow bandwidth, hence it eliminates all the harmonics multiple of $\omega_0$ generated by the amplifier nonlinearity.  Though the harmonics can be present at the amplifier output, where the signal can be distorted, they do not participate to the regeneration process that entertains the oscillation.  Hence, the only practical effect of the nonlinearity on the loop dynamics is to reduce the gain at the fundamental frequency $\omega_0$.   
Therefore, the quasi-sinusoidal approximation can be used.

Assuming that, as it occurs in practice, the resonator relaxation time $\tau$ is larger than $1/\omega_0$ by a factor of at least $10^2$, the oscillator behavior can be mathematically described in terms of the slow-varying complex envelope, as amplitude and phase were decoupled from the oscillation.
In this representation the oscillator splits into two subsystems, one for phase and one for amplitude, as shown in Fig.~\ref{fig:ele-oscill-models}.  Since phase represents time, which cannot be stretched\footnote{This is no longer true in extreme nonlinear oscillators, like the femtosecond laser, which are out of our scope.}, all the non-linearity goes in the amplitude subsystem.

The main advantage of the slow-varying envelope representation is that amplitude noise and phase noise can be represented as additive noise phenomena, regardless of the physical origin.  This eliminates the difficulty of flicker noise and other parametric processes.  The formalism is simple and tightly connected to the experimentally observable quantities. 

We are interested in the mechanism that governs the noise propagation of the internal components to the oscillator output.  
Virtually all oscillators are stable enough for the noise to be a small perturbation to the stationary oscillations, and consequently for a linear model to be accurate for any practical purpose.   Linearization gives access to the Laplace-Heaviside formalism.  The response\footnote{Here $x(t)$ and $y(t)$ are generic functions of time, thus \emph{not} the phase time and the fractional frequency fluctuation commonly used in the oscillator literature.} $y(t)$ to the input $x(t)$ is therefore given by
\begin{align*}
y(t) = x(t)*h(t)  \quad\leftrightarrow\quad Y(s)=X(s)H(s)
\end{align*}
where $h(t)$ is the \emph{impulse response}, i.e., the response to the Dirac $\delta(t)$ function, $H(s)$ is the \emph{transfer function}, the symbol `$*$' is the convolution operator, the double arrow `$\leftrightarrow$' stands for Laplace transform inverse-transform pair, and $s=\sigma+j\omega$ is the Laplace complex variable.     Given the input power spectral density $S_x(f)$, the output power spectral density is given by
\begin{align*}
S_y(f) = |H(jf)|^2\,S_x(f)~.
\end{align*}
The application of this idea to the oscillator rises some difficulties, which will be solved in the next Sections.

\section{Amplifier saturation and noise}\label{sec:ele-amplifiers}
%==================================================
\subsection{Gain compression}\label{ssec:ele-saturation-models}
%-----------------------------------------------------------------------------------
\begin{figure}[t]
\centering\namedgraphics{scale=0.5}{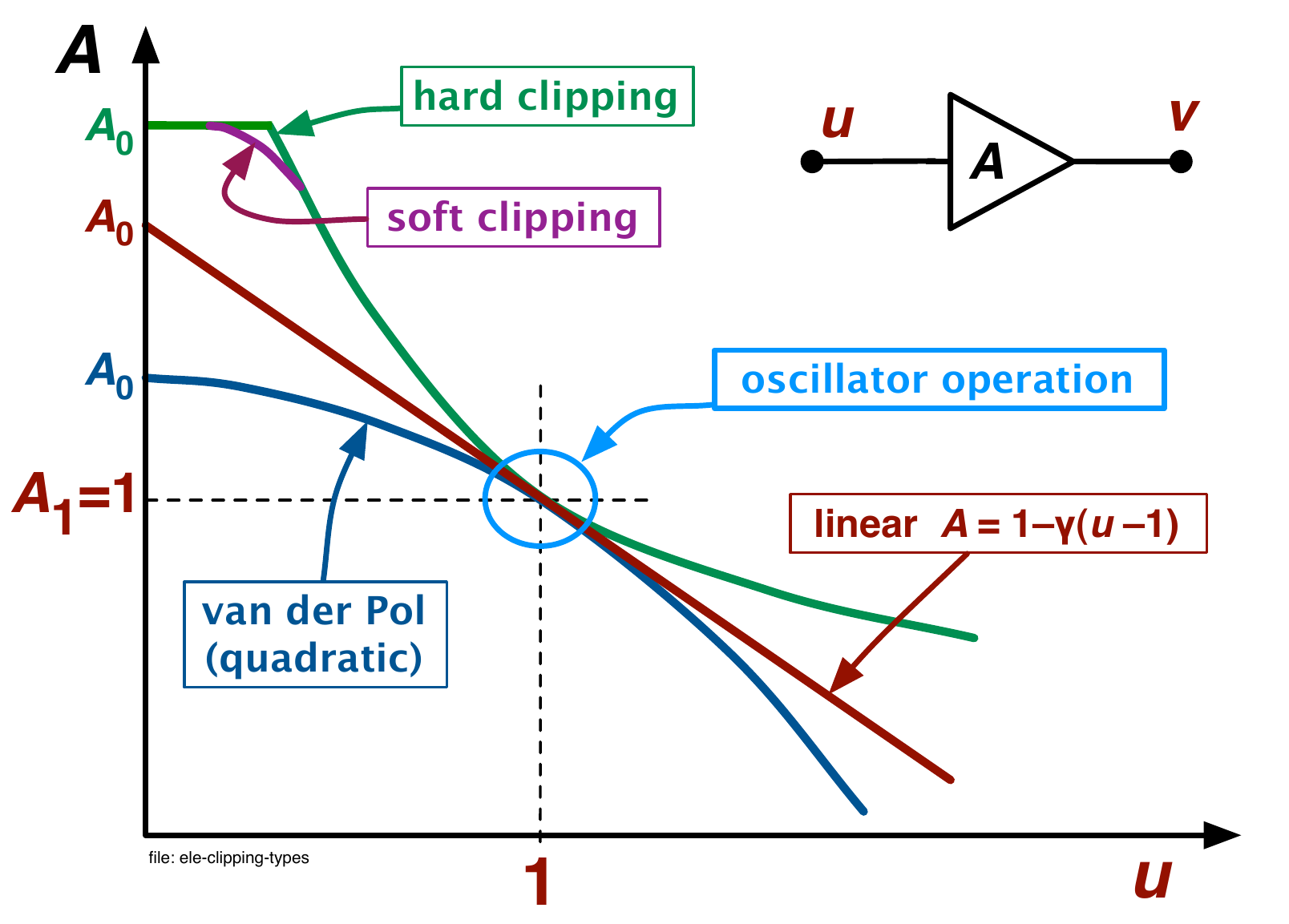}{\columnwidth}
\caption{Most common types of gain saturation.  The quantities $u$ and $v$ are the rms amplitude at the carrier frequency.}
\label{fig:ele-clipping-types}
\end{figure}
In large signal conditions, all amplifiers have some kind of nonlinearity that limits the maximum output power.  Neglecting the band limitation, when a sinusoidal signal $U_\text{rf}(t)=U_1\cos(\omega_0t)$ is present at the input of an amplifier, the saturated output can be written as the Fourier series $V_\text{rf}(t)=\sum_{n=1}^\infty V_n\cos(n\omega_0t)$.  The $n=1$ term is the fundamental and the $n>1$ terms are the harmonics generated by nonlinearity.  The effect of the band limitation is to change (in most cases to reduce) the amplitudes $V_n$ and to introduce a phase in each sinusoidal term.  In a linear amplifier only the fundamental is present at the output, thus $V_n=0$, $\forall n\ge2$.

In the case of the oscillator, the resonator allows only the fundamental to be fed back to the input, for the harmonics can be neglected.  Hence, the oscillation amplitude is described using the slow-varying signals $u$ and $v$ instead of the instantaneous peak amplitudes $U_1$ and $V_1$.  Let us define the amplifier gain as 
\begin{align*}
A = \frac{v}{u} \qquad\text{(definition of $A$)}~, 
\end{align*}
which of course is equivalent to $A=V_1/U_1$.  The gain $A$ should \emph{not} be mistaken for the differential gain $\partial v/\partial u$.

Figure \ref{fig:ele-clipping-types} shows the gain-saturation types most frequently encountered and described underneath.  The small-signal gain is denoted with $A_0$ and the gain at the oscillator nominal amplitude $u=1$ is denoted with $A_1$.   Figure \ref{fig:ele-clipping-types} is normalized for $A_1=1$.   
Around $u=1$ the gain can be linearized as 
\begin{align*}
A = A_1\left[1-\gamma(u-1)\right] \qquad\text{whith}\qquad0\le\gamma<1~, 
\end{align*}
which rewrites as
\begin{align}
A = 1-\gamma(u-1) \qquad\text{whith}\qquad0\le\gamma<1
\label{eqn:ele-gain-linearized}
\end{align}
after normalizing for $A_1=1$.

The slope $-\gamma$ deserves some comments.  The condition $\gamma>0$ is obvious because the gain $A(u)$ must decrease monotonically (Fig.~\ref{fig:ele-clipping-types}).  This is the amplitude-stabilization mechanism.  A second obvious condition is that in the regular-operation region (i.e., around $u=1$) the output $v(u)$ must increase monotonically.
We show that this second condition is equivalent to $\gamma<1$ by substituting \req{eqn:ele-gain-linearized} in $v=Au$ 
\begin{figure}[t]
\centering\namedgraphics{scale=0.5}{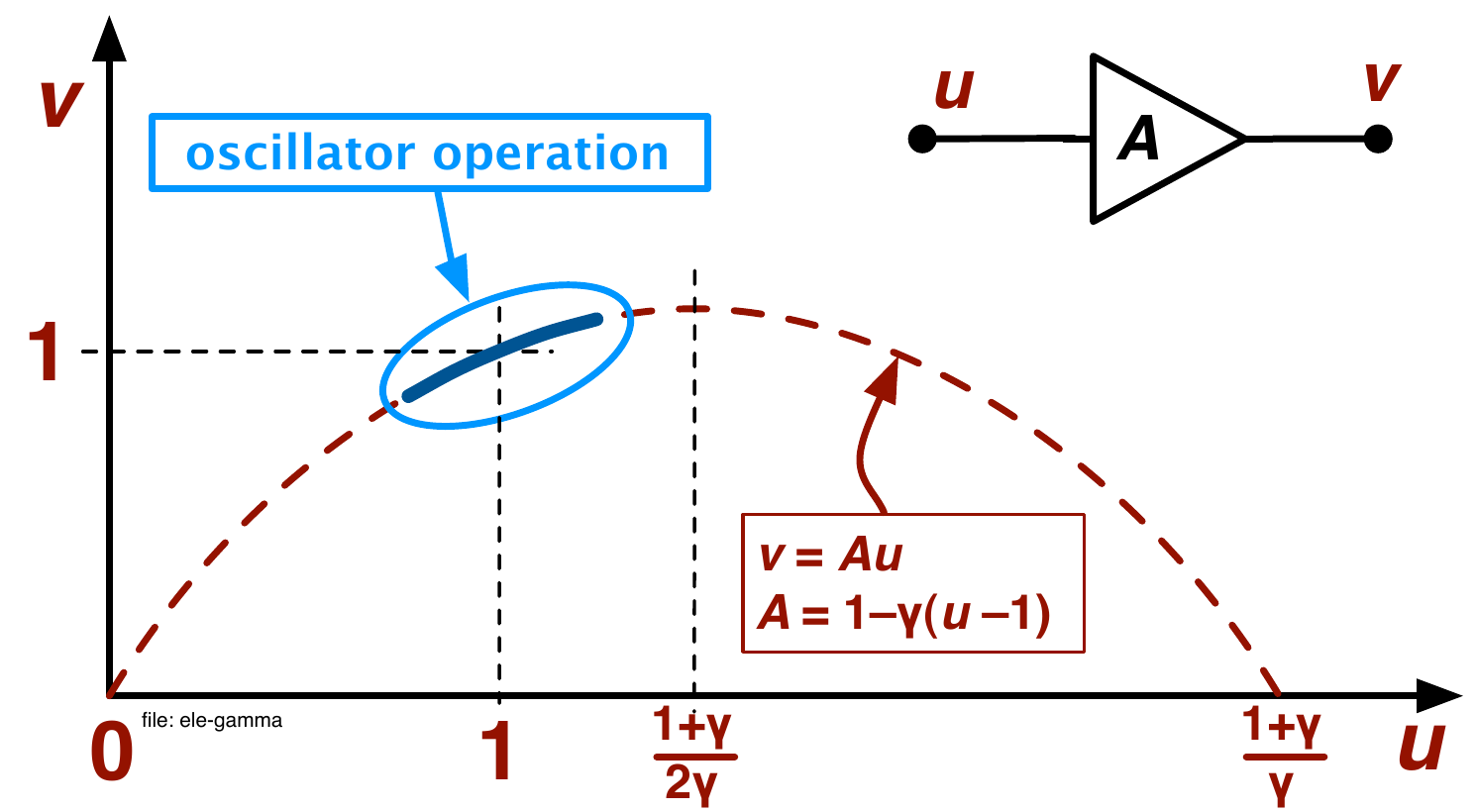}{\columnwidth}
\caption{Most common types of gain saturation.  The quantities $u$ and $v$ are the rms amplitude at the carrier frequency.}
\label{fig:ele-gamma}
\end{figure}
\begin{align*}
v = -\gamma u^2 + (1-\gamma) \qquad\text{(Fig.~\ref{fig:ele-gamma})}~,
\end{align*}
The latter is the `cap' parabola shown in Fig.~\ref{fig:ele-gamma}.
For $v(u)$ to increase monotonically at $u=1$ it is necessary that the point $u=1$ is on the left-side of the maximum, thus 
\begin{align*}
\frac{1-\gamma}{2\gamma}>1~,
\end{align*}
whose solution is $\gamma<1$.

The property that $v(u)$ increases monotonically holds for virtually all amplifiers.  Only a few exceptions are found, the most remarkable of which is a channel that includes a Mach Zehnder electro-optic modulator.  In such cases the Barkhausen $A\beta=1$ may be met (at least) at two different amplitude levels, the first with $\gamma<1$ and the second with $\gamma>1$.  In one of such cases, it has been mathematically proved and experimentally observed that the oscillation amplitude flips between these two levels, producing an amplitude oscillation at half the frequency determined by the loop roundtrip time \cite{chembo2007oll-oeo}.

\subsection{Gain saturation in real amplifiers}

\subsubsection{Quadratic (van der Pol) amplifier}  In the classical van der Pol oscillator \cite{vanderpol1927nature}, the amplifier  input-output function is defined as $y=a_1x-a_3x^3$, with $a_1>0$ and $a_3>0$.  In mathematical treatises the coefficients $a_1$ and $a_3$ are sometimes set to one.
Feeding the signal $U_\text{rf}(t)=U_1\cos(\omega_0t)$ in such amplifier and taking only the fundamental frequency, the output is $V_\text{rf}(t)=U_1\left[a_1-a_3\frac34U_1^2\right]\cos(\omega_0t)$.  Accordingly, the gain becomes $A=a_1-a_3\frac34U_1^2$, which is a `cap' parabola. 

\subsubsection{Hard-clipping amplifier} In small-signal condition the gain is $A_0$, independent of the signal level.  Increasing the input level the output is clipped when it hits a threshold, where the sinusoid progressively turns into a square wave.  The asymptotic amplitude of the fundamental is $4/\pi$ (2.1 dB) higher than the threshold.  
This behavior is often encountered in amplifiers linearized by a strong feedback, as most circuits based on operational amplifiers.  Of course the feedback is no longer effective when the output is expected to exceed the supply voltage.

\subsubsection{Soft-clipping amplifier}  With moderate feedback, the output clipping starts gradually when the output approaches the dynamic-range boundary.  This behavior is typical of microwave amplifiers.  The knee of the gain curve occurs approximately at the 1\,dB compression power.

\subsubsection{Linear-compression amplifier}  The gain law $A = 1-\gamma(u-1)$ holds in the whole dynamic range.  However this model may seem a mere academic exercise, it provides useful results in a simple and compact form.

\subsection{Amplitude and phase noise}\label{ssec:ele-ampli-am-pm-noise}
%============================================

The contents of this Section is extensively discussed in \cite{Boudot2010arxiv-amplifier-noise}, and briefly summarized here.

\subsubsection{Additive noise}
%------------------------------------------------------------
\begin{figure}[t]
\centering\namedgraphics{scale=0.6}{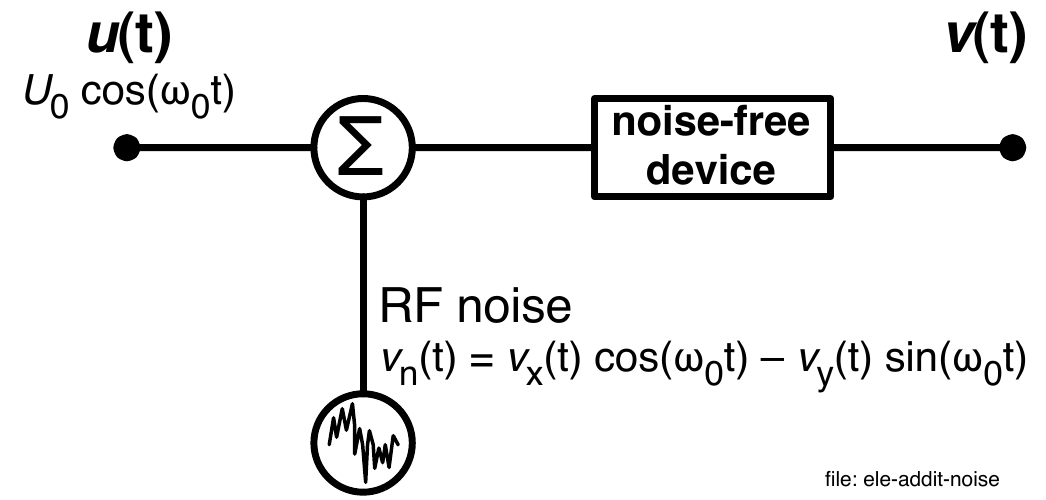}{\columnwidth}
\caption{Additive noise.}
\label{fig:ele-addit-noise}
\end{figure}
Let us consider a quasi-perfect device that adds a noise term $v_n(t)$ to the sinusoidal input signal, as shown in Fig.~\ref{fig:ele-addit-noise}.  Assuming that the device gain is equal to one, the output signal is 
\begin{align}
v(t) &= V_0\cos\omega_0t + v_n(t)~, \nonumber
\intertext{or equivalently}
v(t) &= V_0\cos\omega_0t + v_x(t)\cos\omega_0t - v_y(t)\sin\omega_0t~.
\label{fig:ele-Cartesian}
\end{align}
The random variables $v_x(t)$ and $v_y(t)$, called in-phase and quadrature component of noise, represent the noise $v_n(t)$ in the bandwidth of interest.

Though the Cartesian representation \req{fig:ele-Cartesian} is the closest to the physics of additive noise, polar coordinates can also be used
\begin{align}
v(t) = V_0\left[1+\alpha(t)\right]\cos\left[\omega_0t+\varphi(t)\right]~, 
\label{fig:ele-polar}
\end{align}
where in low-noise conditions it holds that
\begin{align*}
\alpha(t)=\frac{v_x(t)}{V_0} \qquad\text{and}\qquad \varphi(t)=\frac{v_y(t)}{V_0}~.
\end{align*}

The most relevant feature of the additive noise is that all the statistical properties of $v_n(t)$, thus of $v_x(t)$ and $v_y(t)$, are not affected by the input signal.  There follow some relevant properties
\begin{enumerate}
\item Referred to the input, there is an equal amount of AM and PM noise.  Yet, AM/PM asymmetry can show up at the output if the amplitude non-linearity compresses the AM noise.
\item AM and PM noise are statistically independent.
\item The shape of the noise spectrum is independent of the carrier frequency $\omega_0$.  Therefore the noise spectrum cannot have a term $1/f$, $1/f^2$ etc., centered at an arbitrary carrier frequency $\omega_0$.
\item The AM noise and the PM noise scale down with the carrier power. 
\end{enumerate}
The additive noise is generally white, though it can have bumps due to device internal structure and it rolls off out of the bandwidth.
In the case of thermal noise, including the noise figure $F$ (defined only at the temperature $T_0=290$ K), the noise PSD of a generic white-noise process $n(t)$ is
\begin{align}
S_n(f) = FkT_0~.
\end{align}
In polar coordinates the noise PSD is $S_\alpha(f)=h_0$ and $S_\varphi(f)=b_0$, with 
\begin{align}
h_0=\frac{FkT_0}{P_0}
\qquad\text{and}\qquad
b_0=\frac{FkT_0}{P_0}~,
\end{align}
where $P_0$ is the carrier power.

In the case of cascaded amplifiers, the Friis formula \cite{Friis1944ire} applies, by which the noise contribution of each stage is divided by the gain of the preceding stages
\begin{align}
h_0=\frac{kT_0}{P_0}\left[F_1+\frac{F_2}{A_1^2}+\frac{F_3}{A_1^2A_2^2}+\ldots\right]\qquad\text{(Friis, AM noise)}\label{eqn:ele-Friis-AM}\\[1ex]
b_0=\frac{kT_0}{P_0}\left[F_1+\frac{F_2}{A_1^2}+\frac{F_3}{A_1^2A_2^2}+\ldots\right]
\qquad\text{(Friis, PM noise)}\label{eqn:ele-Friis-PM}
\end{align}
where $P_0$ is the carrier power.

\subsubsection{Parametric noise}\label{ssec:ele-param-noise}
%------------------------------------------------------------
\begin{figure}[t]
\centering\namedgraphics{scale=0.6}{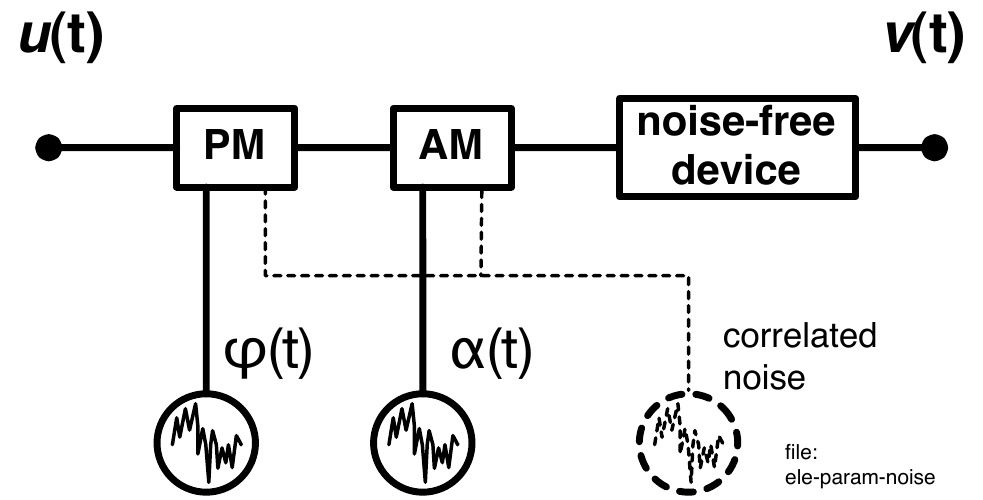}{\columnwidth}
\caption{Parametric noise.}
\label{fig:ele-param-noise}
\end{figure}
The parametric noise originates from a near-dc process that modulates the carrier, as shown in Fig.~\ref{fig:ele-param-noise}.  Accordingly, the polar-coordinate representation  \req{fig:ele-polar} is the closest to the physical mechanism.
The most important parametric random phenomenon is flicker noise, whose PSD is proportional to $1/f$ over several decades.  Other types of parametric noise, with PSD proportional to $1/f^i$, $i=2,~3,~\ldots$, can only exist in a limited frequency region.
For example, $1/f^5$ noise in the region between 1 mHz and 1 Hz has been observed as the phase noise of radio-frequency amplifiers, and also as the offset fluctuation of operational amplifiers.  If these high-slope phenomena would be allowed to span over many decades at low frequencies, the amplitude or the group delay would diverge in the long run, which does not fit the experience about two-port devices.  

As a realistic approximation, one can assume that the near-dc process and the modulation efficiency are independent of the carrier power $P_0$, hence the the statistical properties of $\alpha(t)$ and $\varphi(t)$ tend to be constant in a wide range of power
\begin{align}
h_{-1} = C_1 \qquad\text{and}\qquad b_{-1} = C_2 
\qquad\text{constant, independent of $P_0$}~.
\label{eqn:ele-flicker}
\end{align}
Nonetheless, a too large input power may affect the dc working point, and in turn the amount of parametric noise.

That the parametric noise is independent of $P_0$ has the amazing consequence that the noise of a \emph{multistage amplifier} is the sum of the individual contributions
\begin{align}
S_\alpha(f)~&=~\left[S_\alpha(f)\right]_1 + \left[S_\alpha(f)\right]_2 + \ldots~\\[1ex]
S_\varphi(f)~&=~ \left[S_\varphi(f)\right]_1 + \left[S_\varphi(f)\right]_2 + \ldots~,
\end{align}
independently of the order of the single stages in the chain.

Generally, a parametric process affects both amplitude and phase with separate coefficients, as the dashed noise generator in Fig.~\ref{fig:ele-param-noise} does.  This introduces some correlation between AM and PM noise.  Evidence of this statement is provided by the following examples.
\begin{itemize}
\item In a bipolar transistor, a noise source may affect the thickness of the base region.  Such a process modulates simultaneously the gain (AM noise) and the BE BC capacitances (PM noise), which turns in fully-correlated AM and PM noise.
%\sidenote{\color{red}R\'emi, you provide a simulation}
\item In a laser medium, the pump power affects the partition between excited atoms and ground-state atoms.  Two noise phenomena are simultaneously driven by the power fluctuation of the pump.  
The first and more obvious phenomenon is the fluctuation of gain and of saturation power, which shows up as AM noise  --- referred to as RIN in the jargon of laser optics.
The second phenomenon results from the fact that the contribution of an atom to the refraction index changes if the atom is excited.  This produces phase noise inside the loop, thus frequency noise in the laser beam.

\item The third example is provided by the fluctuation of cathodic emission in vacuum tubes, like triodes, klystrons, magnetrons, TWTs, etc.  Beside the obvious effect on gain, the electron emission impacts on the space charge, and in turn on the capacitance seen by the signal.
\end{itemize}
In all the above examples, a single phenomenon yields fully correlated amplitude and phase noise.

\subsubsection{Phase and amplitude noise spectrum}
%------------------------------------------------------------
\begin{figure}[t]
\centering\namedgraphics{scale=0.6}{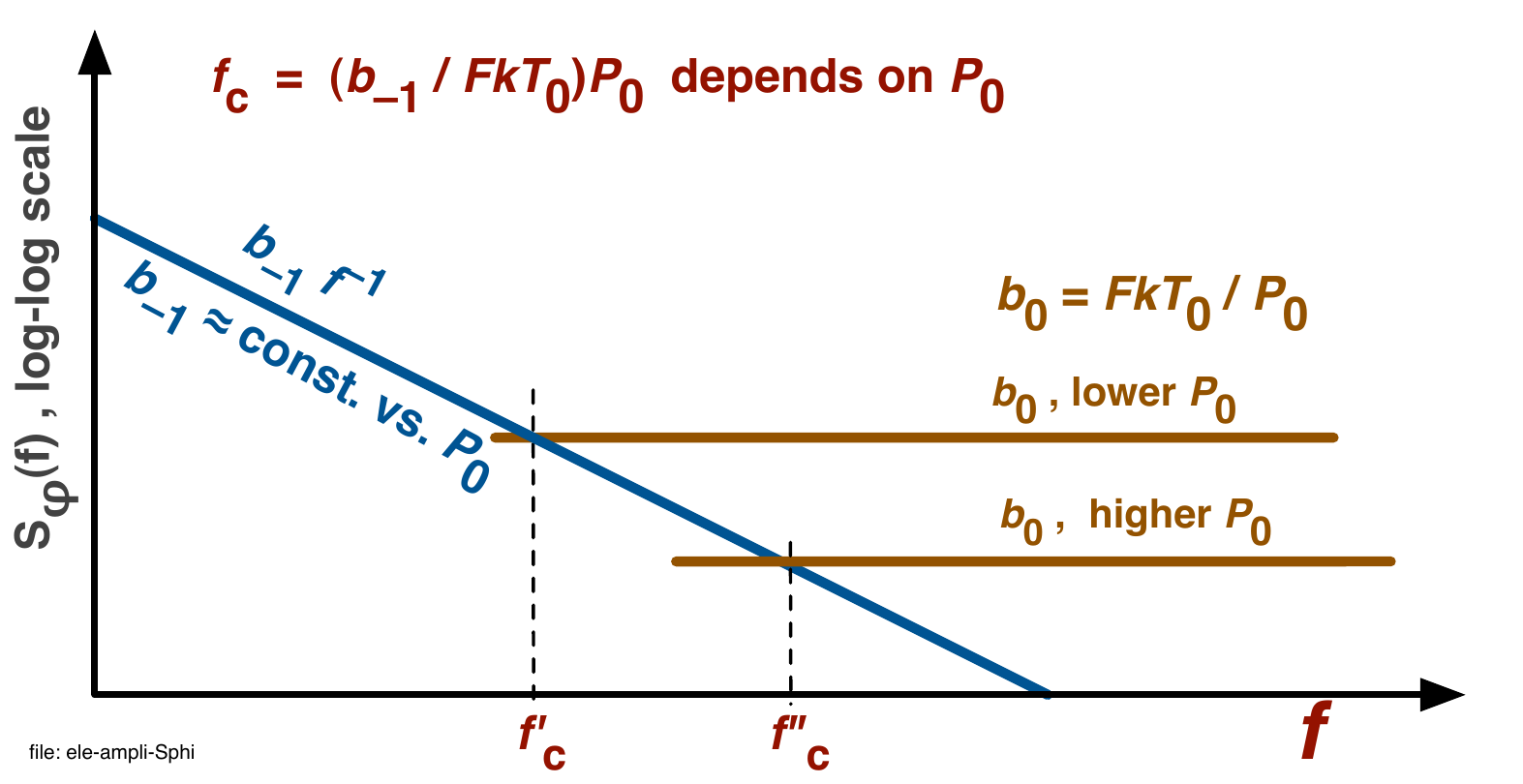}{\columnwidth}
\caption{Amplifier phase (or amplitude) noise power-spectrum density.}
\label{fig:ele-ampli-Sphi}
\end{figure}
It has been seen that the AM noise and the PM noise spectra are
\begin{align}
S_\alpha(f)&=h_0+\frac{h_{-1}}{f}+\ldots	& h_0&=\frac{FkT}{P_0}\qquad h_{-1}=C_1~~\text{(constant)}\\[1ex]
S_\varphi(f)&=b_0+\frac{b_{-1}}{f}+\ldots	& b_0&=\frac{FkT}{P_0}\qquad b_{-1}=C_2~~\text{(constant)}
\end{align}
An example of phase noise spectrum is shown in Figure~\ref{fig:ele-ampli-Sphi}.  This Figure emphasizes the fact that the flicker noise is constant and that the white (additive) noise scales down as the power increases.  The obvious consequence is that the corner frequency $f_c$ also scales with power.
A common mistake found in CAD software is that the flicker is described by a fixed corner frequency, independent of power.   The reader is strongly encouraged to check before trusting a CAD program.

\subsubsection{Gain fluctuation}\label{sssec:ele-gain-fluctuation}
\begin{figure}[t]
\centering\namedgraphics{scale=0.5}{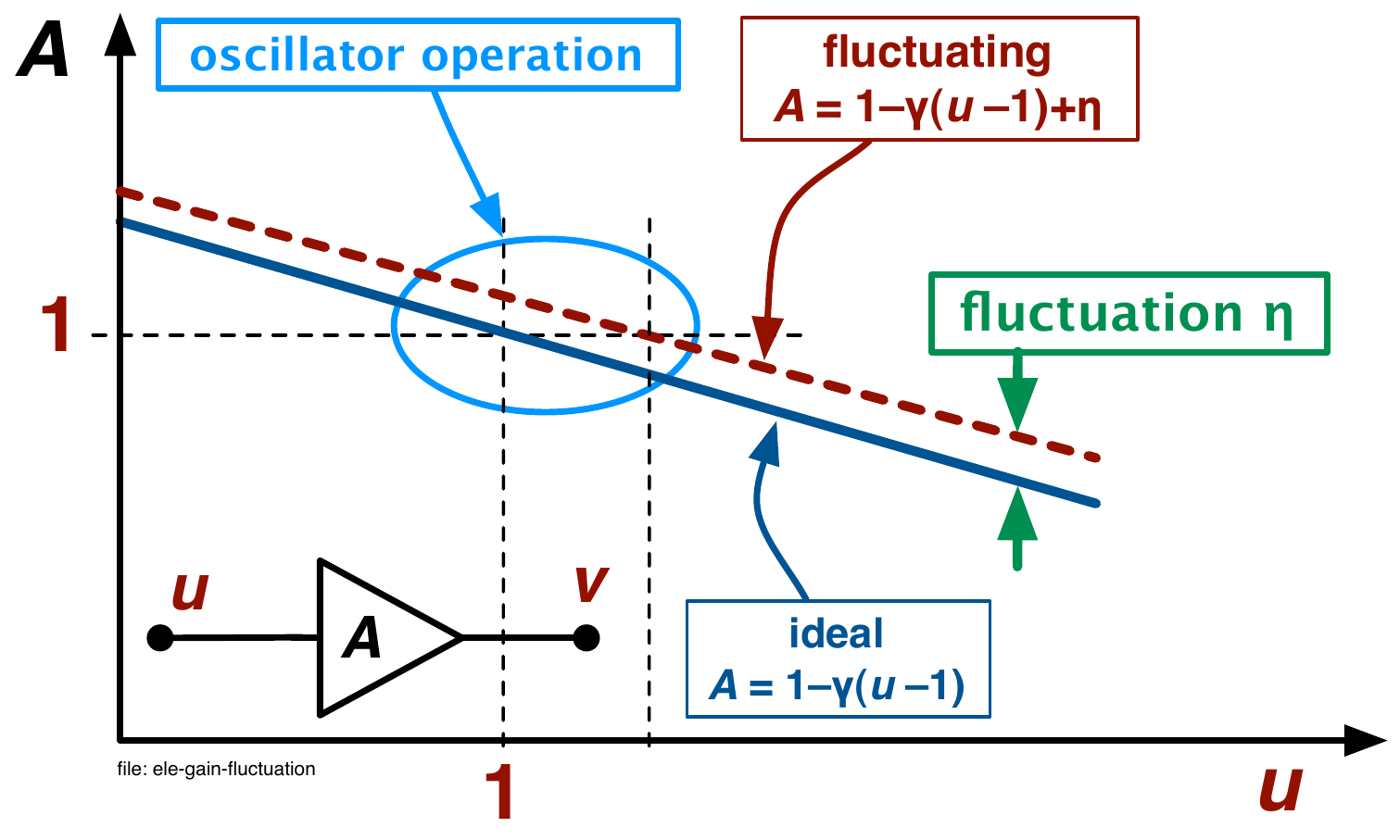}{\columnwidth}
\caption{Parametric fluctuation of the amplifier gain.}
\label{fig:ele-gain-fluctuation}
\end{figure}

%-------------------------------------------------------------------------------------
Modeling the oscillator in the frequency domain, the gain is $A=A_1[1-\gamma(u-1)]$ around the oscillation point $u=1$.  
Introducing a slow fluctuation, $A$ turns into the \emph{slow varying} function of time
\begin{align}
A(t)&=A_1\left[1-\gamma(u-1)\right]\left[1+\eta(t)\right]e^{j\psi(t)}\nonumber
\intertext{approximated as}
A(t)&\simeq A_1\left[1-\gamma(u-1)+\eta(t)\right]e^{j\psi(t)}
\qquad\text{(Fig.~\ref{fig:ele-gain-fluctuation})}~,
\end{align}
where
\begin{align}
&\eta(t)\leftrightarrow\mathcal{N}(s) \qquad\text{(amplitude fluctuation, i.e., AM noise)}\nonumber\\
&\psi(t)\leftrightarrow\Psi(s)\qquad\text{(phase fluctuation, i.e., PM noise)}\nonumber
\end{align}
are the amplitude an phase gain fluctuations, respectively.
Figure~\ref{fig:ele-gain-fluctuation} shows the combined effect of the gain amplitude fluctuation and  compression.

Flicker noise, which results from a parametric effect, impacts directly on the gain.  It can be described by 
\begin{align}
\left[S_\eta(f)\right]_\text{flicker}=\frac{h_{-1}}{f}
\qquad\text{and}\qquad
\left[S_\psi(f)\right]_\text{flicker}=\frac{b_{-1}}{f}
\qquad\text{(constant vs.~$P_0$)}~.\nonumber
\end{align}

Additive noise, albeit of quite different origin, can still be seen as a gain fluctuation because it affects the input/output relationship.  Hence
\begin{align}
\left[S_\eta(f)\right]_\text{additive}~=~\left[S_\psi(f)\right]_\text{additive}~=~\frac{FkT_0}{P_0}
\qquad\text{(constant vs.~$f$)}~.\nonumber
\end{align}

\section{The resonator}\label{sec:ele-resonator}
%==================================================
\begin{figure*}[t]
\centering\namedgraphics{scale=0.55}{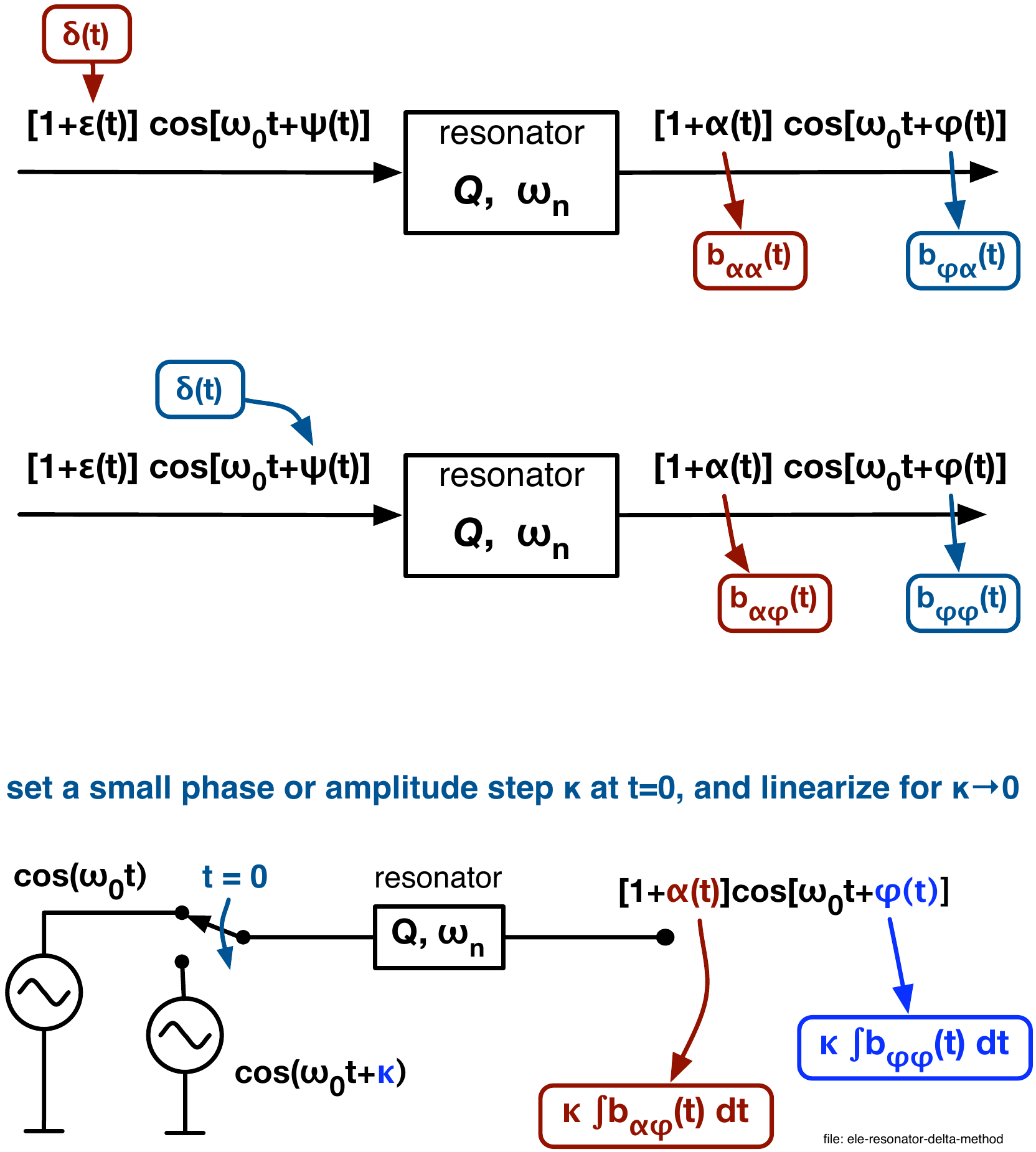}{\textwidth}
\caption{AM and PM response of a resonator.}
\label{fig:ele-resonator-delta-method}
\end{figure*}
The resonator in \emph{actual load conditions}\footnote{Whoever has worked seriously in the field of oscillators, may have in mind three sets of parameters like `$\omega_n$ and `$Q$.'  These sets refer (1) to the unloaded resonator, which is a mathematical abstraction not accessible to the physical experiment;  (2) to the resonator loaded by the measurement test set, from which the unloaded parameters are estimated;  and (3) to the resonator loaded by the oscillator circuit.  The external circuit, either the test set or the resonator, increases the dissipation and affects the natural frequency.
That said, it is to be made clear that here \emph{the resonator is always loaded by the oscillator circuit}, and therefore that \emph{there is no point in discussing the other conditions}.  On the other hand, the process of getting $\omega_n$ and $Q$ from experimental data may be tricky or difficult.  We skip this discussion because it depends on the specific resonator and oscillator, while we aim at a general theory.} is governed, or locally well approximated by the differential equation
\begin{align}
\ddot{v}_o+\frac{\omega_n}{Q}\dot{v}_o+\omega^2_nv_o=\mathrm{L}\{v_i(t)\}~,
\label{eqn:ele-resonator-de-gen}
\end{align}
where $\omega_n$ is the natural frequency, $Q$ is the quality factor, $v_i(t)$ is the external force, and $\mathrm{L}$ is an operator.
The most interesting form of the force term is $\mathrm{L}\{v_i(t)\}=\frac{\omega_n}{Q}\dot{v}_i(t)$ because it is homogeneous with the dissipative term.  This occurs with the series (parallel) RLC resonator driven by a voltage (current) source, and in other relevant cases.  Accordingly, \req{eqn:ele-resonator-de-gen} becomes
\begin{align}
\ddot{v}_o+\frac{\omega_n}{Q}\dot{v}_o+\omega^2_nv_o=\frac{\omega_n}{Q}\dot{v}_i(t)~.
\label{eqn:ele-resonator-de}
\end{align}
Using the Laplace transform, the resonator transfer function $\beta(s)=V_o/V_i$ is 
\begin{align}
\beta(s) = \frac{\omega_n}{Q} \frac{s}{s^2+ \frac{\omega_n}{Q}s +\omega_n^2}~.
\label{eqn:ele-resonator-laplace}
\end{align}
Equations \req{eqn:ele-resonator-de} and \req{eqn:ele-resonator-laplace} are normalized for the resonator to respond to a sinusoid at the exact resonant frequency $\omega_n$ with a sinusoid of the same frequency, phase and amplitude.

We analyze the impulse response of the resonator phase and amplitude in stationary-oscillation conditions.  The phase response is the response to a perturbation $\delta(t)$ in the argument of the driving signal, as shown in Fig.~\ref{fig:ele-resonator-delta-method}.  Similarly, the amplitude response is the response to a perturbation $\delta(t)$ in the amplitude of the driving signal.  
In general literature the impulse response is denoted with $h(t)$, and its Laplace transform with $H(s)$.  Since we use $h(t)\leftrightarrow H(s)$ for the oscillator response, the phase or amplitude impulse response of the resonator is denoted with $b(t)\leftrightarrow B(s)$.
It turns out that the resonator response is the same for amplitude and phase.

In our analysis we replace the impulse $\delta(t)$ with a small phase or amplitude step $\kappa\mathfrak{u}(t)$, where $\mathfrak{u}(t)$ is  the Heaviside function 
\begin{align*}
\mathfrak{u}(t)
=\int_{-\infty}^\infty\delta(t)\;dt
~=~\begin{cases} 0 & t<0\\ 1 & t>0
\end{cases}
\end{align*}
and we linearize for $\kappa\rightarrow0$. 
Then we use the general property of linear systems that the response to $\mathfrak{u}(t)$ is $\int \mathrm{b}(t)\:dt$.  Notice that $\mathfrak{u}(t)$ can be seen as a switch that changes state from off to on at $t=0$; and that $\mathfrak{u}(-t)$ switches from on to off at $t=0$.

\subsection{Sinusoidal transients}
%-------------------------------------------------------------------------------------

\subsubsection{Switch-off transient}
%-------------------------------------------
Let us consider the resonator driven by the signal
\begin{align}
v_i(t)=\frac{1}{\beta_0}\cos(\omega_0t-\theta)
\quad\text{(probe)}
\label{eqn:ele-test-cosine-swoff}
\end{align}
where $\beta_0$ and $\theta$ are chosen for the asymptotic output to be $v_o(t)=\cos(\omega_0t)$ for $t\rightarrow\infty$, i.e., amplitude is equal one and phase equal zero in the general case $\omega_0\neq\omega_n$.  If the probe signal $v_i(t)$ is switched off at the time $t=0$, the output is
\begin{align}
v_o(t)=\cos(\omega_pt)\,e^{-t/\tau}
\quad t>0~,
\quad\text{(response)}
\label{eqn:ele-test-cosine-swoff-resp}
\end{align}
where
\begin{align}
\tau=\frac{2Q}{\omega_n}
\qquad\text{(relaxation time)}
\end{align}
is the resonator relaxation time and 
\begin{align}
\omega_p=\omega_n\sqrt{1-\frac{1}{4Q^2}}
\qquad\text{(free-decay pseudo-frequency)}
\end{align}
is the free-decay pseudo-frequency.
For $Q\gg1$, we can approximate $\omega_p\simeq\omega_n$.   This%\sidenote{\color{magenta}R\'emi said that this explanation is not necessary} 
is justified by the fact that the phase error $\zeta$ accumulated during the relaxation time $\tau$ is
\begin{align*}
\zeta=(\omega_n-\omega_p)\tau=\frac{1}{4Q}~.
\end{align*}
This is seen by replacing $\tau=\smash{\frac{2Q}{\omega_n}}$ and $\omega_p=\omega_n\sqrt{1-1/4Q^2}$ in $\zeta$, and by expanding in series truncated at the first order.

\subsubsection{Switch-on transient}
%------------------------------------------
The response to a sinusoid switched on at the time $t=0$ takes the general form
\begin{align}
v_o(t)&=
\mathscr{A}\cos(\omega_pt)\,e^{-\frac{t}{\tau}}+
\mathscr{B}\sin(\omega_pt)\,e^{-\frac{t}{\tau}}+
\mathscr{C}\cos(\omega_0t)+
\mathscr{D}\sin(\omega_0t)
\quad t>0~,\nonumber
\end{align}
where $\mathscr{A}$, $\mathscr{B}$, $\mathscr{C}$, and $\mathscr{D}$ are  constants determined by the boundary conditions.  

We use the probe signal \req{eqn:ele-test-cosine-swoff}.  This yields immediately $\mathscr{C}=1$ and $\mathscr{D}=0$.  The constants $\mathscr{A}$ and $\mathscr{B}$ are found by assessing the continuity of $v_o(t)$ at $t=0$, which gives $\mathscr{A}=-1$ and $\mathscr{B}=0$. 
Approximating $\omega_p\simeq\omega_n$ for $Q\gg1$, the output is
\begin{align}
v_o(t)&=-\cos(\omega_nt)\,e^{-t/\tau}+\cos(\omega_0t) \quad t>0~.
\label{eqn:ele-test-cosine-swon-resp}
\end{align}

Similarly, using a probe signal $v_i(t)=\frac{1}{\beta_0}\sin(\omega_0t-\theta)$, the output is 
\begin{align}
v_o(t)&=-\sin(\omega_nt)\,e^{-t/\tau}+\sin(\omega_0t) \quad t>0
\label{eqn:ele-test-sinus-swon-resp}
\end{align}

\subsection{Impulse response at the exact natural frequency}
%-------------------------------------------------------------------------------------
%\sidenote{\color{magenta}Use this section only in the short version, with no AM-PM coupling}
When the resonator is used at the exact natural frequency,  it holds that $\omega_0=\omega_n$, $\beta_0=1$, and $\theta=0$.  

A \emph{phase step} $\kappa$ at $t=0$ is described as the probe signal
\begin{align*}
v_i(t)
&=\underbrace{\cos(\omega_0t)\,\mathfrak{u}(-t)}%
_{\substack{\text{switched off at $t=0$}}} + 
\underbrace{\cos(\omega_0t+\kappa)\,\mathfrak{u}(t)}%
_{\substack{\text{switched on at $t=0$}}}~,
\end{align*}
By virtue of linearity, the response is the sum of \req{eqn:ele-test-cosine-swoff-resp} plus \req{eqn:ele-test-cosine-swon-resp}, that is, 
\begin{align}
v_o(t)=\cos(\omega_pt)\,e^{-t/\tau} + 
    \cos(\omega_pt+\kappa)\bigl[1-e^{-t/\tau}\bigr]
    \qquad t>0~.
\label{eqn:ele-vout-phase}
\end{align}
Expanding and using the approximations $\cos(\kappa)\simeq1$ and $\sin(\kappa)\simeq\kappa$ for $\kappa\rightarrow0$, and $\omega_p\simeq\omega_n$ for large $Q$, thus $\omega_p\simeq\omega_0$, we get
\begin{align*}
v_o(t)&=\cos(\omega_0t)-\kappa\sin(\omega_0t)\bigl[1-e^{-t/\tau}\bigr]
\qquad t>0~, 
\end{align*}
This can be seen as a slowly varying phasor
\begin{math}
\mathbf{V_o}(t)=\smash{\frac{1}{\sqrt{2}}}\:\left\{1+j\kappa\left[1-e^{-t/\tau}\right]\right\},
\end{math}
whose angle 
\begin{align*}
\arctan\left(\frac{\Im\{\mathbf{V_o}(t)\}}{\Re\{\mathbf{V_o}(t)\}}\right)
&\simeq \kappa\bigl[1-e^{-t/\tau}\bigr]\qquad t>0
\end{align*}
is the response to $\kappa\mathfrak{u}(t)$.
After deleting $\kappa$ and differentiating, we obtain the impulse response $\mathrm{b}(t)=\smash{\frac{1}{\tau}}e^{-t/\tau}$.

An \emph{amplitude step} $\kappa$ at $t=0$ is described as the probe signal
\begin{align*}
v_i(t)
&=\underbrace{\cos(\omega_0t)\,\mathfrak{u}(-t)}%
_{\substack{\text{switched off at $t=0$}}} + 
\underbrace{(1+\kappa)\cos(\omega_0t)\,\mathfrak{u}(t)}%
_{\substack{\text{switched on at $t=0$}}}~,
\end{align*}
Once again the response is the sum of \req{eqn:ele-test-cosine-swoff-resp} plus \req{eqn:ele-test-cosine-swon-resp}
\begin{align}
v_o(t)=\cos(\omega_pt)\,e^{-t/\tau} +
(1+\kappa)\cos(\omega_pt+\kappa)\bigl[1-e^{-t/\tau}\bigr]\qquad t>0~.
\label{eqn:ele-vout-ampliude}
\end{align}
Expanding under the same approximations as above, i.e.,  $\cos(\kappa)\simeq1$ and $\sin(\kappa)\simeq\kappa$ for $\kappa\rightarrow0$, and $\omega_p\simeq\omega_n$ for large $Q$, and $\omega_p\simeq\omega_0$, we get
\begin{align}
v_o(t)&=\cos(\omega_0t)+\kappa\bigl[1-e^{-t/\tau}\bigr]\cos(\omega_0t)
\qquad t>0~.\nonumber
\end{align}
This is a slowly varying phasor
\begin{math}
\mathbf{V_o}(t)=\smash{\frac{1}{\sqrt{2}}}\:\left\{1+\kappa\left[1-e^{-t/\tau}\right]\right\}, 
\end{math}
whose amplitude swing
\begin{align*}
\mathbf{V_o}(t)-\mathbf{V_o}(0)
&\simeq \kappa\bigl[1-e^{-t/\tau}\bigr]
\qquad t>0
\end{align*}
is the response to $\kappa\mathfrak{u}(t)$.
After deleting $\kappa$ and differentiating, we obtain the impulse response $\mathrm{b}(t)=\smash{\frac{1}{\tau}}e^{-t/\tau}$, the same already found for the phase impulse.

\begin{figure}[t]
\centering\namedgraphics{scale=0.6}{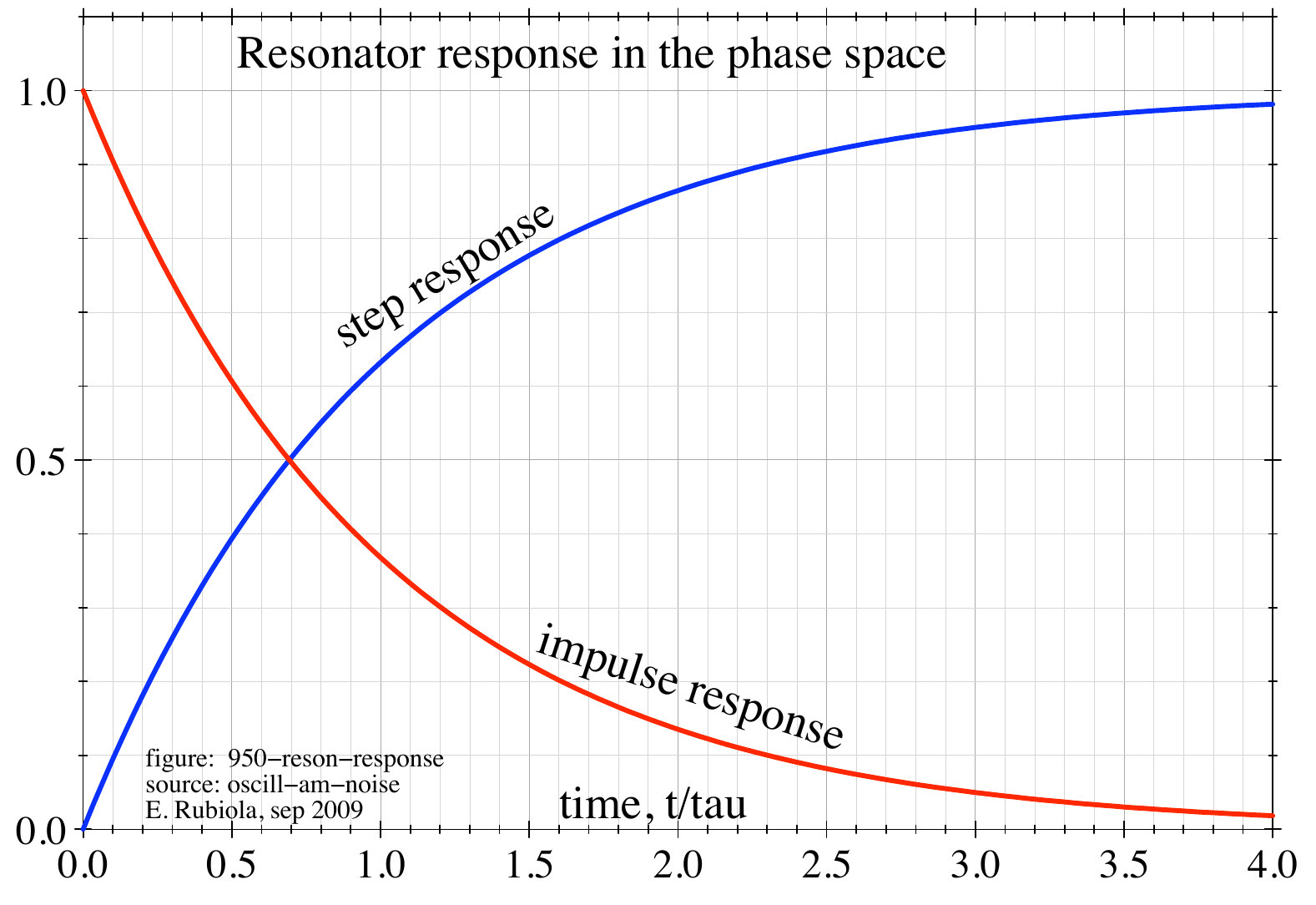}{\columnwidth}
\caption{Resonator response to the step and to the impulse.}
\label{fig:950-reson-response}
\end{figure}
In conclusion, the response to either a \emph{phase impulse} or to an \emph{amplitude impulse} is (Fig.~\ref{fig:950-reson-response})
\begin{align}
\mathrm{b}(t) = \frac1\tau e^{-s\tau}
 \quad\leftrightarrow\quad 
\mathrm{B}(s) = \frac{1/\tau}{s+1/\tau}~.
\label{eqn:eln-resonator-response-B}
\end{align}
Eq.~\req{eqn:eln-resonator-response-B} is that of a simple $RC$ low-pass filter, which we will use in all block diagrams.
The inverse of $\tau$ is known as the Leeson (cutoff) frequency of the resonator
\begin{align*}
\omega_L=\frac1\tau=\frac{\omega_n}{2Q}
\qquad\text{or}\qquad
f_L=\frac{1}{2\pi\tau}=\frac{\nu_n}{2Q}~.
%\qquad\text{(Leeson frequency)}
\end{align*}

Finally, it is to be remarked that \req{eqn:ele-vout-phase} contains no amplitude terms at first order, and that \req{eqn:ele-vout-ampliude} contains no phase terms at first order.  This means that at the exact resonant frequency there is no phase-amplitude coupling.

\subsection{Off-resonance impulse response}\label{ssec:ele-off-freq-resonator}
%-------------------------------------------------------------------------------------
%\sidenote{\color{magenta}Keep this section for the long version, which includes the AM-PM coupling}
\begin{figure}[t]
\centering\namedgraphics{scale=0.6}{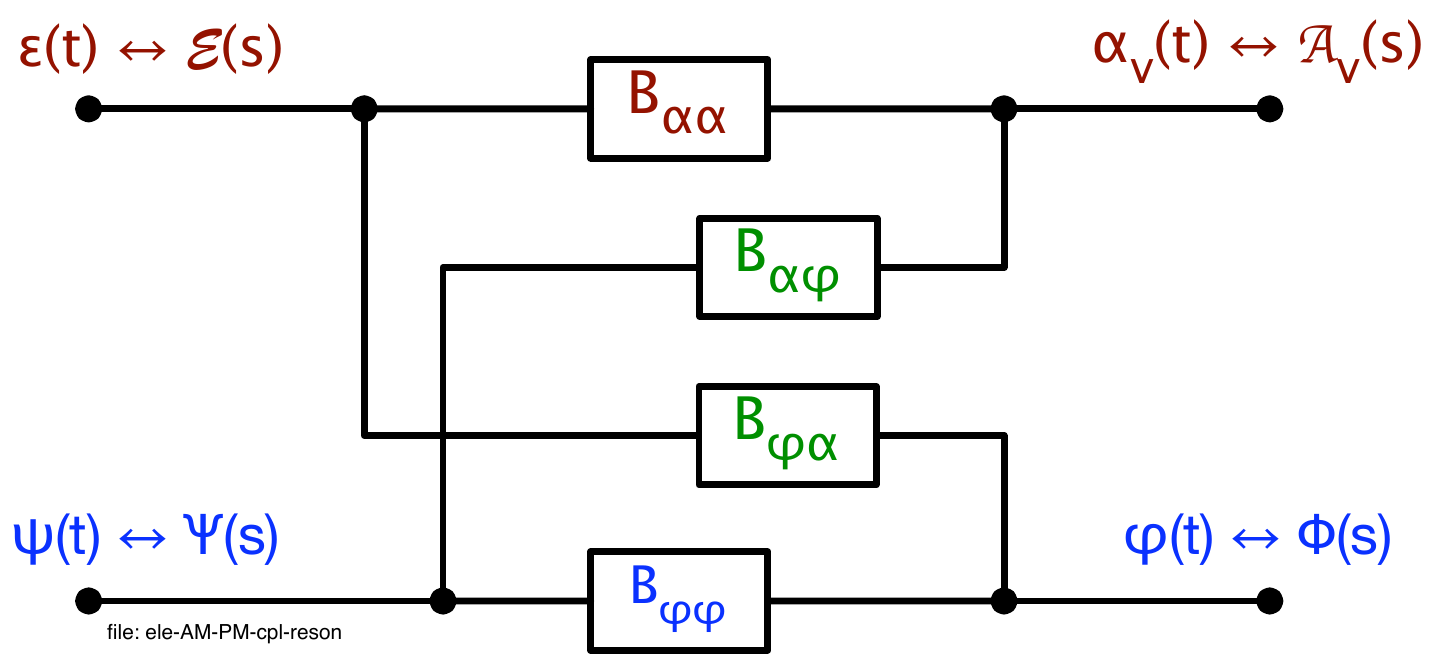}{\columnwidth}
\caption{Detuning the resonator results in coupling AM to PM.}
\label{fig:ele-AM-PM-cpl-reson}
\end{figure}
\begin{figure}[t]
\centering\namedgraphics{scale=0.6}{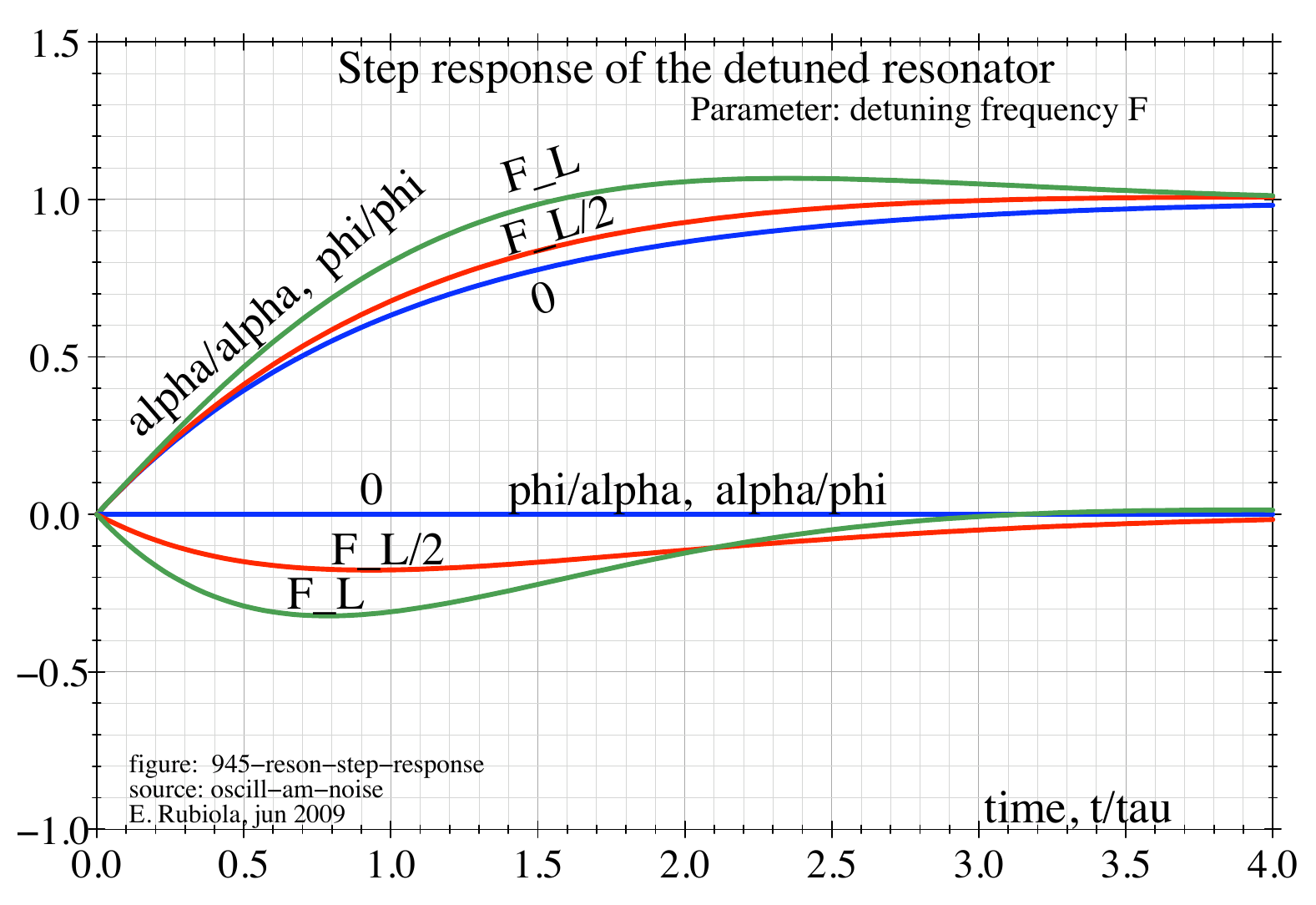}{\columnwidth}
\caption{Step response of the resonator off the natural frequency.}
\label{fig:945-reson-step-response}
\end{figure}
\begin{figure}[t]
\centering\namedgraphics{scale=0.6}{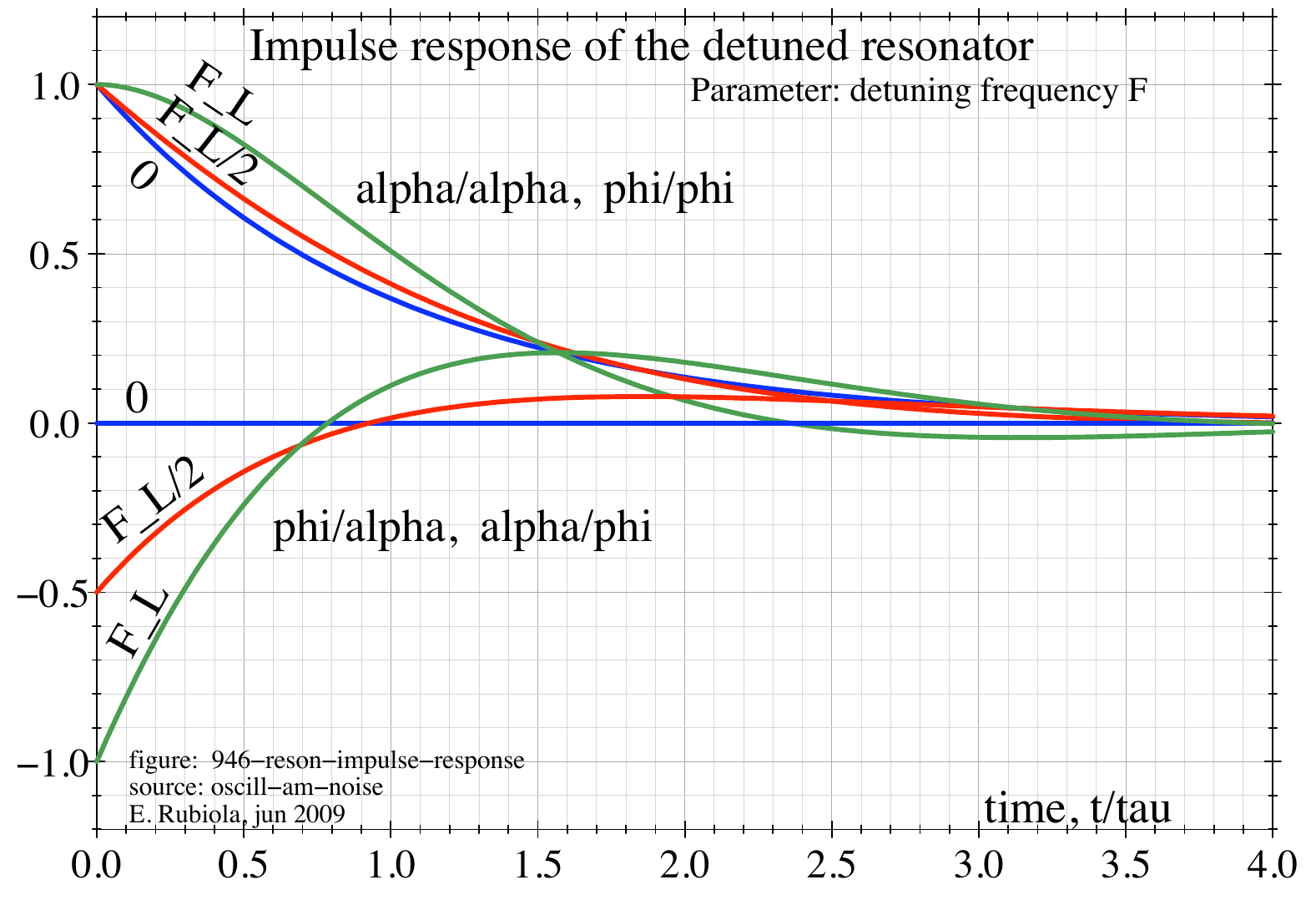}{\columnwidth}
\caption{Impulse response of the resonator off the natural frequency.}
\label{fig:946-reson-impulse-response}
\end{figure}

In this Section we analyze the impulse response of the resonator when the carrier frequency is $\omega_0\ne\omega_n$, with an offset
\begin{align*}
\Omega=\omega_0-\omega_n
\qquad\text{(detuning)}~.
\end{align*}
An amplitude perturbation $\varepsilon(t)$ in the resonator driving signal results in an amplitude fluctuation $\alpha(t)=\mathrm{b}_{\alpha\alpha}(t)*\varepsilon(t)$ plus a phase fluctuation $\varphi(t)=\mathrm{b}_{\varphi\alpha}(t)*\varepsilon(t)$.  Similarly, the resonator responds to a phase perturbation $\psi(t)$ with a phase fluctuation $\varphi(t)=\mathrm{b}_{\varphi\varphi}(t)*\psi(t)$ plus an amplitude fluctuation $\alpha(t)=\mathrm{b}_{\alpha\varphi}(t)*\psi(t)$.  This is written in matrix form as 
\begin{gather*}
\begin{bmatrix}\alpha\\\varphi\end{bmatrix} = 
\begin{bmatrix}
\mathrm{b}_{\alpha\alpha} & \mathrm{b}_{\alpha\varphi}\\
\mathrm{b}_{\varphi\alpha} & \mathrm{b}_{\varphi\varphi}\\
\end{bmatrix} *
\begin{bmatrix}\varepsilon\\\psi\end{bmatrix}
\quad\leftrightarrow\quad
\begin{bmatrix}\mathcal{A}\\\Phi\end{bmatrix} = 
\begin{bmatrix}
\mathrm{B}_{\alpha\alpha} & \mathrm{B}_{\alpha\varphi}\\
\mathrm{B}_{\varphi\alpha} & \mathrm{B}_{\varphi\varphi}
\end{bmatrix} 
\begin{bmatrix}\mathcal{E}\\\Psi\end{bmatrix}~,
\end{gather*}
and shown in Fig.~\ref{fig:ele-AM-PM-cpl-reson}.  
In the following sections we will prove that the step response is (Fig.~\ref{fig:945-reson-step-response})
\begin{align}
\int\begin{bmatrix}\mathrm{b}\end{bmatrix}\!\!(t)\,dt&=
\begin{bmatrix}
\sin(\Omega t)\,e^{-t/\tau}&[1-\cos(\Omega t)]\,e^{-t/\tau}\\[1ex]
[1-\cos(\Omega t)]\,e^{-t/\tau}&\sin(\Omega t)\,e^{-t/\tau}
\end{bmatrix} 
\end{align}
and that the impulse response is (Fig.~\ref{fig:946-reson-impulse-response})
\begin{align}
\begin{bmatrix}\mathrm{b}\end{bmatrix}(t)&=
\begin{bmatrix}
\left(\Omega\sin\Omega t+\frac1\tau\cos\Omega t\right)e^{-t/\tau}
&\left(-\Omega\cos\Omega t+\frac1\tau \sin\Omega t\right)e^{-t/\tau}\\[1ex]
\left(-\Omega\cos\Omega t+\frac1\tau \sin\Omega t\right)e^{-t/\tau} 
&\left(\Omega\sin\Omega t+\frac1\tau\cos\Omega t\right)e^{-t/\tau}
\end{bmatrix} 
\label{eqn:eln-resonator-b-matrix}
\end{align}
\begin{align}
\begin{bmatrix}\mathrm{B}\end{bmatrix}(s)&=
\begin{bmatrix}
\displaystyle
\frac{1}{\tau}\:\frac{s+\frac1\tau+\Omega^2\tau}{s^2+\frac{2}{\tau}s+\frac{1}{\tau^2}+\Omega^2}
&\displaystyle
\frac{-\Omega s}{s^2+\frac{2}{\tau}s+\frac{1}{\tau^2}+\Omega^2}\\[2ex]
\displaystyle
\frac{-\Omega s}{s^2+\frac{2}{\tau}s+\frac{1}{\tau^2}+\Omega^2}
&\displaystyle
\frac{1}{\tau}\:\frac{s+\frac1\tau+\Omega^2\tau}{s^2+\frac{2}{\tau}s+\frac{1}{\tau^2}+\Omega^2}
\end{bmatrix}~.
\label{eqn:eln-resonator-B-matrix}
\end{align}
The resonator response has diagonal symmetry
\begin{align}
\begin{split}
\mathrm{b}_{\alpha\alpha}(t)=\mathrm{b}_{\varphi\varphi}(t)
&\quad\leftrightarrow\quad
\mathrm{B}_{\alpha\alpha}(s)=\mathrm{B}_{\varphi\varphi}(s)\\
\mathrm{b}_{\alpha\varphi}(t)=\mathrm{b}_{\varphi\alpha}(t)
&\quad\leftrightarrow\quad
\mathrm{B}_{\alpha\varphi}(s)=\mathrm{B}_{\varphi\alpha}(s)~.
\end{split}
\label{eqn:eln-resonator-matrix-symmetry}
\end{align}
The proof is given in Sections \ref{sssec:eln-res-phase-impulse}, \ref{sssec:eln-res-ampl-impulse}, and \ref{sssec:eln-res-am-pm-laplace}

\subsubsection{Response to the phase impulse}\label{sssec:eln-res-phase-impulse}
%----------------------------------------------------------------------
A \emph{phase step} $\kappa$ at $t=0$ is described as the probe signal
\begin{align*}
v_i(t)
&=\underbrace{\frac{1}{\beta_0}\cos(\omega_0t-\theta)\,\mathfrak{u}(-t)}%
_{\substack{\text{switched off at $t=0$}}} + 
\underbrace{\frac{1}{\beta_0}\cos(\omega_0t-\theta+\kappa)\,\mathfrak{u}(t)}%
_{\substack{\text{switched on at $t=0$}}}
\\
&=\frac{1}{\beta_0}\cos(\omega_0t-\theta)\,\mathfrak{u}(-t) + 
\frac{1}{\beta_0}\bigl[\cos(\omega_0t-\theta)\cos\kappa
	-\sin(\omega_0t-\theta)\sin\kappa\bigr]\,\mathfrak{u}(t)\\
&\simeq\frac{1}{\beta_0}\cos(\omega_0t-\theta)\,\mathfrak{u}(-t) +
  \frac{1}{\beta_0}\bigl[\cos(\omega_0t-\theta)
	-\kappa\sin(\omega_0t-\theta)\bigr]\,\mathfrak{u}(t)
\quad\kappa\ll1.
\end{align*}
Using \req{eqn:ele-test-cosine-swoff-resp}, \req{eqn:ele-test-cosine-swon-resp} and \req{eqn:ele-test-sinus-swon-resp} under the large-$Q$ approximation ($\omega_p=\omega_n$), the above yields the output
\begin{align*}
v_o(t)&=\cos(\omega_nt)\,e^{-\frac{t}{\tau}}
        +\bigl[-\cos(\omega_nt)\,e^{-\frac{t}{\tau}} + \cos(\omega_0t)\bigr]\\
	&\qquad+\kappa\bigl[\sin(\omega_nt)\,e^{-\frac{t}{\tau}} - \sin(\omega_0t)\bigr]
	\qquad\qquad(t>0)
\intertext{which simplifies to}
v_o(t)&=\cos(\omega_0t) - \kappa\sin(\omega_0t)
	+ \kappa\sin(\omega_nt)\,e^{-t/\tau}\qquad t>0~.
\end{align*}
Introducing the detuning frequency $\Omega=\omega_0-\omega_n$, we get $\sin(\omega_nt)=\sin(\omega_0t-\Omega t)$, thus $
\sin(\omega_nt)=\sin(\omega_0t)\cos(\Omega t)-\cos(\omega_0t)\sin(\Omega t)$.
Hence, the output signal can be rewritten as
\begin{align*}
v_o(t)&=\cos(\omega_0t)-\kappa\sin(\omega_0t)
+\kappa\sin(\omega_0t)\cos(\Omega t)\,e^{-\frac{t}{\tau}}
	-\kappa\cos(\omega_0t)\sin(\Omega t)\,e^{-\frac{t}{\tau}}~,
\end{align*}
which simplifies to
\begin{align}
v_o(t)&=\cos(\omega_0t)\Bigl[1-\kappa\sin(\Omega t) e^{-t/\tau}\Bigr]
	-\kappa\sin(\omega_0t)\Bigl[1-\cos(\Omega t) e^{-t/\tau}\Bigr]~.
\end{align}
Freezing the oscillation $\omega_0t$, the above turns into the slow-varying phasor
\begin{align*}
\mathbf{V_o}(t)&=\frac{1}{\sqrt{2}}
\left\{1 - \kappa\sin(\Omega t)e^{-t/\tau}+ j\kappa\bigl[1-\cos(\Omega t) e^{-t/\tau}\bigr]\right\}
~\qquad\kappa\ll1
\end{align*}
of angle
\begin{align*}
\arctan\frac{\Im\{\mathbf{V_o}(t)\}}{\Re\{\mathbf{V_o}(t)\}} = \kappa\Bigl[1-\cos(\Omega t)e^{-t/\tau}\Bigr]
\end{align*}
and amplitude\sidenote{Fix this}
\begin{align*}
\left|\mathbf{V_o}(t)\right| = \left|\mathbf{V_o}(0)\right| -\kappa\sin(\Omega t)e^{-t/\tau}
\end{align*}
After deleting $\kappa$ and differentiating, we obtain the impulse response
\begin{align}
\mathrm{b}_{\varphi\varphi}(t)
&=\Bigl[\Omega\sin(\Omega t) + \frac1\tau \cos(\Omega t)\Bigr] e^{-t/\tau}
&&\text{phase}
\label{eqn:ele-b-phi-phi}\\
\mathrm{b}_{\alpha\varphi}(t)
&=\Bigl[-\Omega\cos(\Omega t) + \frac1\tau \sin(\Omega t)\Bigr] e^{-t/\tau}
&&\text{amplitude}
\label{eqn:ele-b-alpha-phi}
\end{align}

\subsubsection{Response to the amplitude impulse}\label{sssec:eln-res-ampl-impulse}
%----------------------------------------------------------------------
An \emph{amplitude step} $\kappa$ at $t=0$ is described as the probe signal
\begin{align*}
v_i(t)
&=\underbrace{\frac{1}{\beta_0}\cos(\omega_0t-\theta)\,\mathfrak{u}(-t)}%
_{\substack{\text{switched off at $t=0$}}} + 
\underbrace{(1+\kappa)\frac{1}{\beta_0}\cos(\omega_0t-\theta)\,\mathfrak{u}(t)}%
_{\substack{\text{switched on at $t=0$}}}
\end{align*}
Using \req{eqn:ele-test-cosine-swoff-resp}, \req{eqn:ele-test-cosine-swon-resp} and \req{eqn:ele-test-sinus-swon-resp}, under the approximation $\omega_p=\omega_n$ the above yields the output
\begin{align*}
v_o(t)&=\cos(\omega_nt)\,e^{-t/\tau}
+(1+\kappa)\bigl[-\cos(\omega_nt)\,e^{-t/\tau} + \cos(\omega_0t)\bigr]
\qquad t>0\\
&=\cos(\omega_0t) + \kappa\cos(\omega_0t)
	+ \kappa\cos(\omega_nt)\,e^{-t/\tau}~.
\end{align*}
Using $\cos(\omega_nt)=\cos(\omega_0t)\cos(\Omega t)+\sin(\omega_0t)\sin(\Omega t)$, the output is
\begin{align}
v_o(t)&=
\cos(\omega_0t)\Bigl\{ 1+\kappa\Bigl[1-\cos(\Omega t)\,e^{-t/\tau}\Bigr]\Bigr\}
-\kappa\sin(\omega_0t)\sin(\Omega t)e^{-t/\tau}
\end{align}
Freezing the oscillation $\omega_0t$, the above turns into the slow-varying phasor
\begin{align*}
\mathbf{V_o}(t)&=\frac{1}{\sqrt{2}}
\left\{1 + \kappa\bigl[1-\cos(\Omega t) e^{-t/\tau}\bigr]
+j\kappa\sin(\Omega t)e^{-t/\tau}\right\}
\end{align*}
of angle
\begin{align*}
\arctan\frac{\Im\{\mathbf{V_o}(t)\}}{\Re\{\mathbf{V_o}(t)\}} =
\kappa\sin(\Omega t)e^{-t/\tau}
\qquad\kappa\ll1
\end{align*}
and amplitude swing\sidenote{Fix this}
\begin{align*}
\left|\mathbf{V_o}(t)-\mathbf{V_o}(0)\right| =
\kappa\Bigl[1-\cos(\Omega t)e^{-t/\tau}\Bigr]
\qquad\kappa\ll1
\end{align*}
After deleting $\kappa$ and differentiating, we obtain the impulse response
\begin{align}
\mathrm{b}_{\alpha\alpha}(t)
&=\Bigl[\Omega\sin(\Omega t) + \frac1\tau \cos(\Omega t)\Bigr] e^{-t/\tau}
&&\text{amplitude}
\label{eqn:ele-b-alpha-alpha}\\
\mathrm{b}_{\varphi\alpha}(t)
&=\Bigl[-\Omega\cos(\Omega t) + \frac1\tau \sin(\Omega t)\Bigr] e^{-t/\tau}
&&\text{phase}
\label{eqn:ele-b-phi-alpha}
\end{align}

\subsubsection{Remark}
%----------------------------------------------------------------------
%\sidenote{\color{magenta}This Section can be omitted}
Interestingly, the phase noise bandwidth of the resonator increases when the resonator is detuned.  This is related to the following facts.
\begin{enumerate}

\item When the resonator is detuned, it holds that 
\begin{align}
\left|\frac{d\,\arg[\beta(j\omega)]}{d\omega}\right|_{\omega_0} <
~~
\left|\frac{d\,\arg[\beta(j\omega)]}{d\omega}\right|_{\omega_n}~.
\end{align}
With lower slope, the oscillator phase noise is higher.

\item Detuning the resonator, the symmetry of $\arg[\beta(j\omega)]$ around the oscillation frequency is broken.  This explains the frequency overshoot seen in Fig.~\ref{fig:946-reson-impulse-response} for $\Omega\ne0$.

\item The step response decays faster when the resonator is detuned. 

\end{enumerate}

\subsubsection{Appendix: Formal derivation of \boldmath$[\mathrm{B}(s)]$ from $[\mathrm{b}(t)]$}\label{sssec:eln-res-am-pm-laplace}
%----------------------------------------------------------------------
%\sidenote{\color{magenta}This Section can be omitted}
For the sake of completeness, we derive the full expression of $[\mathrm{B}(s)]$ from $[\mathrm{b}(t)]$, that is \req{eqn:eln-resonator-B-matrix} from \req{eqn:eln-resonator-b-matrix}.  Thanks to the symmetry properties \req{eqn:eln-resonator-matrix-symmetry}, we only need to derive $\mathrm{B}_{\alpha\alpha}(s)$ and $\mathrm{B}_{\alpha\varphi}(s)$.

Using the well known properties 
\begin{align*}
e^{-t/\tau} &\quad\leftrightarrow\quad \frac{1}{s+1/\tau}\\
e^{at}f(t)   &\quad\leftrightarrow\quad F(s-a)~,
\end{align*}
we notice that it holds
\begin{align}
e^{\pm j\Omega t} e^{-t/\tau} \quad\leftrightarrow\quad \frac{1}{s \mp j\Omega+\frac1\tau}
\label{eqn:eln-laplace-exp-exp}~.
\end{align}

$\mathrm{B}_{\alpha\alpha}(s)=\mathcal{L}\left\{\mathrm{b}_{\alpha\alpha}(t)\right\}$ is found using \req{eqn:eln-laplace-exp-exp}, after expanding \req{eqn:ele-b-alpha-alpha} with the Euler formulae
\begin{align}
\cos(\Omega t)&=\frac12\left(e^{j\Omega t} + e^{-j\Omega t}\right) &
\sin(\Omega t)&=\frac{1}{j2}\left(e^{j\Omega t} - e^{-j\Omega t}\right)~.
\label{eqn:eln-euler}
\end{align}
Thus, 
\begin{align}
\mathrm{B}_{\alpha\alpha}(s)
&=\mathcal{L}\Bigl\{\Bigl[
   \Omega\:\frac{1}{j2}\left(e^{j\Omega t} - e^{-j\Omega t}\right)
  +\frac1\tau\: \frac12\left(e^{j\Omega t} + e^{-j\Omega t}\right)
  \Bigr] e^{-t/\tau}\Bigr\}
  \nonumber\\
&=\frac{\Omega}{j2}\left[\frac{1}{s-j\Omega+\frac1\tau}-\frac{1}{s+j\Omega+\frac1\tau}\right]
+\frac{1}{2\tau}\left[\frac{1}{s-j\Omega+\frac1\tau}-\frac{1}{s+j\Omega+\frac1\tau}\right]\nonumber\\
&=\frac{\Omega^2+s/\tau+1/\tau^2}{%
    \left(s+\frac1\tau-j\Omega\right)\left(s+\frac1\tau+j\Omega\right)}
    \nonumber\\
&=\frac{1}{\tau}\:\frac{s+\frac1\tau+\Omega^2\tau}{%
    \left(s+\frac1\tau-j\Omega\right)\left(s+\frac1\tau+j\Omega\right)}~,
\nonumber
\end{align}
and finally
\begin{align}
\mathrm{B}_{\alpha\alpha}(s)
&=\frac{1}{\tau}\:\frac{s+\frac1\tau+\Omega^2\tau}{%
    s^2+\frac{2}{\tau}s+\frac{1}{\tau^2}+\Omega^2}
    \qquad\text{qed}~.
\end{align}

Similarly, $\mathrm{B}_{\alpha\varphi}(s)=\mathcal{L}\left\{\mathrm{b}_{\alpha\varphi}(t)\right\}$ is found using \req{eqn:eln-laplace-exp-exp}, after expanding \req{eqn:ele-b-alpha-phi} with the the Euler formulae \req{eqn:eln-euler}.  Thus, 
\begin{align}
\mathrm{B}_{\alpha\varphi}(s)
&=\mathcal{L}\Bigl\{\Bigl[
   -\frac{\Omega}{2}\left(e^{j\Omega t} + e^{-j\Omega t}\right)
  +\frac{1}{j2\tau}\left(e^{j\Omega t} - e^{-j\Omega t}\right)
  \Bigr] e^{-t/\tau}\Bigr\}
  \nonumber\\
&=-\frac{\Omega}{2}\left[\frac{1}{s-j\Omega+\frac1\tau}+\frac{1}{s+j\Omega+\frac1\tau}\right]
+\frac{1}{j2\tau}\left[\frac{1}{s-j\Omega+\frac1\tau}-\frac{1}{s+j\Omega+\frac1\tau}\right]\nonumber\\
&=\frac{-\Omega s -\omega/\tau+\omega/\tau^2}{%
    \left(s+\frac1\tau-j\Omega\right)\left(s+\frac1\tau+j\Omega\right)}~,
    \nonumber\\
\nonumber
\end{align}
and finally
\begin{align}
\mathrm{B}_{\alpha\varphi}(s)
&=\frac{-\Omega s}{s^2+\frac{2}{\tau}s+\frac{1}{\tau^2}+\Omega^2}
    \qquad\text{qed}~.
\end{align}

\section{The Leeson effect}\label{sec:ele-leeson}
%==================================================
\begin{figure}[t]
\centering\namedgraphics{scale=0.6}{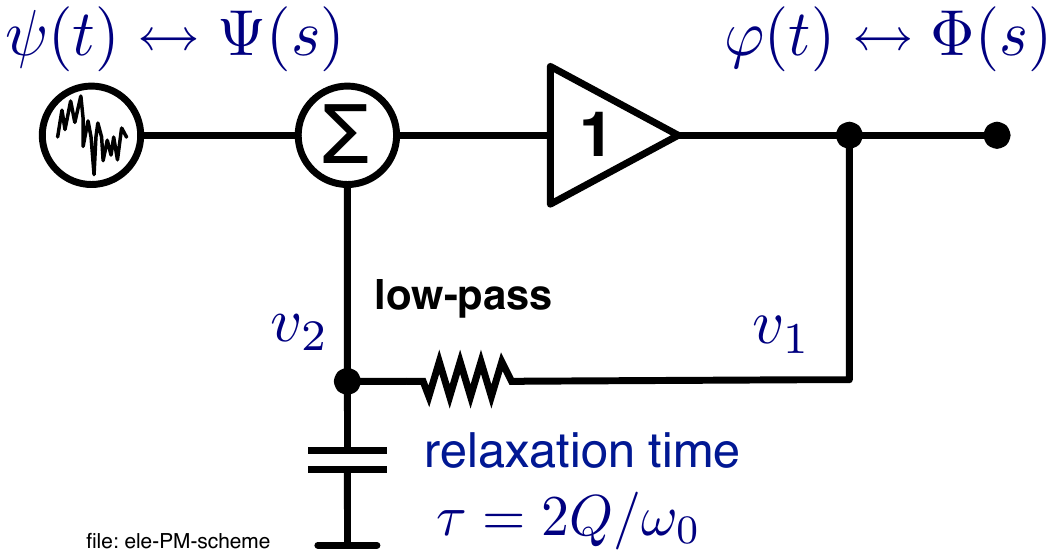}{\columnwidth}
\caption{Phase-noise model of the feedback oscillator.}
\label{fig:ele-PM-scheme}
\end{figure}
Figure \ref{fig:ele-PM-scheme} shows the phase-noise model of the oscillator.  In this figure, all signals are the phase fluctuation of the oscillator sinusoidal signal.  Here, the resonator turns into a lowpass filter of time constant $\tau$, as explained in Section~\ref{sec:ele-resonator}.  A noise-free amplifier has a gain exactly equal to one because the amplifier repeats the phase of the input signal.  The real amplifier introduces the random phase $\psi(t)$, which in this representation is additive noise, regardless of the physical origin.  For the sake of simplicity, we put in $\psi(t)$ all the phase-noise sources.

\begin{figure}[t]
\centering\namedgraphics{scale=0.6}{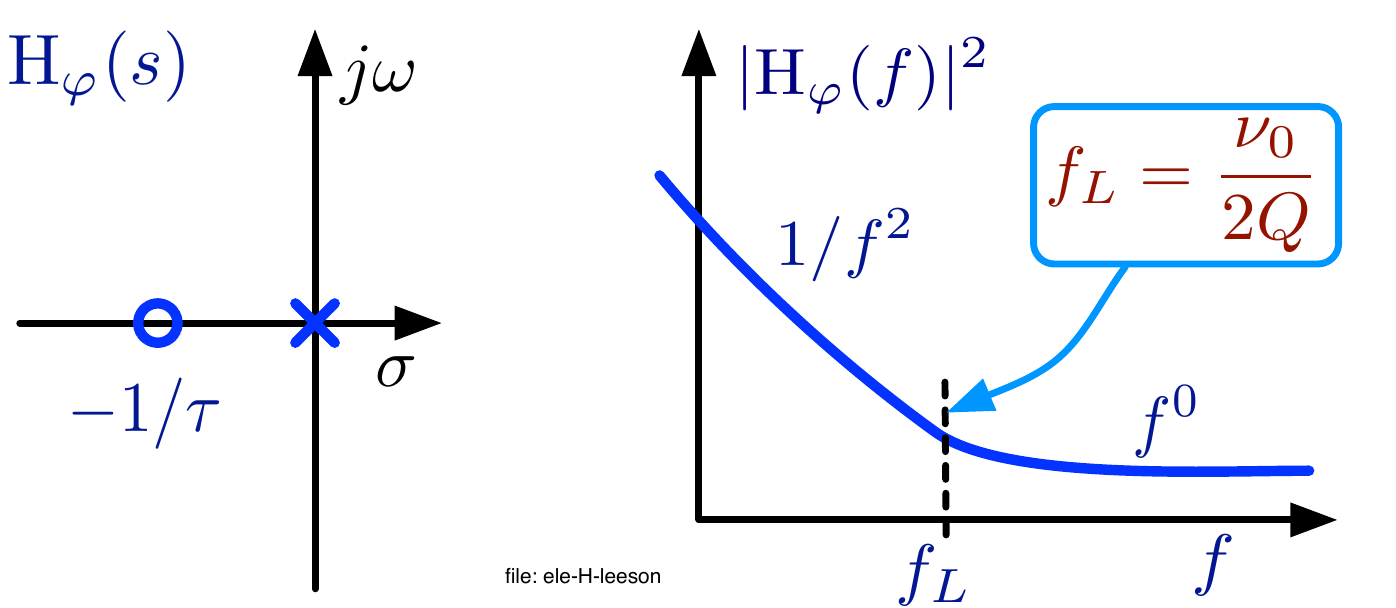}{\columnwidth}
\caption{Phase-noise transfer function.}
\label{fig:ele-H-leeson}
\end{figure}

We define the phase-noise transfer function as
\begin{align*}
\mathrm{H}(s)=\frac{\Phi(s)}{\Psi(s)}~.
\end{align*}
Applying the elementary feedback theory to the circuit of Fig.~\ref{fig:ele-PM-scheme} we find
\begin{align*}
\mathrm{H}(s)
&=\frac{1}{1+\mathrm{B}(s)}~,
\end{align*}
where $\mathrm{B}(s)$ is the resonator transfer function \req{eqn:eln-resonator-response-B}, and therefore
\begin{align}
\mathrm{H}(s)&=\frac{s+1/\tau}{s}
\label{eqn:ele-leeson-effect}
\qquad\text{(Fig.~\ref{fig:ele-H-leeson})}~.
\end{align}
This is the Leeson effect, by which the oscillator integrates the slow phase fluctuation, turning it into frequency fluctuation.  The phase-noise transfer function is plotted in Fig.~\ref{fig:ele-H-leeson}.

\section{Low-pass model of the oscillator amplitude}\label{sec:ele-am-model}
%==================================================
\begin{figure}[t]
\centering\namedgraphics{scale=0.6}{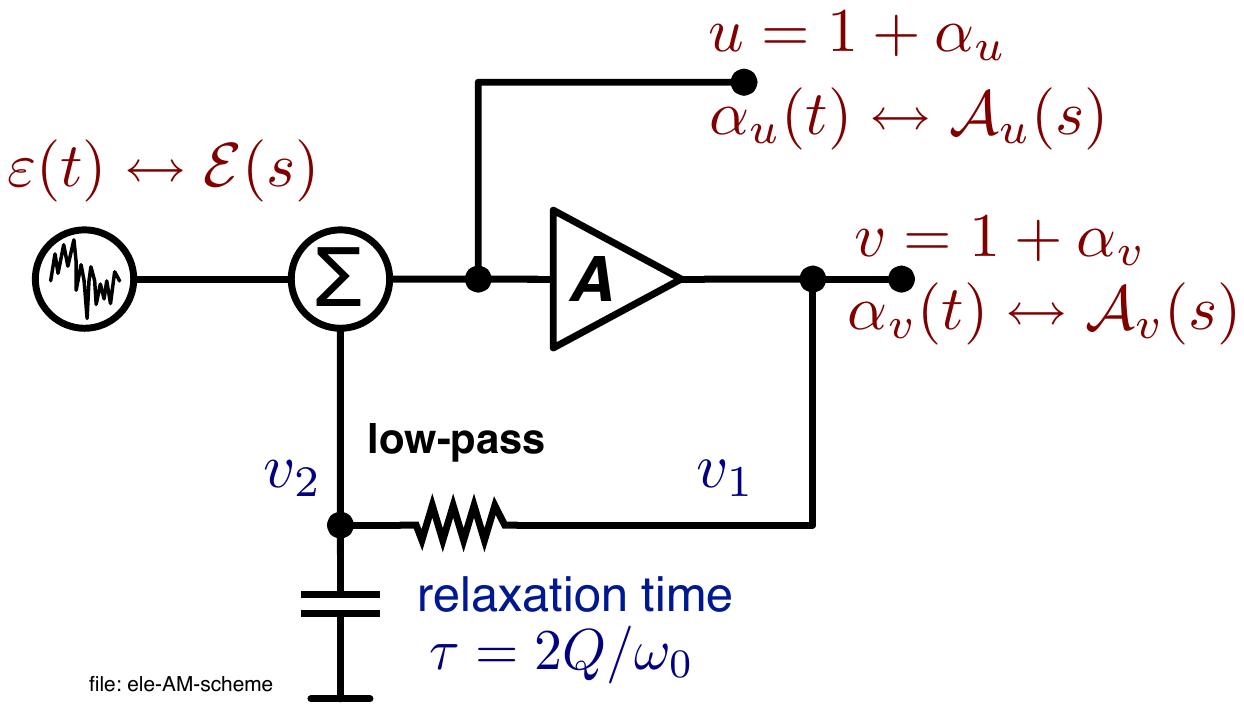}{\columnwidth}
\caption{Amplitude-noise model of the feedback oscillator.}
\label{fig:ele-AM-scheme}
\end{figure}

Figure \ref{fig:ele-AM-scheme} shows the low-pass model that describes the oscillator amplitude.  Since the gain $A$ depends on amplitude, the Laplace/Heaviside formalism cannot be used directly.  We first need to linearize the system in the appropriate conditions.

%-----------------------------------------------------------------------------------
\subsection{Differential equation}\label{ssec:ele-am-de}
%----------------------------------------------------------------------------------
Cutting the feedback loop at the amplifier input, we get 
\begin{align*}
u = \varepsilon + v_2~,
\end{align*}
where $v_2$ results from the lowpass filter
\begin{align*}
v_2 = \frac1\tau \int(v_1-v_2) \;dt~.
\end{align*} 
Combining the above equations and replacing $v_1=Au$  and $v_2=u-\varepsilon$, we get 
\begin{align}
u-\frac1\tau\int(A-1)u \:dt &=  \varepsilon + \frac1\tau\int\varepsilon\:dt
&&\text{(general IE)}~,
\label{eqn:ele-general-am-ie}
\intertext{thus}
\dot{u}-\frac1\tau\left(A-1\right)u &=  \dot{\varepsilon} + \frac1\tau\varepsilon
&&\text{(general DE)}~.
\label{eqn:ele-general-am-de}
\end{align} 
Notice that \req{eqn:ele-general-am-ie}-\req{eqn:ele-general-am-de} are general because $A$ is still unspecified.
Substituting $A=1-\gamma(u-1)$, as in Fig.~\ref{fig:ele-clipping-types}, 
\req{eqn:ele-general-am-de} becomes\sidenote{Fixed a sign}
\begin{align}
\dot{u}+\frac{\gamma}{\tau} \bigl(u-1\bigr)u &= \dot{\varepsilon} + \frac1\tau\varepsilon
&&\text{with $A=1-\gamma(u-1)$}~.
\label{eqn:ele-am-de-compression}
\end{align}
The system free running is governed by the homogeneous equation\sidenote{Consequently}
\begin{align}
\dot{u}+\frac{\gamma}{\tau} \bigl(u-1\bigr)u &= 0~.
\label{eqn:ele-am-de-hom}
\end{align}
The solution is
\begin{align}
u(t) = \frac{1}{1+Ce^{-\gamma t/\tau}}
&&\text{(solution of \req{eqn:ele-am-de-hom})}~,
\label{eqn:ele-am-sol-hom}
\end{align}
where $C$ is a constant determined by the initial conditions.

%-----------------------------------------------------------------------------------
\subsection{Simplified oscillator model}
%-----------------------------------------------------------------------------------
A simplified model for the oscillator\sidenote{Moved here and rewritten} is obtained by assuming that the linear approximation $A=1-\gamma(u-1)$ holds in the whole amplitude range.  
One can object that this case is only of academic interest because in real amplifiers the parameter $\gamma$ is constant only in a narrow region around $u=1$, as shown in Fig.~\ref{fig:ele-clipping-types}.
Nonetheless, the general description that follows can be easily adapted to practical cases.  

\begin{figure}[t]
\centering\namedgraphics{scale=0.6}{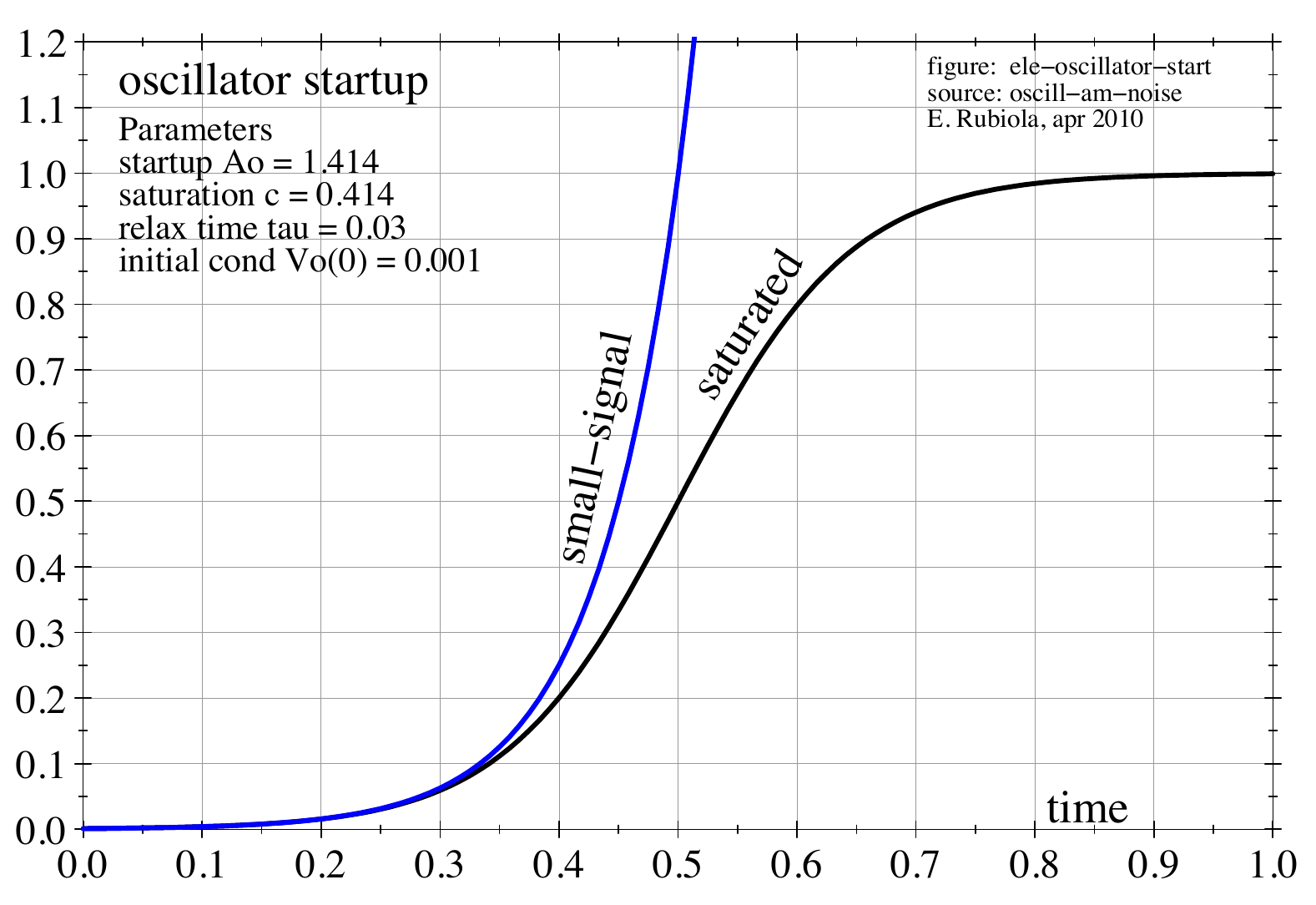}{\columnwidth}
\caption{Oscillator startup.}
\label{fig:ele-oscillator-start}
\end{figure}

\begin{figure}[t]
\centering\namedgraphics{scale=0.9}{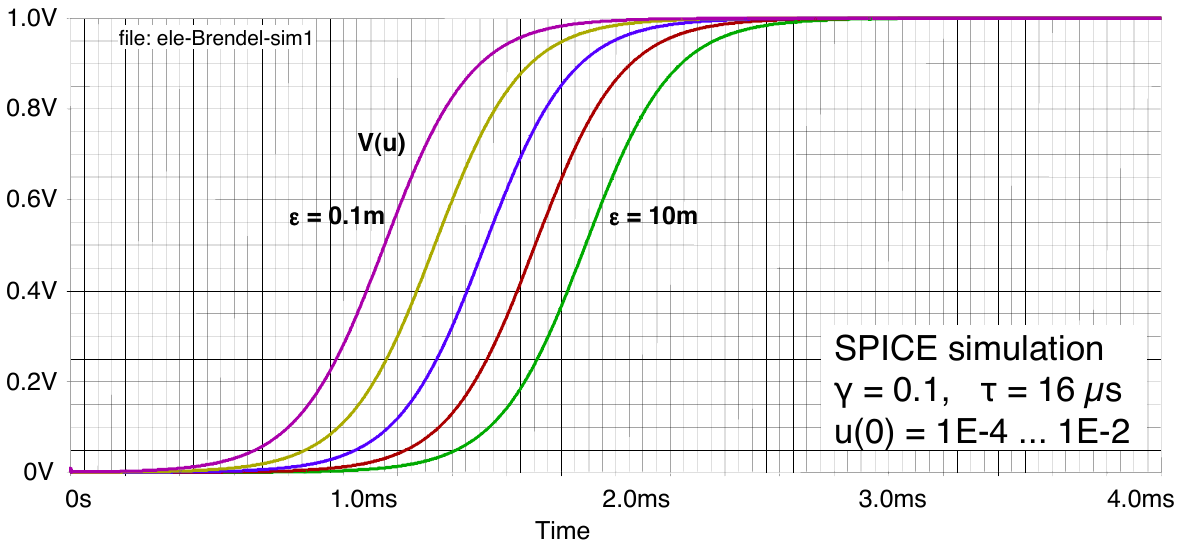}{\columnwidth}
\caption{Simulation of the startup of the model shown in Fig.~\ref{fig:ele-AM-scheme}, assuming that the amplifier law $A=1-\gamma(u-1)$ holds in the full range.}
\label{fig:ele-Brendel-sim1}
\end{figure}

Assuming that $A=1-\gamma(u-1)$, the oscillator amplitude is fully described by \req{eqn:ele-am-sol-hom},
where $C$ is related to the initial $u(0)$ by
\begin{align*}
u(0)=\frac{1}{1+C} \qquad\rightarrow\qquad
C=\frac{1}{u(0)}-1~.
\end{align*} 
Thus, 
\begin{align}
u(t)=\frac{1}{1 + \left(\frac{1}{u(0)}-1\right) \:e^{-\gamma t/\tau}}~.
\label{eqn:ele-am-simplif-model}
\end{align}
In the absence of a switch-on transient, oscillation starts from noise.  Thus, $u(0)$ is a small positive quantity, hence $1/u(0)-1\simeq1/u(0)$ and
\begin{align}
u(t)\simeq\frac{1}{1 + \frac{1}{u(0)}\:e^{-\gamma t/\tau}}
&&0<u(0)\ll1~.
\label{eqn:ele-am-simplif-model-small-u0}
\end{align}
Accordingly the following asymptotic expression hold
\begin{align}
u(t) &= u(0) \, e^{\gamma t/\tau}	&& t\rightarrow0 \label{eqn:ele-am-simplif-start}\\
u(t) &= 1	&& t\rightarrow\infty
\end{align} 
Figure~\ref{fig:ele-oscillator-start} shows the complete oscillation start \req{eqn:ele-am-simplif-model},\sidenote{Added Fig.~\ref{fig:ele-Brendel-sim1}} which saturates to $u=1$, and the small-signal approximation \req{eqn:ele-am-simplif-start}.
Figure~\ref{fig:ele-Brendel-sim1} shows a Spice simulation. 

%-----------------------------------------------------------------------------------
\subsection{Oscillation soft-start}\label{ssec:ele-soft-start}
%-----------------------------------------------------------------------------------
Actual oscillators\sidenote{Moved here and rewritten} differ from the above simplified model in that the small-signal gain follows the law $A=1-\gamma(u-1)$ only in the vicinity of $u=1$, as shown in Fig.~\ref{fig:ele-clipping-types}.  In the absence of a general model, we denote the small-signal gain with $A_0$, which is a circuit-specific parameter that we assume to be constant for $u\rightarrow0$.
Replacing $A=A_0$ (constant), the homogeneous equation \req{eqn:ele-am-de-hom} becomes
\begin{align*}
&\dot{u}-\frac1\tau \left(A_0-1\right)u = 0~.
\end{align*} 
The solution is
\begin{align}
u(t)=u(0)\,e^{(A_0-1) \, t/\tau}~,
\label{eqn:ele-soft-start}
\end{align}
where $u(0)$ is the small initial condition set by noise.  This solution is similar to \req{eqn:ele-am-simplif-start}, but for the different value of the time constant.

\begin{figure}[t]
\centering\namedgraphics{scale=0.70}{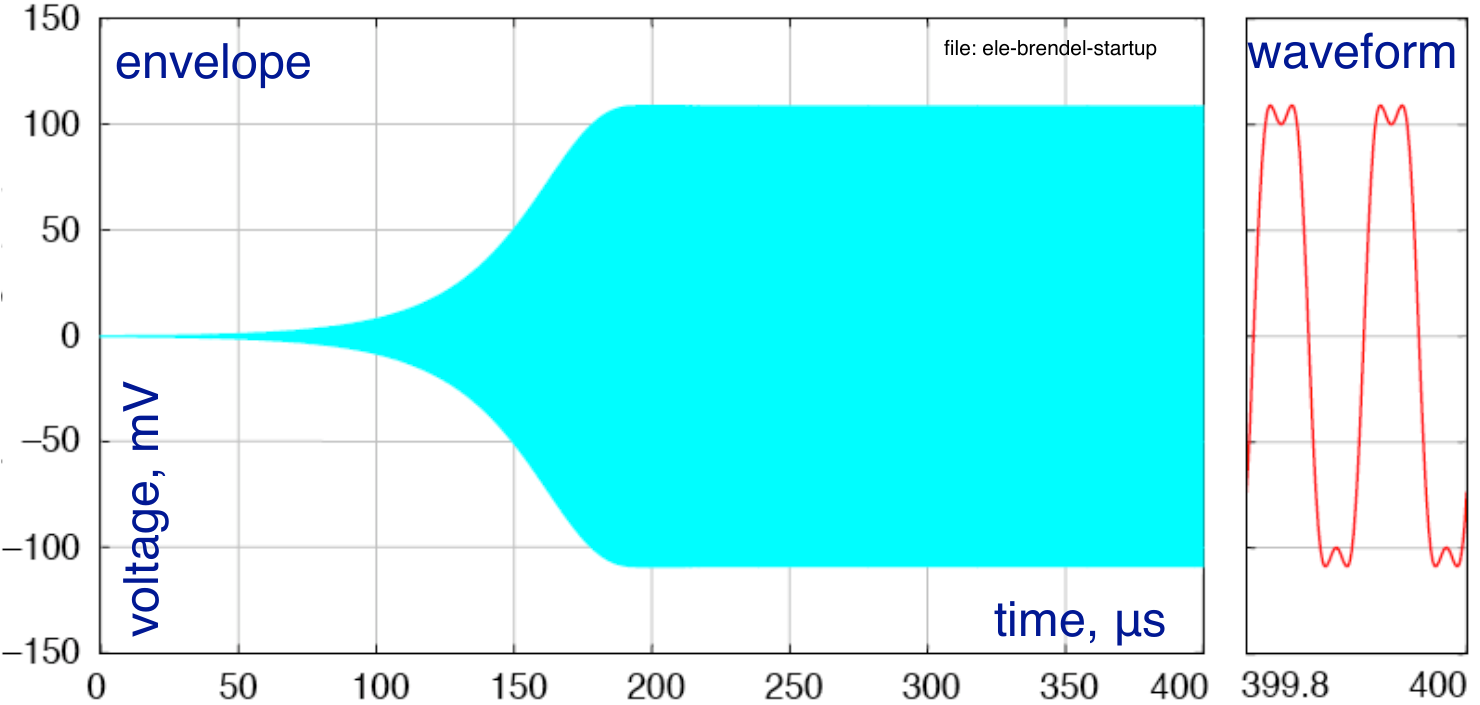}{\columnwidth}
\caption{Simulated oscillator startup.}
\label{fig:ele-Brendel-startup}
\end{figure}

A number of computer simulations were done independently by R. B. well before the approach presented here was developed \cite{Addouche:phd,Brendel2003uffc}.  This led to the preliminary work published in \cite{brendel2007fcs}.  Figure \ref{fig:ele-Brendel-startup}\sidenote{Fig.~\ref{fig:ele-Brendel-startup} moved here} shows the simulated startup.  The left-hand side of the envelope, until $t\approx100$ $\mu$s, fits well the theoretical prediction \req{eqn:ele-soft-start}.

%--------------------------------------------------------------------------
\subsection{Restoring mechanism}\label{ssec:ele-restoring}
%--------------------------------------------------------------------------
It is interesting\sidenote{Rewritten} to study the amplitude free running when the initial condition is set close to the steady state, thus at $u(0)=1+\kappa$ with $0<\kappa\ll1$.  In this conditions the approximation $A=1-\gamma(u-1)$ holds, and therefore the amplitude is given by \req{eqn:ele-am-simplif-model} with $u(0)=1+\kappa$
\begin{align}
u(t) &= \frac{1}{1-\dfrac{\kappa}{1+\kappa}\,e^{-\gamma t/\tau}}
&& \text{($u(0)=1+\kappa$)}~,
\label{eqn:ele-am-sol-hom-kappa}
\end{align} 
For small $\kappa$, this is linearized as 
\begin{align}
u(t) &= 1+\kappa e^{-\gamma t/\tau}~,
\label{eqn:ele-am-sol-hom-kappa-lin}
\intertext{or equivalently}
\alpha_u(t) &= \kappa e^{-\gamma t/\tau}
\label{eqn:ele-am-sol-hom-alpha-lin}
\end{align} 
because $\alpha_u=u-1$ (Fig.~\ref{fig:ele-AM-scheme}).
The time constant 
\begin{align}
\tau_r=\frac{\tau}{\gamma}
\qquad\text{(restoring time)}
\end{align} 
is the oscillator restoring time for amplitude perturbations.  Since in virtually all amplifiers it holds that $0<\gamma<1$, as widely discussed in Section \ref{ssec:ele-saturation-models}, it holds that $\tau_r>\tau$.

%-----------------------------------------------------------------------------------
\subsection{Amplitude impulse response}
%-----------------------------------------------------------------------------------
Now, we study\sidenote{Rewritten} the oscillator response to the amplitude impulse $\varepsilon(t)=\delta(t)$ occurring when the oscillator is in the steady state $u=1$, $v=1$.  The impulse response is the derivative of the step response, linearized for small perturbation.   Thus, referring to Fig.~\ref{fig:ele-AM-scheme}, we apply at the input the small step 
\begin{align*}
\varepsilon(t)=\kappa\mathfrak{u}(t)\qquad 0<\kappa\ll1~.
\end{align*}
We know from Section~\ref{ssec:ele-restoring} that the small-signal response is a decaying exponential $e^{-\gamma t/\tau}$.  Hence the response is completely determined by the initial and final values
\begin{align}
u(t)=u(\infty) + \bigl[u(0_+)-u(\infty)\bigr] \, e^{-\gamma t/\tau} \qquad t>0~.
\label{eqn:ele-general-e-step-resp}
\end{align}
Since the perturbation takes time to propagate through the lowpass filter, it holds that 
\begin{math}u(0_+)=1+\kappa\end{math}.  The final value is obtained by inspection on Fig.~\ref{fig:ele-AM-scheme} after bypassing the lowpass filter and setting $\epsilon(t)=\kappa$.  Thus
\begin{align*}
u&=v_2+\kappa &&\text{(adder)}\\
v_2&=\left[1-\gamma(u-1)\right]u &&\text{(amplifier and filter)}~,
\intertext{hence}
u&=\left[1-\gamma(u-1)\right]u+\kappa &&\text{($t\rightarrow\infty$)}~.
\end{align*}
The algebraic solutions 
\begin{align*}
u_1&=\frac{\gamma-\sqrt{\gamma^2+4\gamma\kappa}}{2\gamma}
&u_2&=\frac{\gamma+\sqrt{\gamma^2+4\gamma\kappa}}{2\gamma}~,
\intertext{and for small $\kappa$}
u_1&\asymp-\frac\kappa\gamma
&u_2&\asymp1+\frac\kappa\gamma&&\kappa\rightarrow0~.
\end{align*}
It is immediately seen that $u_1<0$ and $u2>0$.  Hence $u_2$ is the physical solution while $u_1$ is discarded.  Setting $\kappa=1$ (unit step) and using \req{eqn:ele-general-e-step-resp}, we find the step response
\begin{gather}
\int\mathrm{h}^{(\varepsilon)}_u(t)\,dt=1 + \frac{1}{\gamma}\,\mathfrak{u}(t) -
	\left(\frac{1}{\gamma}-1\right)\,e^{-\gamma t/\tau} \,\mathfrak{u}(t)
\label{eqn:ele-e-step-resp}~.
\end{gather}
Notice that the term `$1+$' is the steady state before the step is applied.
The subscript $u$ and the superscript $(\varepsilon)$, which refer to the input $\varepsilon$ and to the output $u$, are introduced to emphasize the difference versus other response functions.

The impulse response is found by differentiating \req{eqn:ele-e-step-resp}
\begin{gather}
\mathrm{h}^{(\varepsilon)}_u(t) = \delta(t) + 
\frac\gamma\tau \left(\frac{1}{\gamma}-1\right)\,e^{-\gamma t/\tau} \,\mathfrak{u}(t)
\label{eqn:ele-e-impulse-resp}~.
\end{gather}

The Laplace transform $H^{(\varepsilon)}_u(s) = \mathcal{L}\{h^{(\varepsilon)}_u(t)\}$ is found immediately using the properties $\mathcal{L}\{\delta(t)\}=1$ and $\mathcal{L}\{e^{-at}\}=\frac{1}{s+a}$
\begin{align}
\mathrm{H}^{(\varepsilon)}_u(s) &= 1 + \frac\gamma\tau \left(\frac1\gamma-1\right)\frac{1}{s+\gamma/\tau}~,\nonumber
\intertext{which simplifies to}
\mathrm{H}^{(\varepsilon)}_u(s) &= \frac{s+1/\tau}{s+\gamma/\tau}~.
\end{align}
So, the transfer function is completely determined by the roots
\begin{align*}
&\text{(pole)}	&	s_p&=-\gamma/\tau	&	f_p&=\gamma f_L < f_L\\
&\text{(zero)}	&	s_z&=-1/\tau		&	f_z&=f_L~~.
\end{align*}

\section{Extension of the Leeson effect to AM noise}\label{sec:ele-extended-leeson}
%==================================================
In this Section we study the effect of the parametric fluctuation of the gain by introducing the random variable $\eta(t)$, as anticipated in Section~\ref{sssec:ele-gain-fluctuation} and Fig.~\ref{fig:ele-gain-fluctuation}
\begin{gather}
A=1-\gamma(u-1)+\eta
\label{eqn:ele-gain-fluctuation}\\
\eta(t)\leftrightarrow\mathcal{N}(s)\nonumber~.
\end{gather}
We linearize the system for low noise, and we search for the transfer functions
\begin{gather*}
\mathrm{H}^{(\eta)}_u(s)=\frac{\mathcal{A}_u(s)}{\mathcal{N}(s)}
\qquad\text{and}\qquad
\mathrm{H}^{(\eta)}_v(s)=\frac{\mathcal{A}_v(s)}{\mathcal{N}(s)}~,
\end{gather*}
where
\begin{gather*}
\alpha_u(t)\leftrightarrow\mathcal{A}_u(s)
\qquad\text{and}\qquad
\alpha_v(t)\leftrightarrow\mathcal{A}_v(s)~,
\end{gather*}
are the amplitude fluctuations at the amplifier input and output, respectively.

\subsection{Noise at the amplifier input}
\begin{figure}[t]
\centering\namedgraphics{scale=0.6}{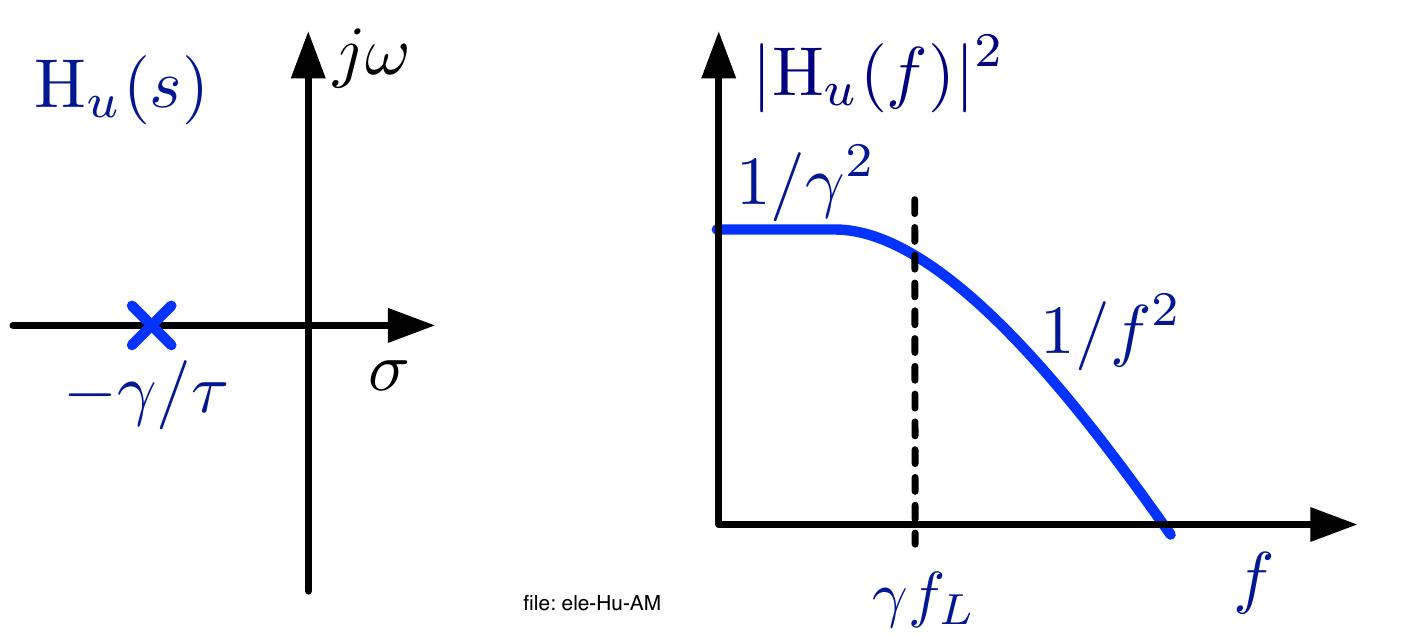}{\columnwidth}
\caption{Amplitude-noise transfer function (amplifier input).}
\label{fig:ele-Hu-AM}
\end{figure}

%-----------------------------------------------------------------------------------
By replacing $A=1-\gamma(u-1)+\eta$ in the homogeneous equation \req{eqn:ele-am-de-hom}, we get
\begin{gather*}
\dot{u}+\frac{\gamma}{\tau}(u-1)u=\frac\eta\tau u~.
\end{gather*}
Since $u=1+\alpha_u$, it holds that $\dot{u}=\dot{\alpha}_u$ and $u-1=\alpha_u$, thus
\begin{gather*}
\dot{\alpha}_u+\frac{\gamma}{\tau}\alpha_u\,u=\frac\eta\tau u~.
\end{gather*}
For small fluctuations $\alpha_u$ and $\eta$, we linearize the above using $u\simeq1$
\begin{gather*}
\dot{\alpha}_u+\frac{\gamma}{\tau}\alpha_u=\frac1\tau\eta~.
\end{gather*}
The linearized system can now be described in using the Laplace transforms
\begin{gather}
\left(s+\frac{\gamma}{\tau}\right)\mathcal{A}_u(s)=\frac1\tau\mathcal{N}(s)~,
\label{eqn:eln-AuN}
\end{gather}
which gives the transfer function (Fig.~\ref{fig:ele-Hu-AM})
\begin{gather}
\mathrm{H}^{(\eta)}_u(s)=\frac{1/\tau}{s+\gamma/\tau}~.
\end{gather}

\subsection{Noise at the amplifier output}
\begin{figure}[t]
\centering\namedgraphics{scale=0.6}{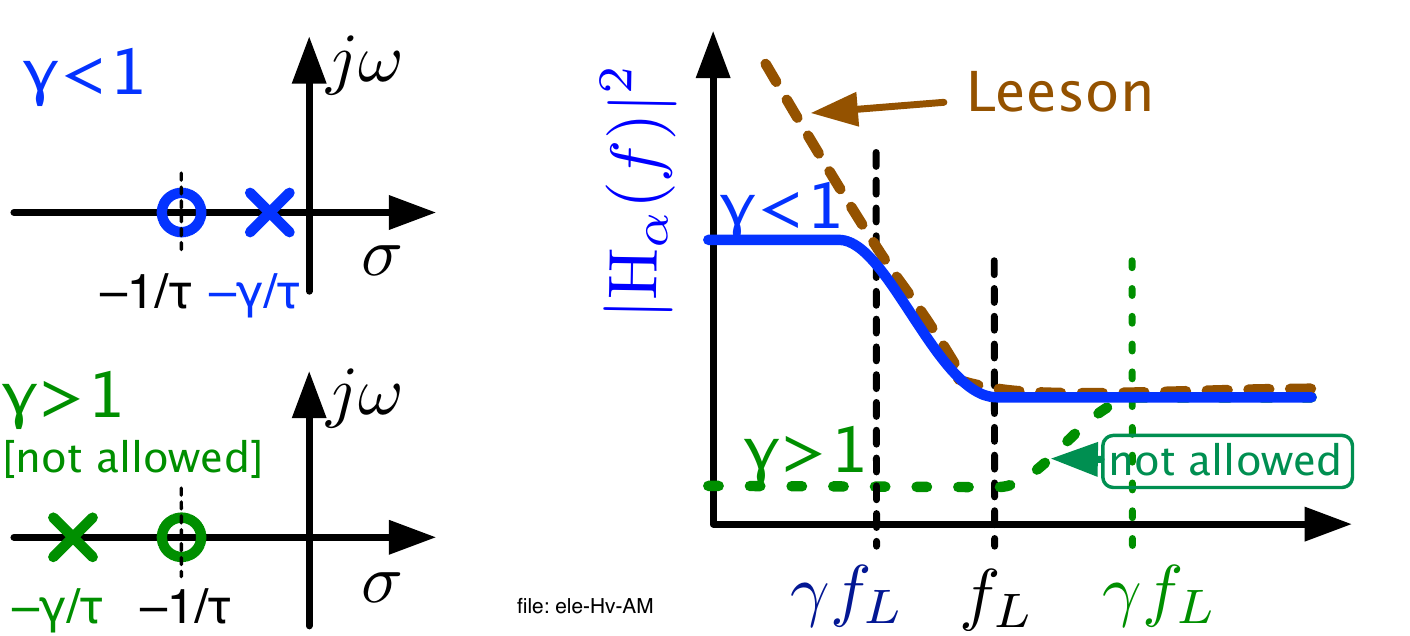}{\columnwidth}
\caption{Amplitude-noise transfer function (oscillator output).}
\label{fig:ele-Hv-AM}
\end{figure}

%-----------------------------------------------------------------------------------
We first need to relate $\alpha_v$ to $\alpha_u$.  This is done by replacing $A=-\gamma(u-1)+1+\eta$ in $v=Au$
\begin{align*}
v&=\left[-\gamma(u-1)+1+\eta\right]u~,
%\\&=\left[-\gamma\alpha_u+1+\eta\right]\,\left[1+\alpha_u\right]
\end{align*}
and by expanding using $v=1+\alpha_v$ and $u=1+\alpha_u$
\begin{align*}
1+\alpha_v=1+\eta-\gamma\alpha_u+\alpha_u-\alpha_u\eta-\gamma\alpha_u^2~.
\end{align*}
Neglecting the second-order noise terms $\alpha_u\eta$ and $\alpha^2$
\begin{align*}
\alpha_v=(1-\gamma)\alpha_u+\eta~,
\end{align*}
we get 
\begin{align*}
\alpha_u=\frac{\alpha_v-\eta}{1-\gamma}
\qquad\leftrightarrow\qquad
\mathcal{A}_u(s)=\frac{\mathcal{A}_v(s)-\mathcal{N}(s)}{1-\gamma}~.
\end{align*}
Then, by replacing the above $\mathcal{A}_u(s)$ in Eq.~\req{eqn:eln-AuN} we get
\begin{align*}
\Bigl(s+\frac{\gamma}{\tau}\Bigr)\mathcal{A}_v(s)=
\Bigl(s+\frac{1}{\tau}\Bigr)\mathcal{N}(s)~,
\end{align*}
and finally
\begin{align}
\mathrm{H}^{(\eta)}_v(s)=\frac{s+1/\tau}{s+\gamma/\tau}~.
\label{eqn:ele-am-Hv}
\end{align}
The transfer function $\mathrm{H}^{(\eta)}_v(s)$ is shown in Fig.~\ref{fig:ele-Hv-AM}
Notice that the case $\gamma>1$ (dashed green curve) is not allowed by the condition $0<\gamma<1$ for the amplifier gain.

\begin{figure}[t]
\centering\namedgraphics{scale=0.9}{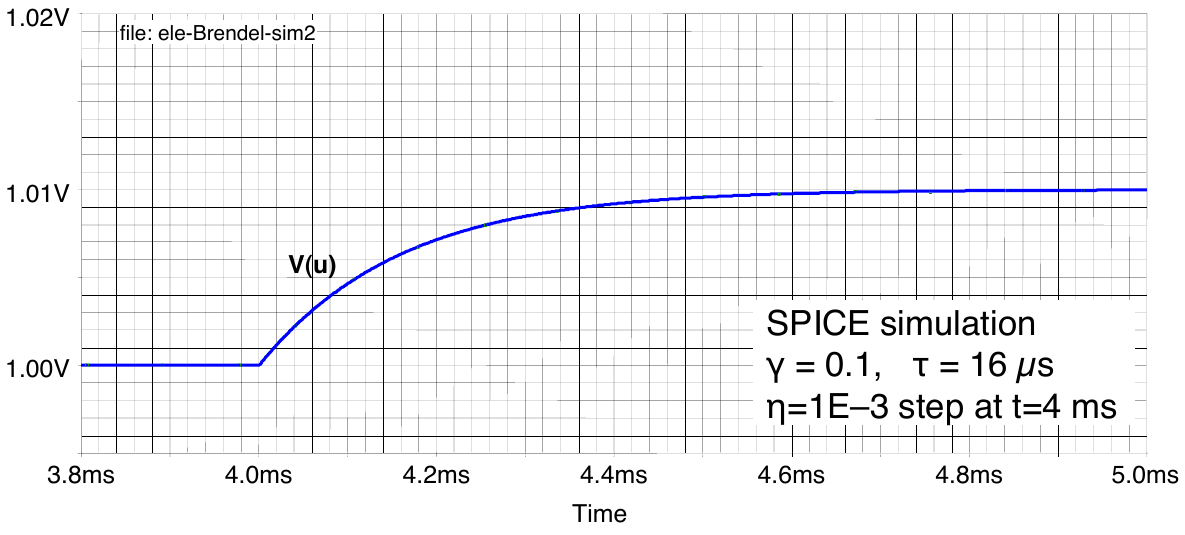}{\columnwidth}
\caption{Simulation of the step response of the AM model shown in Fig.~\ref{fig:ele-AM-scheme}.}
\label{fig:ele-Brendel-sim2}
\end{figure}

Figure~\ref{fig:ele-Brendel-sim2} show a SPICE simulation\sidenote{Fig.~\ref{fig:ele-Brendel-sim2} moved here} of the oscillator response to a gain step $\eta=10^{-3}$, assuming that the gain-compression parameter is $\gamma=0.1$.  The rising exponential reaches the final value $1+\eta/\gamma$ with a time constant $\tau/\gamma$, which confirms Eq.~\req{eqn:ele-am-Hv}.\sidenote{This was a full section at end}

\subsection{Predicted spectra}
%----------------------------------------------------------------------------
We calculate the oscillator AM and PM spectra due to the Leeson effect alone.
The fluctuation of the resonator natural frequency, not accounted in this Section, may be added afterwards.
With the remarkable exception of the laser, virtually all practical oscillators are followed by a buffer, which contributes with its own noise.
Referring to first plot of Fig.~\ref{fig:ele-spectrum-PM} (amplifier PM noise), we notice that the output buffer has higher flicker and lower white noise than the sustaining amplifier.  The buffer flicker is higher because the buffer has higher number of stages, each of which adds its $1/f$ phase noise independent of the carrier power [Eq.~\req{eqn:ele-flicker}].  Conversely, the buffer white noise is lower because this type of noise is additive and the input power is higher at the buffer input [Equations \req{eqn:ele-Friis-AM}--\req{eqn:ele-Friis-PM}].  The same is seen on the first plot of Fig.~\ref{fig:ele-spectrum-AM}, which refers to the amplifier AM noise.

\subsubsection{Phase noise}
%----------------------------------------------------------------------------
\begin{figure}[t]
\centering\namedgraphics{scale=0.6}{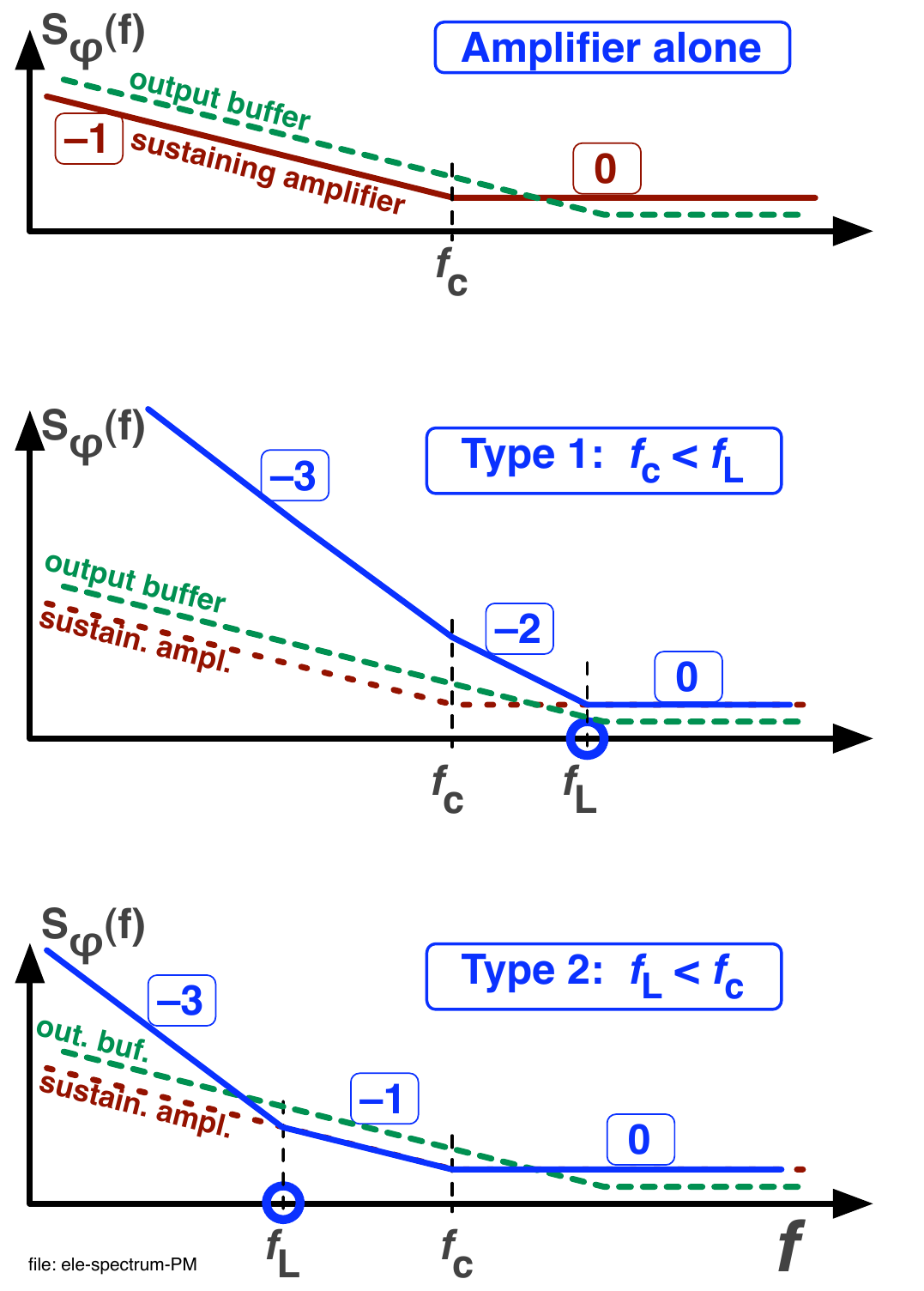}{\columnwidth}
\caption{Phase-noise spectrum in log-log scale.  The cutoff frequency generated by the zero of the transfer function at $s=-1/\tau$ is shown on the horizontal axis.  The pole at $s=0$ cannot be shown on a logarithmic scale.}
\label{fig:ele-spectrum-PM}
\end{figure}
With reference to Fig.~\ref{fig:ele-spectrum-PM}, the analysis starts from the sustaining-amplifier noise, which shows the flicker corner at $f=f_c$.  This noise is turned into the oscillator noise by the transfer function $\mathrm{H}_\varphi(f)$ [Eq.~\req{eqn:ele-leeson-effect}], which is completely described by a pole at $f=0$ and a zero at $f_L=\frac{1}{2\pi\tau}$ on the Bode plot.

With a low-$Q$ resonator we get the spectrum of the Type 1, where $f_L>f_c$.  At the oscillator output, before buffering, only the slopes $1/f^3$, $1/f^2$ and $f^0$ are present.
The buffer noise is generally not visible because it rises with lower slope ($1/f$) on the right hand of the plot.  
Only in a few special cases, when noise special techniques are used to reduce the phase noise of the sustaining amplifier \cite{howe2005fcs,ivanov98mtt}, thus the gap between the flicker of the buffer and of the sustaining amplifier is large, some $1/f$ noise shows up at the buffered output in the region around $f_L$.

If the resonator $Q$ is higher we get the spectrum of the Type 2, where $f_L<f_c$.  Before buffering, only the slopes $1/f^3$, $1/f$ and $f^0$ are present.
The buffer $1/f$ noise shows up because it is higher than the sustaining-amplifier noise and has the same slope.

%-------------------------------------------------------------
\subsubsection{Amplitude noise}
%-------------------------------------------------------------
\begin{figure}[t]
\centering\namedgraphics{scale=0.6}{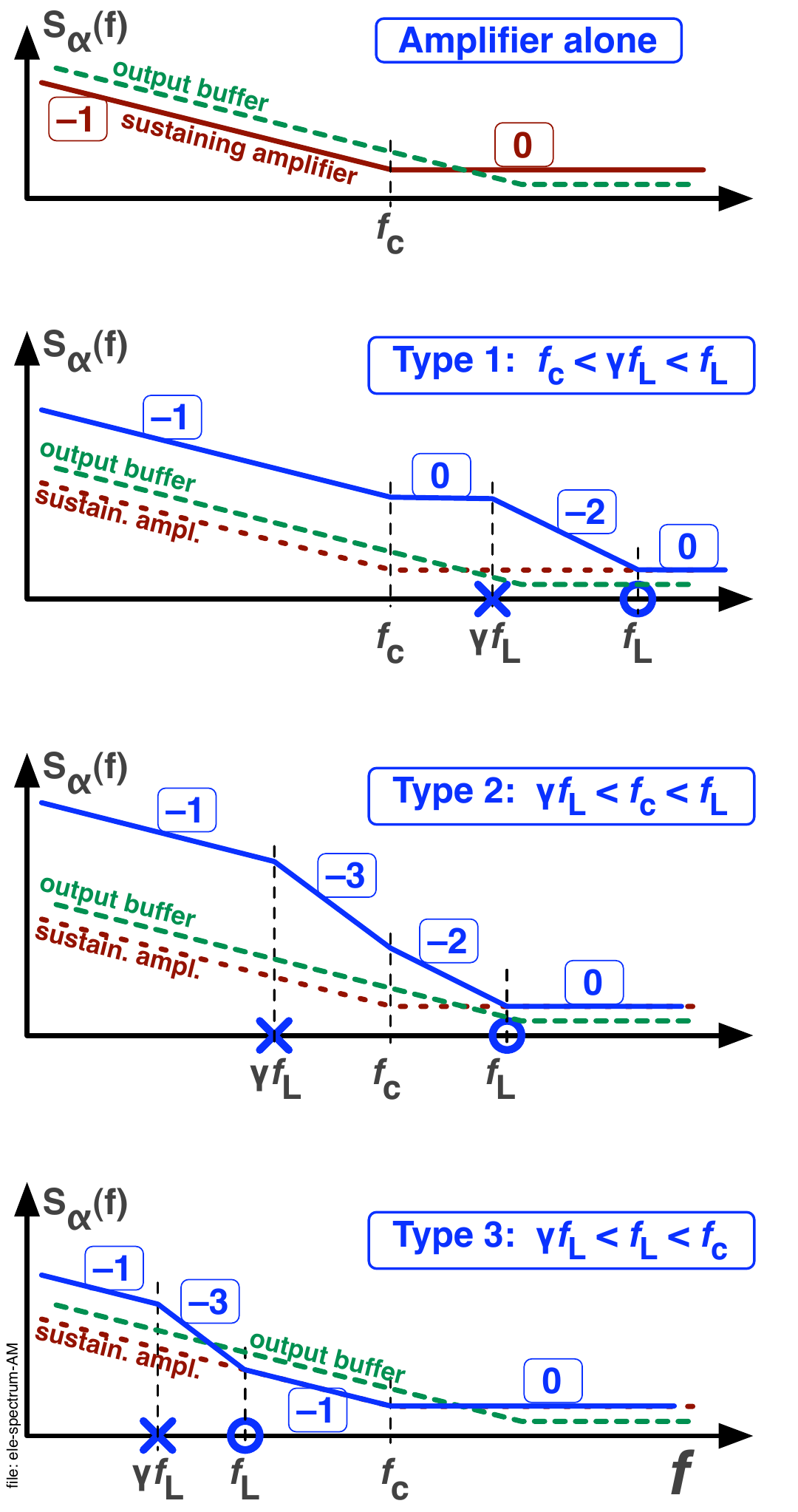}{\columnwidth}
\caption{Amplitude-noise spectrum in log-log scale.  The cutoff frequencies generated by the roots of the transfer function at $s=-\gamma/\tau$ and $s=-1/\tau$ are shown on the horizontal axis.}
\label{fig:ele-spectrum-AM}
\end{figure}
The amplitude noise (Figure~\ref{fig:ele-spectrum-AM}) is more complex than the phase noise because the transfer function $\mathrm{H}_v(f)$ [Eq.~\req{eqn:ele-am-Hv}] is described by two roots, a real pole at $f=\frac{\gamma}{2\pi\tau}$ and a real zero at $f_L=\frac{1}{2\pi\tau}$.  Increasing the resonator $Q$, these roots may occur both on the right-hand of $f_c$ in the Type-1 spectrum (low $Q$), one on the left hand and the other on the right hand of $f_c$ in the Type-2 spectrum (medium $Q$), and both on the left-hand of $f_c$ in the Type-3 spectrum (high $Q$).

Generally, the buffer $1/f$ noise shows up only in the Type-3 spectrum, in the $1/f$ region between $f_L$ and $f_c$.  It may also show up in the Type-2 spectrum around $f_L$ if the gap between the flicker of the buffer and of the sustaining amplifier is made large by the use of a noise-degeneration sustaining amplifier.

%=======================================================
\section{AM-PM coupling in the amplifier}\label{sec:ele-ampli-ampm-correl}
%=======================================================
\begin{figure}[t]
\centering\namedgraphics{scale=0.55}{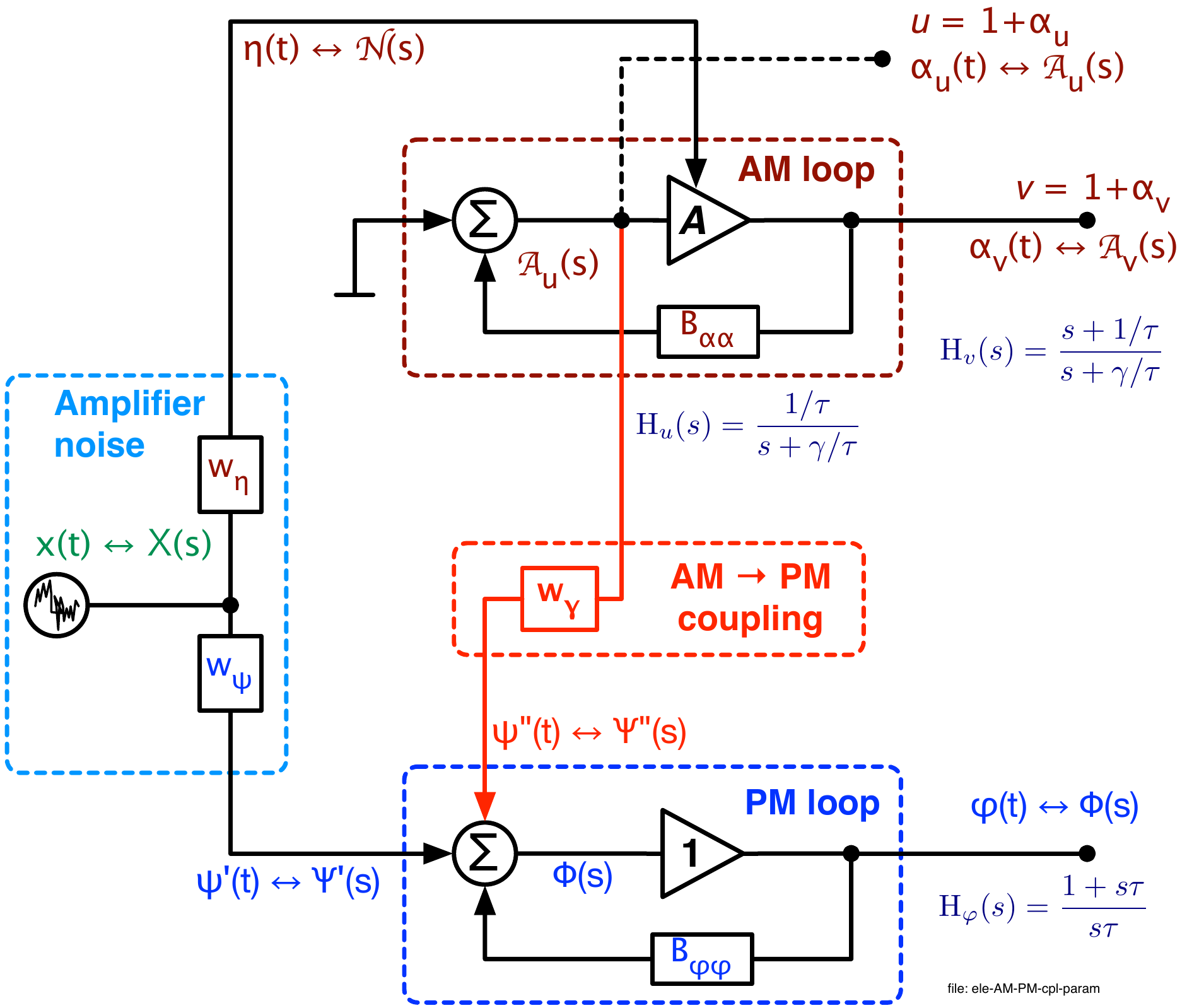}{\columnwidth}
\caption{Combined effect of the AM-PM coupling in amplifier and the loop, which forces the Barkhausen condition.}
\label{fig:ele-AM-PM-cpl-param}
\end{figure}
\begin{figure}[t]
\centering\namedgraphics{scale=0.5}{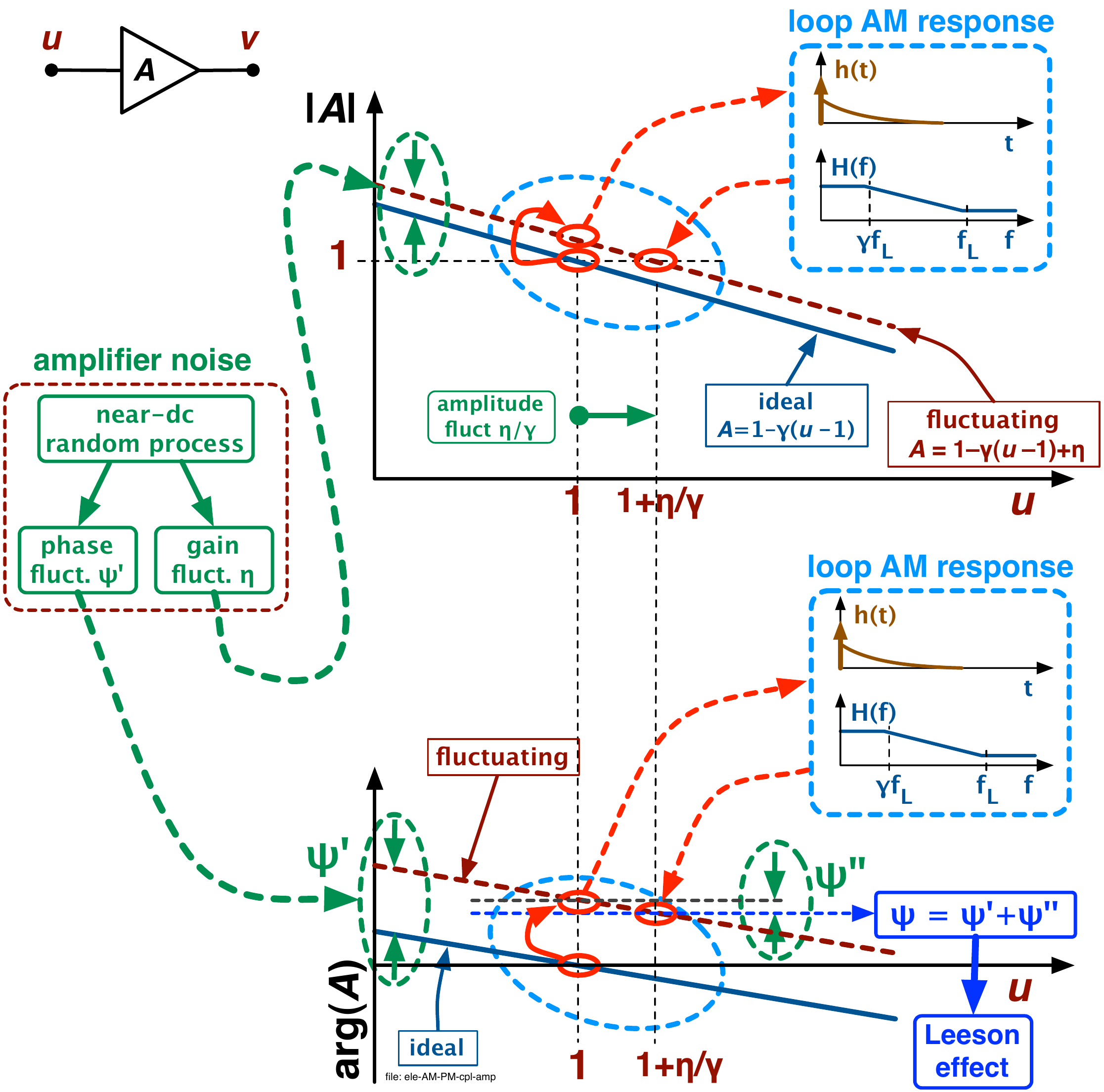}{\columnwidth}
\caption{Combined effect of the AM-PM coupling in amplifier.  The Barkhausen condition turns the gain fluctuation $\eta$ into amplitude fluctuation, and in turn to the phase fluctuation $\psi''$.  The latter adds to the phase fluctuation $\psi'$.}
\label{fig:ele-AM-PM-cpl-amp}
\end{figure}

We turn our attention to the AM-PM noise coupling mechanism\sidenote{Completely rewritten}  shown in Fig.~\ref{fig:ele-AM-PM-cpl-param}.  Noise modulates the gain.  Yet the Barkhausen condition forces the loop gain to be equal to one through the gain-compression mechanism.  The consequence is that the gain fluctuation is transformed into a fluctuation of the oscillation amplitude, and in turn into a fluctuation of the amplifier phase.  The conclusion is that the phase $\psi$ seen by the Leeson effect is the sum of \emph{two contributions}, the first comes straight from the amplifier, and the second results from the effect on the fluctuating amplitude.  The detailed model that follows is shown in Figures~\ref{fig:ele-AM-PM-cpl-param} and \ref{fig:ele-AM-PM-cpl-amp}, and discussed underneath.

For the sake of simplicity, we assume that the oscillator is tuned at the exact natural frequency of the resonator, and we assume that the amplifier is perturbed by one dominant source of noise. 
These hypotheses give a realistic picture of the oscillator.

Denoting with $x(t)\leftrightarrow X(s)$ the near-dc noise process, and introducing the modulation efficiency $w_\eta$ and $w_\psi$, the amplifier gain is perturbed by a factor $(1+w_\eta x)e^{jw_\psi x}$. 
Accounting for compression and neglecting the second-order terms, the complete expression of the gain is  
\begin{align*}
A = \bigl[1-\gamma(u-1)+w_\eta x\bigr] \, e^{jw_\psi x}~.
\end{align*}
The stationary oscillation is ruled by the Barkhausen condition $A\beta=1$.  With the normalization $\beta(\omega_0)=1$, this implies that $|A|=1$.  There follows that the instantaneous gain fluctuation $\eta(t)$ cannot increase $|A|$.   Instead, $\eta(t)$ causes the oscillation amplitude $u(t)$ to change from $1$ to $1+\eta(t)/\gamma$, as shown in the upper plot of Fig.~\ref{fig:ele-AM-PM-cpl-amp}.
In this condition the amplifier introduces a phase term $\psi''\propto\eta/\gamma$, which adds to the `simple' phase noise $\psi'$ of the amplifier.  The superscripts `prime' and `second' are introduced in order to keep the symbols $\psi$ and $\varphi$ for the overall phase fluctuation.
Therefore, the phase fluctuation seen by the Leeson effect (Sec.~\ref{sec:ele-leeson}) is $\psi=\psi'+\psi''$.
After \req{eqn:ele-leeson-effect}, the first phase-noise contribution is 
\begin{align*}
\Phi'(s) &= w_\psi \, \frac{s+1/\tau}{s} \, X(s)~.
\end{align*}
By inspection on Fig.~\ref{fig:ele-AM-PM-cpl-param}, the second phase noise contribution is 
\begin{align*}
\Phi''(s) &= w_\gamma w_\eta \, \frac{s+1/\tau}{s+\gamma/\tau} \, \frac{s+1/\tau}{s} \, X(s)~.
\end{align*}
Since $\Phi'(s)$ and $\Phi''(s)$ depend deterministically on $X(s)$, they must be added.  Thus, the noise transfer function 
\begin{align}
\mathrm{H}^{(c)}(s) &= \frac{\Phi(s)}{X(s)} = \frac{\Phi'(s)+\Phi''(s)}{X(s)} 
\end{align}
is 
\begin{align}
\mathrm{H}^{(c)}(s) &=
w_\psi \left[ 1 + \frac{w_\gamma w_\eta}{w_\psi} \, \frac{s+1/\tau}{s+\gamma/\tau} \right] 
\frac{s+1/\tau}{s} \, X(s)~.
\label{eqn:ele-cpl-H-X}
\end{align}
The term `1' is the simple Leeson effect as introduced in Section~\ref{sec:ele-leeson}, but for the trivial factor $w_\psi$ introduced because \req{eqn:ele-cpl-H-X} refers to the near-dc process $x\leftrightarrow X$ instead of to the phase $\psi\leftrightarrow\Psi$.  The term $\smash{\frac{w_\gamma w_\eta}{w_\psi}}$ is the phase fluctuation induced by AM noise.  

\begin{figure}[t]
\centering\namedgraphics{scale=0.6}{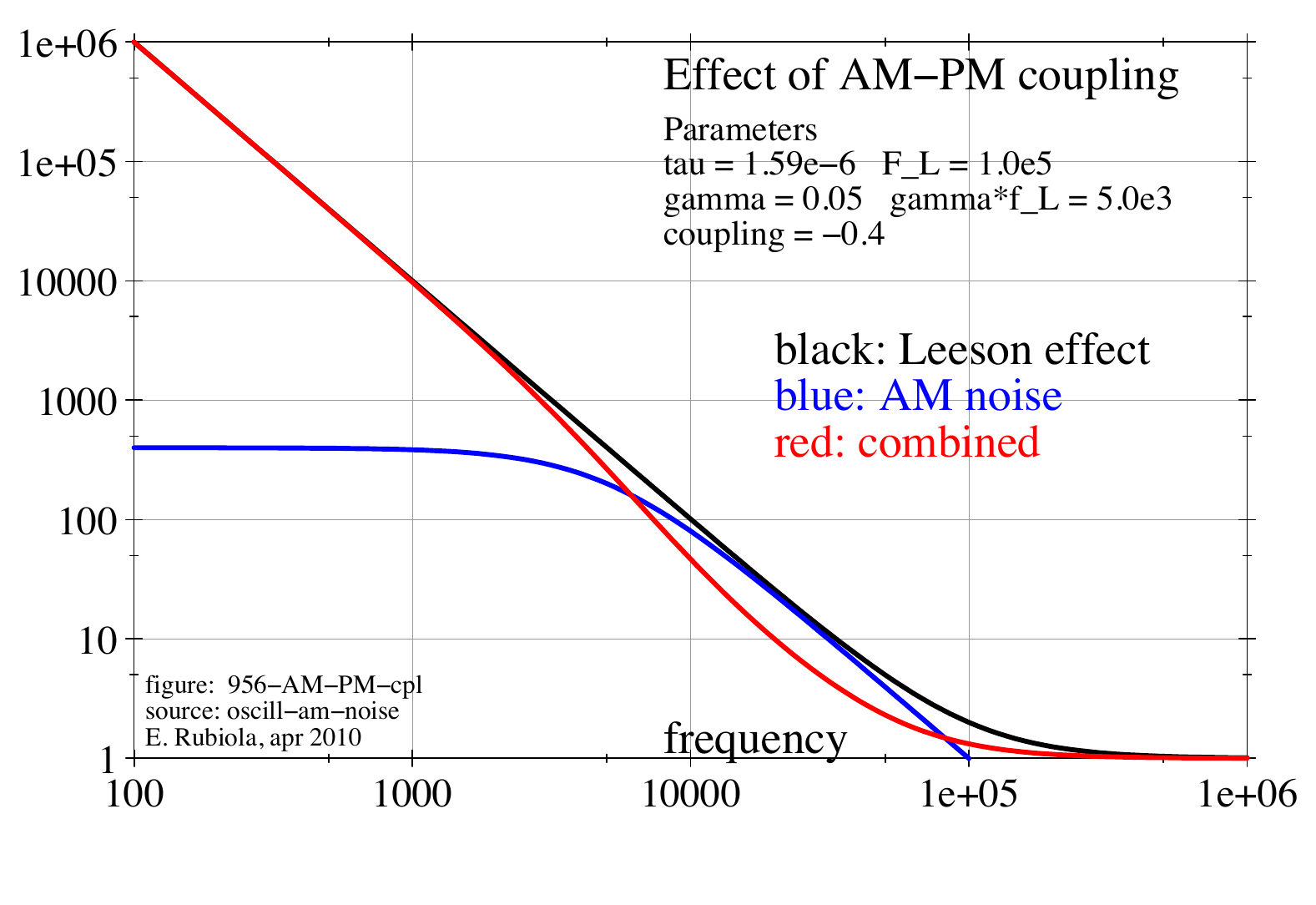}{\columnwidth}
\caption{AM-PM coupling due to parametric noise.  The `coupling' parameter is equal to $w_\gamma w_\eta/w_\psi$.}
\label{fig:956-AM-PM-cpl}
\end{figure}

\begin{figure}[t]
\centering\namedgraphics{scale=0.6}{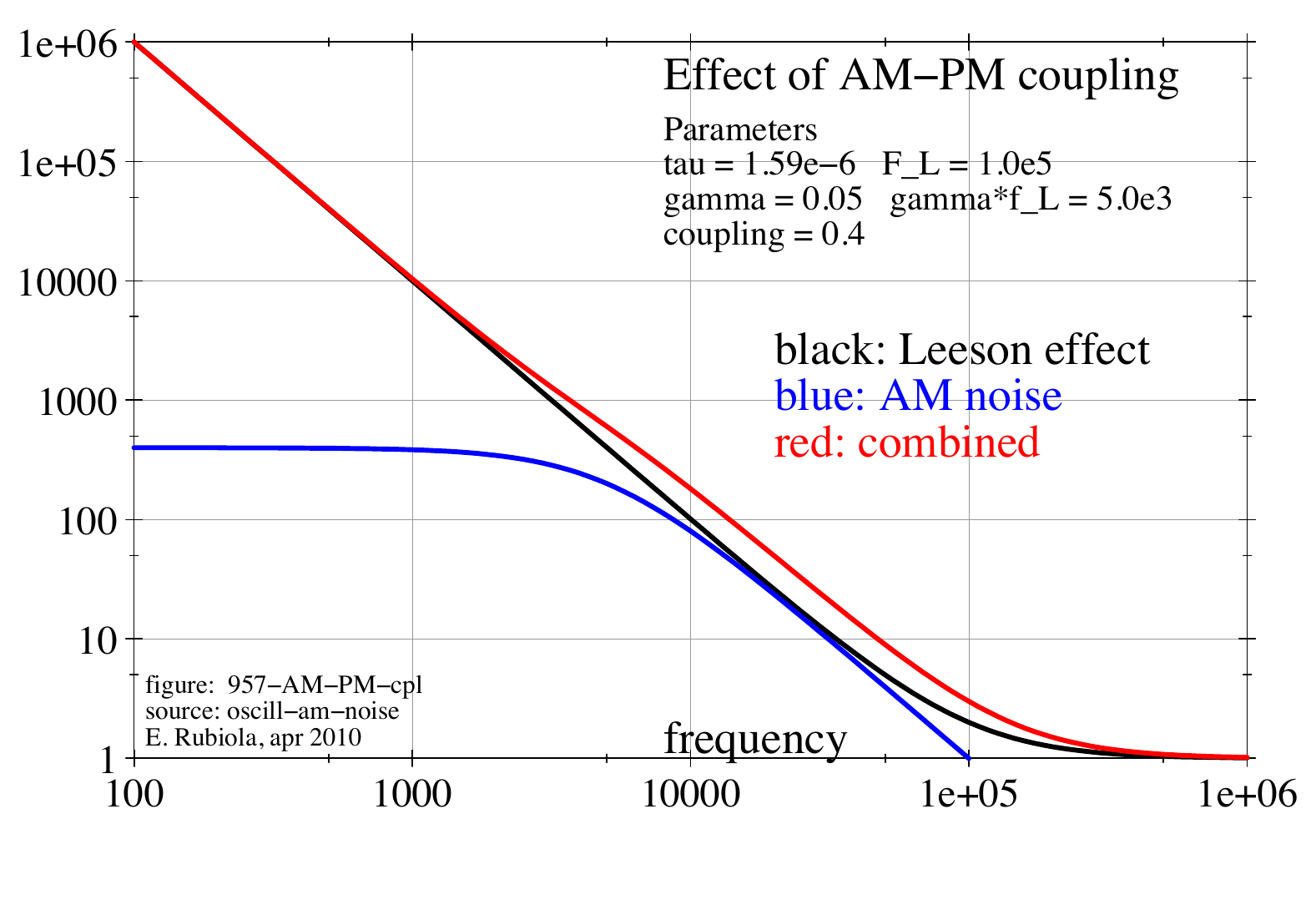}{\columnwidth}
\caption{AM-PM coupling due to parametric noise.  The `coupling' parameter is equal to $w_\gamma w_\eta/w_\psi$.}
\label{fig:957-AM-PM-cpl}
\end{figure}

Figures \ref{fig:956-AM-PM-cpl} and \ref{fig:957-AM-PM-cpl} show the noise transfer function $|\mathrm{H}^{(c)}(f)|^2$ in two cases.  The \emph{signature} of the AM-PM coupling shows up in the frequency range between $\gamma f_L$ and $f_L$.  In this region, the plot is parallel to that of the simple Leeson effect.  Interestingly, the AM-PM coupling can either increase or reduce the noise in the region between $\gamma f_L$ and $f_L$.   Of course, the phase-noise plot accounts for the slope of the near-dc process $X$ by which the transfer function is multiplied.  Thus for example the same signature can be seen in the $1/f^3$ region if $X$ is flicker noise.


%==================================================
\section{Extended Leeson effect for the delay-line oscillator}\label{sec:ele-dly-line}
%==================================================
\begin{figure}[t]
\centering\namedgraphics{scale=0.6}{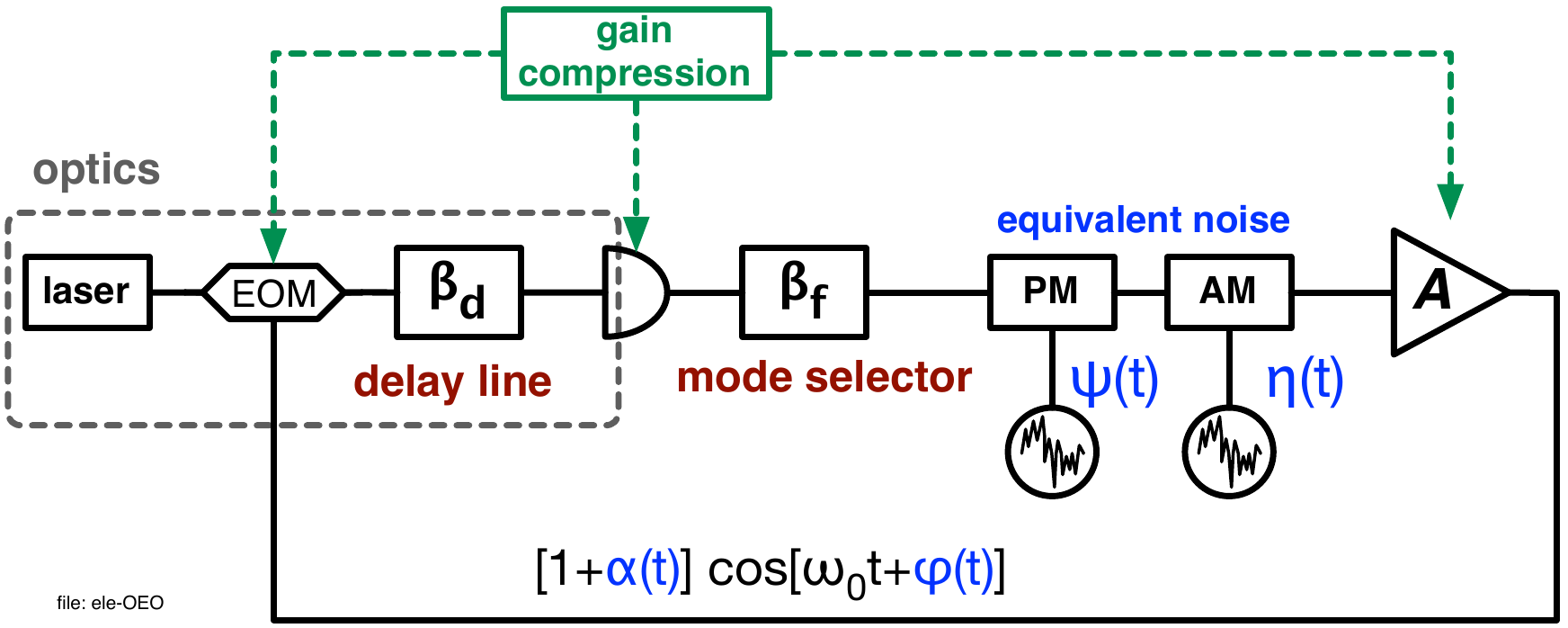}{\columnwidth}
\caption{Opto-electronic oscillator.  The equivalent noise originates in both electronic and optical path.  Gain compression takes place either in the amplifier, in the photodetector or in the electro-optic intensity modulator.}
\label{fig:ele-OEO}
\end{figure}
The delay-line oscillator is a variant of the oscillator in which the resonator is replaced by a delay line of\sidenote{Rewritten from the scratch} delay $\tau_d$, so that the oscillation frequency is an integer multiple of $1/\tau_d$.  To the extent of the Leeson effect, the delay line is equivalent to a resonator of quality factor $Q=\pi\nu_0\tau_d$ because the slope $d\arg(\beta)/d\omega$ is the same.  Of course, longer delay gives access to lower phase noise and higher frequency stability, provided the delay be stable.  For this reason the modern version of the delay-line oscillator, called OEO \cite{Yao1996josab-oeo,volyanskiy2008josab-optical-fiber} and shown in Fig.~\ref{fig:ele-OEO}, makes use of an optical fiber as the delay unit.  The optical fiber exhibits high thermal stability ($6.85{\times}10^{-6}$/K) and low loss (0.15 dB/km at 1.55 $\mu$m wavelength, equivalent to 0.03 dB/$\mu$s), limited by the Rayleigh scattering.  Other implementations are possible, based on a surface-wave devices and on electrical lines.

The phase noise theory of the delay-line oscillator is widely discussed in \cite[Chap\,5]{Rubiola2008cambridge-leeson-effect}.  Here we give some key element to extend the theory to AM noise and to the AM-PM noise coupling.  

Since the delay line is  a wide-band device, the loop can sustain oscillations at any frequency multiple of $1/\tau_d$.  A mode-selector filter is therefore necessary to choose one frequency by lowering the loop gain at the other frequencies.  For this reason the feedback function of Fig.~\ref{fig:ele-oscill-models} is split into delay and filter, denoted with the subscripts $d$ and $f$, respectively.  It is important to understand that group delay of the mode selector must be orders of magnitude shorter than the delay of the line because the sensitivity to environment parameters is weighted proportionally to the delay.

\begin{figure}[t]
\centering\namedgraphics{scale=0.84}{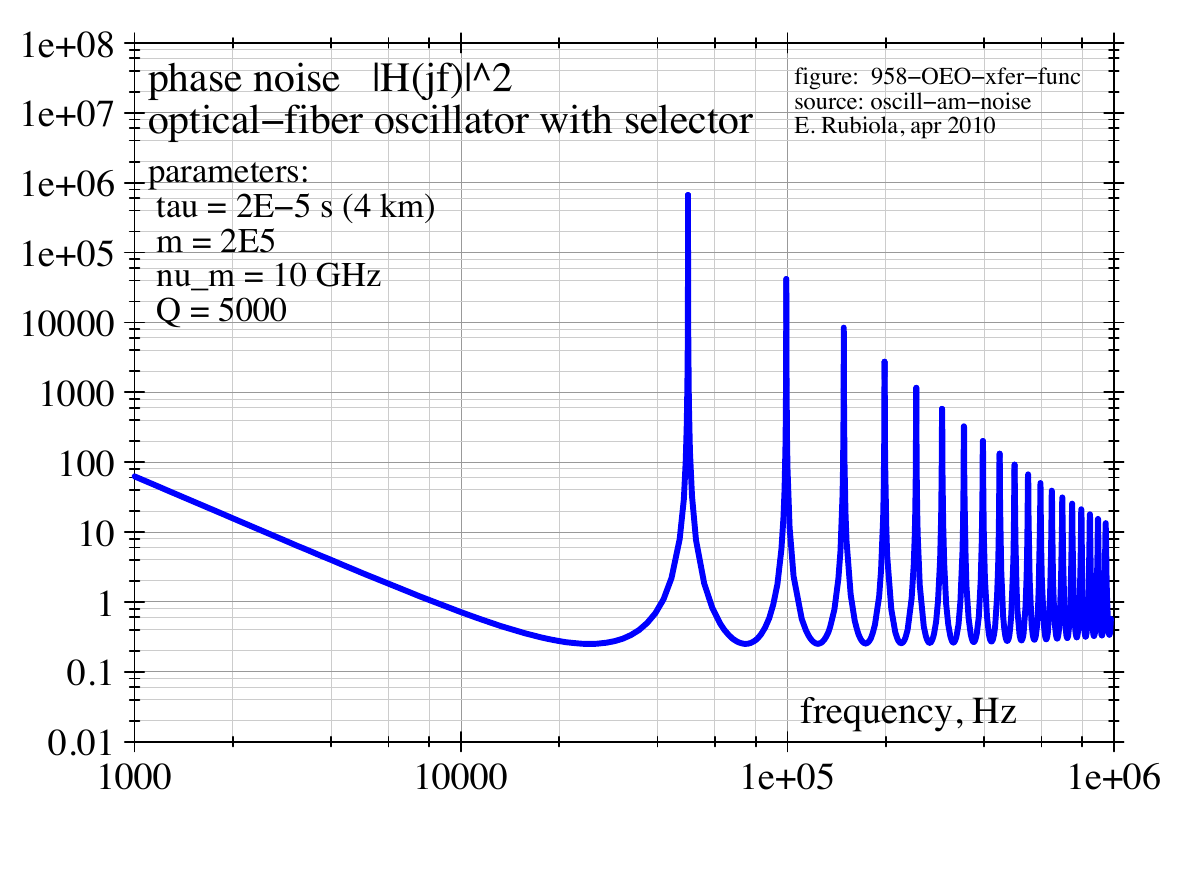}{\columnwidth}\vspace*{-2em}
\caption{Noise transfer function of the opto-electronic oscillator, either for AM noise or PM noise. The parameter $m$ is the mode order.  Hence with 20 $\mu$s delay, the mode of order $2{\times}10^5$ falls at $\nu_0=10$ GHz.}
\label{fig:958-OEO-xfer-func}
\end{figure}

There are practical reasons to use a resonator as the mode selector.  We assume that the delay-line attenuation is independent of frequency, moving the flatness defect to the resonator transfer function.  
Using the elementary theory of the Laplace transform and the material developed in Section \ref{sec:ele-resonator}, the slow-varying envelope representation of the feedback path is 
\begin{align}
\mathrm{b}_f(t)=\frac{1}{\tau_f}\:e^{-t/\tau_{f}}
&\quad\leftrightarrow\quad
\mathrm{B}_f(s)=\frac{1/\tau_f}{s+1/\tau_f}\\[1ex]
\mathrm{b}_d(t)=\delta(t-\tau_d)
&\quad\leftrightarrow\quad
\mathrm{B}_d(s)=e^{-s\tau_d}\\[1ex]
\mathrm{b}(t)=\mathrm{b}_f(t)*\mathrm{b}_d(t)
&\quad\leftrightarrow\quad
\mathrm{B}(s)=\mathrm{B}_f(s) \, \mathrm{B}_d(s)~,
\end{align}
thus
\begin{align}
\mathrm{B}(s)=\frac{1/\tau_f}{s+1/\tau_f}\:e^{-s\tau_d}~.
\end{align}
Inserting such function in the phase-noise feedback loop we get 
\begin{align}
\mathrm{H}(s)&=\frac{1}{1-\mathrm{B}(s)}
\intertext{and therefore}
\mathrm{H}(s)&=\frac{s+1/\tau_f}{s+(1-e^{-s\tau_d})/\tau_f}~,
\end{align}
an example of which is shown in Fig.~\ref{fig:958-OEO-xfer-func}.

The full extension to AM noise, as derived in Section \ref{sec:ele-extended-leeson} for the oscillator based on a simple resonator, takes cumbersome and tedious algebra.  Yet at low frequencies, below the Leeson frequency, the asymptotic approximation of the delay line is a resonator of quality factor $Q=\pi\nu_0\tau_d$.  This simplification gives account for the low-frequency behavior, and at least a qualitative prediction for the AM noise peaks.

%---------------------------------------------------------------------
\subsection{The impact of the laser RIN}
%---------------------------------------------------------------------
Let us start with the analysis of Figure \ref{fig:ele-OEO} in \emph{open loop} conditions.  The path from the amplifier to the EOM is broken.  First, we observe that the bandpass filter, however large, eliminates the harmonics at frequencies multiple of $\omega_0$.  Thus the light power at the photodetector input can be described by
\begin{align}
P_\lambda(t)=\overline{P}_\lambda\bigl[1+m\cos(\omega_0t)\bigr]~,
\end{align}
where $m=J_1(z)$ is the modulation index, $J_1(z)$ is the first-order Bessel function of the first kind, and $z$ is proportional to the microwave voltage at the input of the intensity modulator \cite{Rubiola2005josab-delay-line}.  Though the theoretical maximum is $m\simeq1.164$, in practice we get at most $m=1$.
The current at the photodetector input is 
\begin{align}
I(t) &= \rho P_\lambda(t)\\
&= \rho\overline{P}_\lambda\bigl[1+m\cos(\omega_0t)\bigr]~,
\end{align}
where $\rho=\smash{\frac{\eta q}{\hbar\nu_\lambda}}$ is the photodetector responsivity, $\eta$ the quantum efficiency, and $\hbar\omega_\lambda$ the photon energy.  Assuming a quantum efficiency of 0.6, the responsivity is of 0.75 A/W at 1.55 $\mu$m wavelength, and of 0.64 A/W at 1.31 $\mu$m.  Filtering out the dc component, the rms voltage across a resistor $R_0$ at the photodetector output is 
\begin{align}
v &= \frac{1}{\sqrt{2}} \, R_0 \, \rho \, m \, \overline{P}_\lambda 
\qquad\text{(rms voltage at the photodetector out)}
\label{eqn:ele-photodet-rms-volt}
\end{align}
The path from the EOM to the microwave amplifier output of Fig.~\ref{fig:ele-OEO} can be seen as an `amplifier,' plus a filter function.  The `amplifier' includes intensity modulator, photodetector, and microwave amplifier.  By virtue of \req{eqn:ele-photodet-rms-volt} the laser power affects the gain, thus the relative intensity noise (RIN) makes the gain fluctuate.  Since in open loop $\eta=\delta v/v$, the laser RIN induces a gain fluctuation 
\begin{align*}
\eta&=\frac{\delta\overline{P}_\lambda}{\overline{P}_\lambda}
&S_\eta^{(\text{RIN})}(f) &= S_\text{RIN} 
\end{align*}
Closing the loop, the results of Section \ref{sec:ele-extended-leeson} (AM noise) and Section \ref{sec:ele-ampli-ampm-correl} (AM-PM coupling) apply.

Interestingly, the RIN of some lasers does not follow the polynomial law.  Instead, slopes of $-5$ dB/dec and $-15$ dB/dec appear in the spectrum.  At the beginning of he process of collecting data from the literature, we suspect that this is typical of the distributed-feedback laser.  Anyway, regardless of the physical explanation beyond, the presence of $-5$ dB/dec and $-15$ dB/dec slopes in the RIN spectrum in conjunction with the Leeson effect could explain the slope of $-35$ dB/dec and $-25$ dB/dec observed in the phase noise spectrum of some oscillators.

%\begin{figure}[t]
%\centering\namedgraphics{scale=0.84}{959-rin-em4ff}{\columnwidth}
%\caption{FIGURE RIN}
%\label{fig:959-rin-em4ff}
%\end{figure}

%==================================================
%\section*{\color{blue}Acknowledgements}\addcontentsline{toc}{section}{Acknowledgements}
%==================================================

\clearpage
%==================================================
\appendix
\section{Exotic issues}\label{sec:ele-exotic}
%==================================================
\textsf{\bfseries This Appendix is not a finished work.  We report on some facts intended to be the seed for further analysis.}

%----------------------------------------------------------------------------
\subsection{AM-PM coupling in the off-resonance resonator}
%----------------------------------------------------------------------------
It has been shown in Section \ref{sec:ele-resonator} that the resonator operated at the exact natural frequency responds to a phase perturbation with a decaying exponential of phase, with no effect on the amplitude; and that it responds to an amplitude perturbation with a decaying exponential of amplitude, with no effect on the phase.  It has also been shown that cross terms appear off the resonance, for the resonator response is described by $[\mathrm{b}(t)]\leftrightarrow[\mathrm{B}(s)]$ \req{eqn:eln-resonator-b-matrix}--\req{eqn:eln-resonator-B-matrix}.  

\begin{figure}[b]
\centering\namedgraphics{scale=0.6}{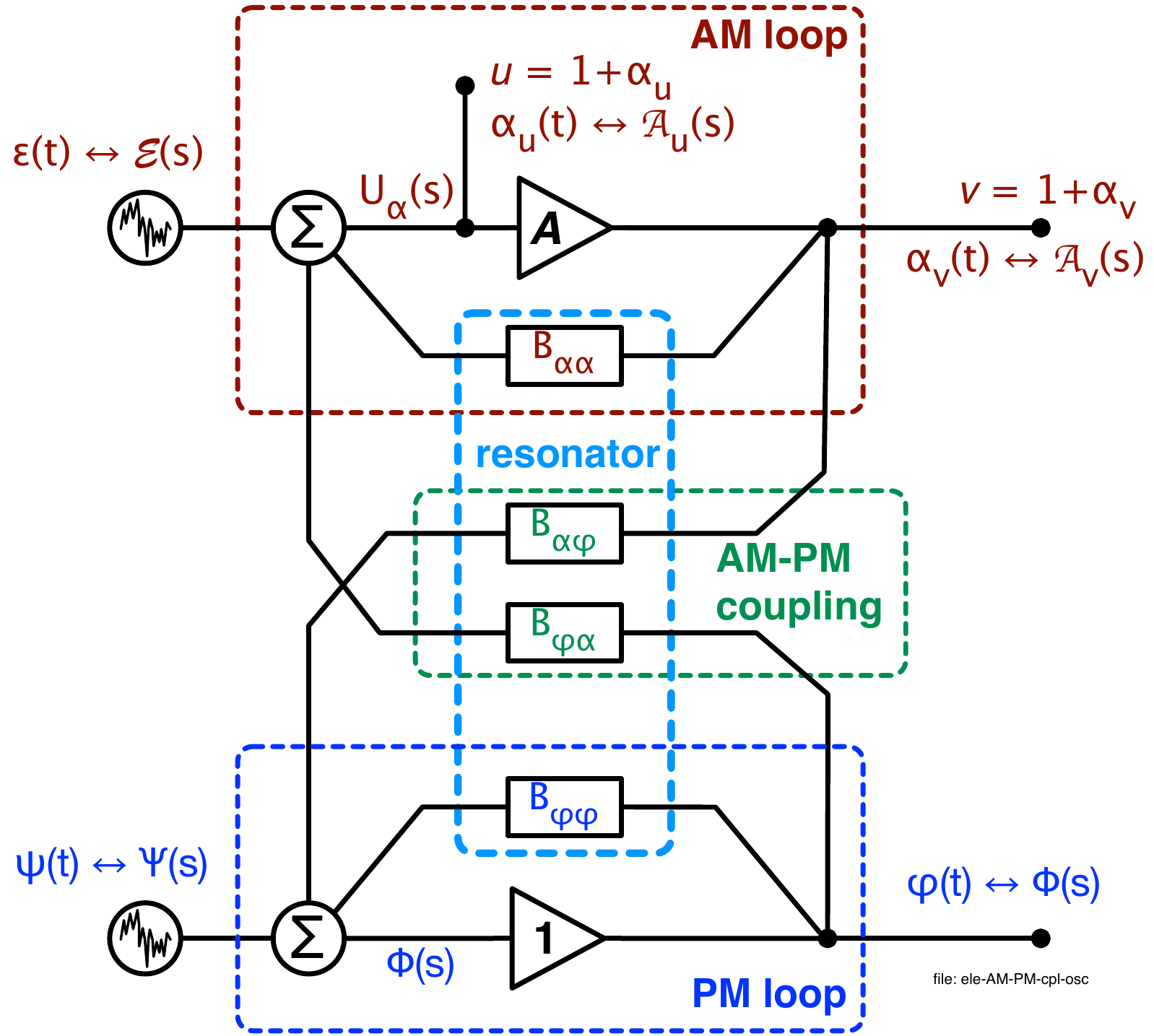}{\columnwidth}
\caption{Detuning the resonator results in coupling AM to PM.}
\label{fig:ele-AM-PM-cpl-osc}
\end{figure}

The AM-PM coupling inside the resonator yields naturally to the oscillator model depicted in Fig.~\ref{fig:ele-AM-PM-cpl-osc}.
In this figure, the symbol $s$ (expressed or implied) cannot be the Laplace complex variable because the system is nonlinear.  Instead, $s$ is to be interpreted as the derivative operator $\frac{d}{dt}$, which is allowed.
Hence, the simplest approach is to derive the resonator equations using the Laplace formalism, and then to convert these equations into regular differential equations by replacing $s\rightarrow\frac{d}{dt}$.  

The lower loop of Fig.~\ref{fig:ele-AM-PM-cpl-osc} yields the Leeson effect, as described in Sec.~\ref{sec:ele-leeson}.   The upper loop models the amplitude noise as discussed in Sec.~\ref{sec:ele-am-model}-\ref{sec:ele-extended-leeson}.  The two loops are coupled by the terms $B_{\alpha\varphi}$ and $B_{\varphi\alpha}$, which are nonzero when the resonator is off resonance.
By inspection on Fig.~\ref{fig:ele-AM-PM-cpl-osc} we get
\begin{align*}
U_\alpha&=\mathcal{E}+AB_{\alpha\alpha}U_\alpha+B_{\varphi\alpha}\Phi\\[0.5ex]
\Phi&=\Psi+B_{\varphi\varphi}\Phi+AB_{\alpha\varphi}U_\alpha~,
\end{align*}
which can be rewritten as
\begin{align*}
U_\alpha&=\frac{1}{1-AB_{\alpha\alpha}}\mathcal{E}
	+\frac{B_{\varphi\alpha}}{1-AB_{\alpha\alpha}}\Phi\\[1ex]
\Phi&=\frac{1}{1-B_{\varphi\varphi}}\Psi
	+\frac{AB_{\alpha\varphi}}{1-B_{\varphi\varphi}}U_\alpha~.
\end{align*}
Combining the above equations, we get
\begin{align*}
U_\alpha&=\frac{1}{1-AB_{\alpha\alpha}}\mathcal{E}
    +\frac{B_{\varphi\alpha}}{(1-AB_{\alpha\alpha})(1-B_{\varphi\varphi})}\Psi
    +\frac{AB_{\varphi\alpha}B_{\alpha\varphi}}{
        (1-AB_{\alpha\alpha})(1-B_{\varphi\varphi})}U_\alpha\\[1ex]
\Phi&=\frac{1}{1-B_{\varphi\varphi}}\Psi
    +\frac{AB_{\alpha\varphi}}{(1-AB_{\alpha\alpha})(1-B_{\varphi\varphi})}\mathcal{E}
    +\frac{AB_{\alpha\varphi}B_{\varphi\alpha}}{
        (1-AB_{\alpha\alpha})(1-B_{\varphi\varphi})}\Phi~, 
\end{align*}
hence
\begin{align*}
U_\alpha&=\frac{(1-AB_{\alpha\alpha})(1-B_{\varphi\varphi})}{%
    (1-AB_{\alpha\alpha})(1-B_{\varphi\varphi})-AB_{\varphi\alpha}B_{\alpha\varphi}}
    \left[ \frac{\mathcal{E}}{1-AB_{\alpha\alpha}} 
        +\frac{B_{\varphi\alpha}\,\Psi}{
        (1-AB_{\alpha\alpha})(1-B_{\varphi\varphi})}\right]\\[1ex]
\Phi&=\frac{(1-AB_{\alpha\alpha})(1-B_{\varphi\varphi})}{%
    (1-AB_{\alpha\alpha})(1-B_{\varphi\varphi})-AB_{\alpha\varphi}B_{\varphi\alpha}}
    \left[ \frac{AB_{\alpha\varphi}\,\mathcal{E}}{
        (1-AB_{\alpha\alpha})(1-B_{\varphi\varphi})}
        +\frac{\Psi}{1-B_{\varphi\varphi}}\right]
\end{align*}
and finally\vspace{0.5ex}%
\begin{align}
\begin{bmatrix}U_\alpha\\\Phi\end{bmatrix}
&=
\frac{1}{%
    (1-AB_{\alpha\alpha})(1-B_{\varphi\varphi})-AB_{\varphi\alpha}B_{\alpha\varphi}}
\begin{bmatrix}
1-B_{\varphi\varphi}&B_{\varphi\alpha}\\
AB_{\alpha\varphi}&1-AB_{\alpha\alpha}
\end{bmatrix}    
\begin{bmatrix}\mathcal{E}\\\Psi\end{bmatrix}~.
\end{align}\vspace{0.5ex}%
The above equation is let in closed form for further analysis.  The following formulae will be useful
\begin{align*}
1-AB_{\alpha\alpha}
&=\frac{s^2+\frac{1}{\tau}(2-A)s+(1-A)\left(\frac{1}{\tau^2}+\Omega^2\right)}{%
    s^2+\frac{2}{\tau}s+\frac{1}{\tau^2}+\Omega^2}\\[1ex]
1-B_{\varphi\varphi}
&=\frac{s^2+\frac{1}{\tau}s}{%
    s^2+\frac{2}{\tau}s+\frac{1}{\tau^2}+\Omega^2}\\[1ex]  
\mathrm{B}_{\alpha\varphi}=\mathrm{B}_{\varphi\alpha}
&=\frac{-\Omega s}{s^2+\frac{2}{\tau}s+\frac{1}{\tau^2}+\Omega^2}~.
\end{align*}

%----------------------------------------------------------------------------
\subsection{Parametric fluctuation of the \boldmath$S$ matrix}
%----------------------------------------------------------------------------
\begin{figure}[t]
\centering\namedgraphics{scale=0.60}{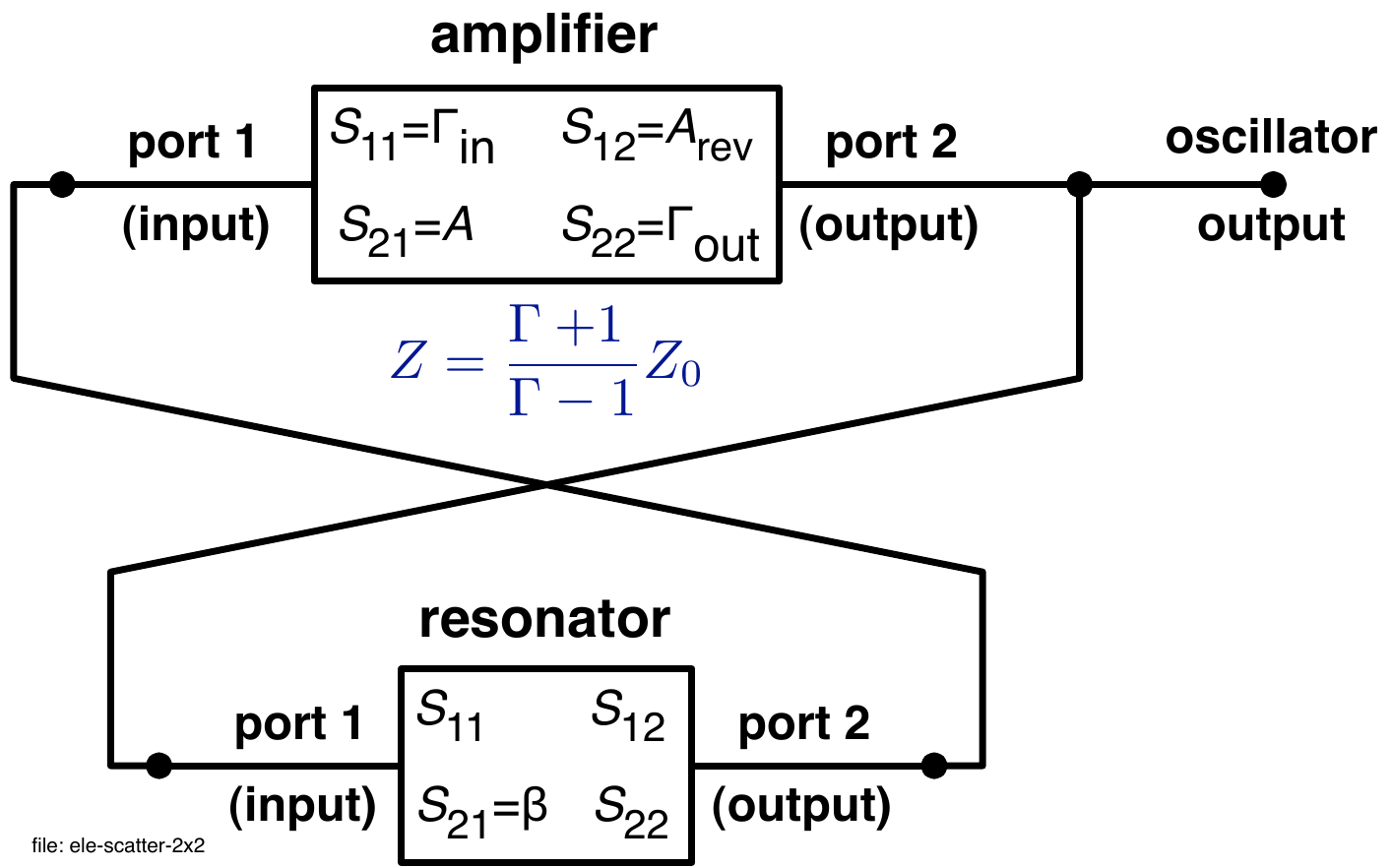}{\columnwidth}
\caption{In most oscillators, the amplifier and the resonator can be described in terms of scatter matrix.}
\label{fig:ele-scatter-2x2}
\end{figure}
All over this report, the oscillator loop is analyzed as a simple block diagram in which the signal flows in one direction only, and there is no interaction due to impedances.  Breaking this assumption, the amplifier and the resonator can be described in terms of the scatter matrix $S$.  

The case of the traditional microwave oscillator, where amplifier and resonator are described by a $2{\times}2$ matrix, is shown in Fig.~\ref{fig:ele-scatter-2x2}.  
The gain $A$ and the resonator transfer function $\beta(s)$ are the element $S_{21}$ of the respective matrix.  Hence, the amplifier AM and PM noise as introduced in the previous Sections, go in $\Re\{S_{21}\}$ and $\Im\{S_{21}\}$, respectively.
The finite isolation of the amplifier is represented as $S_{12}\neq0$.  This effect has little importance because the isolation ratio of actual amplifier is high enough for the reverse signal not to circulate in the loop.
Another effect is due to the amplifier input and output impedances, related to the scatter matrix by $Z=\frac{\Gamma+1}{\Gamma-1}Z_0$, $\Gamma_\text{in}=S_{11}$, and $\Gamma_\text{out}=S_{22}$.  The amplifier input and output impedances interact with the resonator parameters.  Thus, the fluctuations of $S_{11}$ and $S_{22}$ turn into frequency fluctuations.

\begin{figure}[t]
\centering\namedgraphics{scale=0.60}{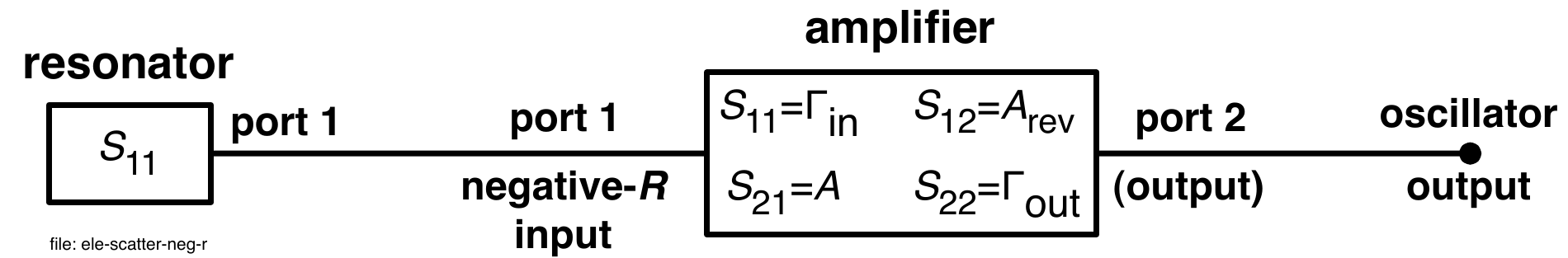}{\columnwidth}
\caption{Scatter matrix representation of the negative-resistance (dipolar) oscillator.}
\label{fig:ele-scatter-neg-r}
\end{figure}

The quartz oscillator and other negative-resistance oscillators can also be described with the scatter matrix formalism (Fig.~\ref{fig:ele-scatter-neg-r}).  In this case, the resonator degenerates into a single-value matrix.  The amplifier $S_{11}$ models the negative resistance that makes the system oscillate.  Strictly, only $S_{11}$ is necessary.  Yet, in most cases the amplifier also acts as a buffer of gain $S_{21}=A_\text{buf}$, reverse gain (isolation) $S_{12}=A_\text{rev}$, and output reflection coefficient $S_{22}=\Gamma_\text{out}$.

Of course, the scattering matrix formalism also apply to optics.
The laser (oscillator) differs from the above analysis in that the laser amplifier is bidirectional and the signal may be a stationary wave.

%----------------------------------------------------------------------------
\subsection{The Miller effect}
%----------------------------------------------------------------------------
\begin{figure}[t]
\centering\namedgraphics{scale=0.45}{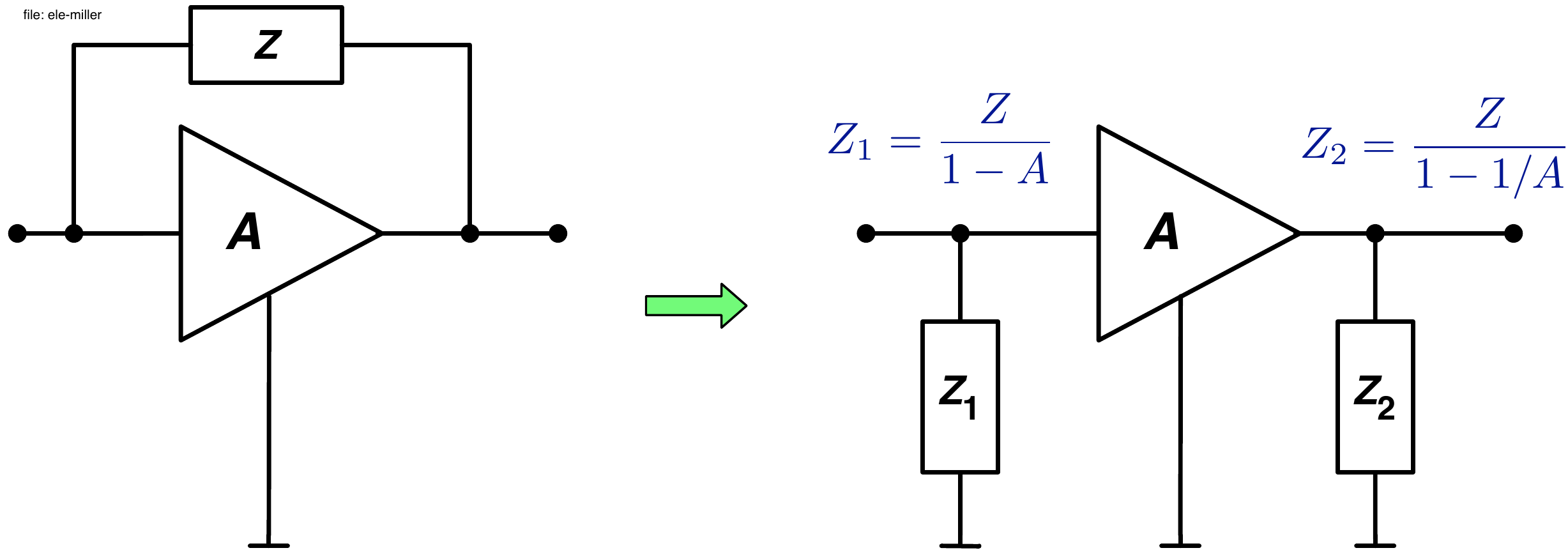}{\columnwidth}
\caption{Miller effect.}
\label{fig:ele-miller}
\end{figure}

The Miller theorem \cite{miller1920spbs} states that an impedance $Z$ in the feedback path of an amplifier of gain $A$ can be replaced by two impedances,
\begin{align*}
Z_1&=\frac{Z}{1-A}\quad\text{(input)}
\qquad\text{and}\qquad
Z_2=\frac{Z}{1-1/A}\quad\text{(output)}~,
\end{align*}
connected at the amplifier input output, respectively (Fig.~\ref{fig:ele-miller}).
For our purposes, the left-hand side of Fig.~\ref{fig:ele-miller} is formally equivalent to the oscillator loop, for we can identify $Z$ as the resonator in the feedback path, and $A$ as the sustaining amplifier.  

Unfortunately the Miller theorem cannot be inverted in the general case because it would be necessary to collapse three degrees of freedom ($A$, $Z_1$ and $Z_2$) into two degrees of freedom ($A$ and $Z$).  The parameters of the specific circuit are needed to get $Z$ from $Z_1$ and $Z_2$.
Nonetheless, the Miller theorem provides evidence that the gain fluctuations affect the impedances of the whole circuit, and that a fluctuating impedance at the amplifier input or output can be turned into a fluctuating impedance in parallel to $Z$, hence to the resonator.  In turn, the oscillator frequency fluctuates.

\clearpage
%==================================
\def\bibfile#1{/Users/rubiola/Documents/articles/bibliography/#1}
%\def\bibfile#1{/home/rubiola/docs/bib/#1}
%==================================
\addcontentsline{toc}{section}{References}
\bibliographystyle{./IEEEtranBST/IEEEtran}
%\bibliography{./IEEEtranBST/IEEEexample}
\bibliography{\bibfile{ref-short},%
              \bibfile{references},%
              \bibfile{rubiola}}

\end{document}